\documentclass[aps,pre,superscriptaddress,longbibliography,nofootinbib]{revtex4-2}

\usepackage[T1]{fontenc}
\usepackage{amsmath,amssymb,bm}
\usepackage{graphicx}
\usepackage{tikz}
\usepackage{booktabs}
\usepackage{etoolbox}
\usepackage[colorlinks=true,linkcolor=blue,citecolor=blue,pdfpagemode=UseNone,pdfstartview=FitH,pdfpagelayout=OneColumn]{hyperref}
\allowdisplaybreaks

\makeatletter
\def\paragraph{%
  \@startsection{paragraph}{4}{\z@}{1.6ex \@plus.5ex \@minus.2ex}{-1em}%
    {\normalfont\normalsize\itshape}%
}%
\makeatother

\newcommand{\lc}{\ln 2\cosh}
\newcommand{\SG}{\mathrm{SG}}
\newcommand{\FM}{\mathrm{FM}}
\newcommand{\PM}{\mathrm{PM}}
\newcommand{\Lam}{\Lambda}
\newcommand{\bc}{\beta_c}
\newcommand{\Mpt}{\mathrm{M}}
\newcommand{\jM}{j_{0\Mpt}}

\begin{document}

\title{Reentrance and temperature chaos in the $p$-spin Ising spin glass}

\author{Hidetoshi Nishimori}
\affiliation{Institute of Integrated Research, Institute of Science Tokyo, Nagatsuta-cho, Midori-ku, Yokohama 226-8501, Japan}
\affiliation{Graduate School of Information Sciences, Tohoku University, Sendai 980-8579, Japan}

\date{\today}

\begin{abstract}
Reentrant transitions and temperature chaos are two unusual properties of spin glasses. We recently established a general logical relation between these two phenomena [Phys.\ Rev.\ E \textbf{112}, 044140 (2025)]. In the present paper, we provide an explicit example of this logical relation in the fully connected Ising $p$-spin glass with a ferromagnetic bias. By expanding the free energy around the triple point, we show analytically that the ferromagnetic--spin glass boundary is reentrant in the vicinity of the triple point throughout the examined range of finite $p>2$. It then follows from the above logical relation that there exists at least one pair of distinct temperatures in the spin glass phase for which the overlap of spin configurations vanishes. This is a necessary condition for temperature chaos, but it would be quite unusual for the overlap to vanish only for selected temperature pairs but not for others within a single spin glass phase. These results therefore strongly suggest the existence of temperature chaos in the sense that the overlaps of spin configurations vanish for all temperature pairs throughout the spin glass phase. Independent evidence for this behavior is provided by a two-temperature replica calculation.
\end{abstract}

\maketitle

\section{Introduction}\label{sec:intro}

Spin glasses are among the most challenging problems in statistical physics, and their applications extend far beyond the original context of disordered magnets \cite{Edwards1975,Sherrington1975,Parisi1980,Talagrand2006,Nishimori2001,Charbonneau2023}. The theoretical framework has also become an important tool for problems beyond the traditional boundaries of physics \cite{Nishimori2001,Charbonneau2023,Zdeborova2016}.

The present paper concerns two unusual properties of spin glasses: reentrant transitions and temperature chaos. A reentrant transition occurs when a phase boundary bends backward as the temperature is lowered. Near the ferromagnetic--spin glass phase boundary, this means that cooling can destroy ferromagnetic long-range order and drive the system into the less ordered spin glass phase. This behavior is counterintuitive because cooling usually promotes order. Nevertheless, numerical studies of the Edwards--Anderson model \cite{Edwards1975} have found evidence of reentrance in finite dimensions \cite{Nobre2001,Wang2003,Amoruso2004,Hasenbusch2007,Toldin2009,Ceccarelli2011,Thomas2011,Liu2025}.

Temperature chaos is a different type of unusual behavior. A small change of temperature can lead to a completely different equilibrium spin configuration, so that typical configurations at two distinct temperatures become uncorrelated in the thermodynamic limit for all temperature pairs in the spin glass phase. In other words, the distribution function of the equilibrium spin-state overlap at two temperatures becomes a delta function concentrated at the origin, that is, at zero overlap. This phenomenon has been predicted and studied by droplet, scaling, mean-field, and numerical approaches \cite{Bray1987,Banavar1987,Fisher1986,Fisher1988,Kondor1989,Ney-Nifle-1997,Ney-Nifle1998,Parisi2010,Mathieu2001,Bouchaud2001,Aspelmeier2002,Rizzo2003,Houdayer2004,Katzgraber2007b,Fernandez-2013,Wang2015,Billoire2018,Baity-Jesi2021}. Its numerical detection is difficult because the asymptotic behavior often appears only at very large system sizes. Rigorous results have also been obtained for certain mixed $p$-spin models under suitable conditions \cite{Chen2013,Panchenko2016}, although these results do not directly apply to the pure $p$-spin spin glass studied in the present paper. Reentrance and temperature chaos therefore arise in rather different settings and have mostly been studied as separate problems, with the exception discussed below.

A direct logical relation between the two was recently established by Nishimori, Ohzeki, and Okuyama \cite{Nishimori2025a} (NOO hereafter). Using a model with correlated disorder \cite{Nishimori2024}, they proved that if a spin glass phase exists at finite temperature and temperature chaos is absent in that phase, the ferromagnetic (FM)--spin glass (SG) boundary cannot be reentrant. The contrapositive of this statement is that if the FM--SG boundary is reentrant, then either no finite-temperature SG phase exists or there is at least one pair of distinct temperatures in the SG phase for which the distribution of the overlap of spin configurations is a delta function at the origin, a necessary condition for temperature chaos. Thus, a reentrant property of the phase diagram imposes a constraint on the relation between equilibrium states at different temperatures.

The converse is not necessarily true: temperature chaos does not by itself imply reentrance. The Sherrington--Kirkpatrick (SK) model provides an example in which temperature chaos exists \cite{Rizzo2003}, whereas the boundary of the ferromagnetic phase is not reentrant \cite{Toulouse1980,GabayToulouse1981}. For the contrapositive of the NOO proposition to have nontrivial content, one needs a standard model of spin glasses in which a reentrant FM--SG boundary can be established with sufficient analytic control, so that a necessary condition for temperature chaos follows from reentrance. The fully connected Ising $p$-spin glass with a ferromagnetic bias introduced by Nishimori and Wong \cite{Nishimori1999} is a natural candidate. Its FM--SG boundary below the triple point, at which the paramagnetic, FM, and SG phases meet, had been drawn as almost vertical \cite{Nishimori1999,Gillin2001}, and both reentrance and temperature chaos have in fact been observed in this model for $p=3$ by Krzakala and Zdeborov\'a \cite{KrzakalaZdeborova2010EPL,ZdeborovaKrzakala2010PRB} in a study whose main subject was the adiabatic evolution of Gibbs states. They developed a generalization of the cavity method that follows a Gibbs state as the temperature is changed slowly (the state-following method), and showed at the level of one-step replica symmetry breaking (1RSB) that temperature chaos exists in the model without ferromagnetic bias: below the static SG transition temperature, the states that dominate the equilibrium distribution at one temperature are overtaken in free energy by other states as soon as the temperature is changed.

They also pointed out a correspondence with the biased model. Between the dynamical transition temperature and the static SG transition temperature, the model without bias is paramagnetic in the thermodynamic sense, but its equilibrium distribution is already split into exponentially many Gibbs states; if a state is taken from equilibrium at a temperature $T_e$ in this range, including the static transition temperature itself, following it is equivalent, through a gauge transformation, to analyzing the ferromagnetic state of the biased model at the point of the Nishimori line (NL) \cite{Nishimori1981,Nishimori2001} with temperature $T_e$ and then changing the temperature at fixed bias. In the numerical solution of the 1RSB equations for the biased model, they noticed that the FM--SG boundary below the triple point bends slightly toward larger bias as the temperature is lowered, and interpreted this small reentrance as a consequence of the level crossings: the state in equilibrium at the triple point is overtaken by other states upon cooling, so that at the bias of the triple point the ferromagnetic phase gives way to the SG phase \cite{ZdeborovaKrzakala2010PRB}. They further remarked that chaos disappears as $p\to\infty$, where the boundary becomes strictly vertical.

The present paper treats the same model from the opposite direction and for a different purpose. We first establish the reentrance analytically by expanding the free energies of the FM and SG phases around the triple point; this quantifies the observation of Ref.~\cite{ZdeborovaKrzakala2010PRB} for $p=3$ and extends it to more general values of $p$. We then derive a necessary condition for temperature chaos from the reentrance through the contrapositive of the NOO proposition. The argument from chaos to reentrance in Ref.~\cite{ZdeborovaKrzakala2010PRB} relies on two ingredients. One is the above mapping of state following onto the biased model, which is valid only for states taken from equilibrium in this paramagnetic temperature range, although such states can then be followed to lower temperatures. The other is the first-order nature of the FM--SG transition, with a finite jump of the magnetization at the triple point. The SK model shows that this direction of argument does not hold in general. The NOO proposition, in contrast, is based on exact gauge identities and requires neither a mean-field treatment nor state following; it applies equally to finite-dimensional Ising spin glasses with the gauge-compatible disorder distributions considered in its derivation. The two routes are therefore complementary: Refs.~\cite{KrzakalaZdeborova2010EPL,ZdeborovaKrzakala2010PRB} computed the level crossings directly and read the shape of the boundary from them, whereas we establish the shape of the boundary and infer a constraint on the two-temperature overlap from it. The two conclusions agree for $p=3$, and the two-temperature replica calculation presented later, which is independent of both routes, is consistent with the level crossings of Refs.~\cite{KrzakalaZdeborova2010EPL,ZdeborovaKrzakala2010PRB} and extends the evidence for chaos to $p$ up to $12$.

A related situation is found on hierarchical lattices, where the Migdal--Kadanoff renormalization recursion relation is exact. Reentrant phase diagrams \cite{Nobre2001,Hinczewski2005,Guven2008} and temperature chaos \cite{Banavar1987,Ney-Nifle-1997,Aspelmeier2002,Sasaki2003} have both been found there, although in separate numerical studies of the recursion relations, and the NOO proposition applies to that case as well, as discussed in Sec.~\ref{sec:conclusion}.

The model of Nishimori and Wong is well suited to this purpose. It has a gauge symmetry compatible with the $p$-body interaction \cite{Nishimori1999,Gillin2001}, and so the exact identities associated with the NL can be used as powerful tools in the analysis. The model is also interesting as it approaches the random energy model (REM) in the limit $p\to\infty$ \cite{Derrida1981,Gross1984}, where temperature chaos is known to be absent \cite{Derrida2021}. 

To achieve our goal, we use the explicit forms of the free energies of the FM and SG phases and expand them around the triple point at which the paramagnetic, FM, and SG phases meet. The leading term representing curvature of the phase boundary near the triple point is then expressed in a closed form in terms of the difference between the specific heats of the SG and FM phases at the triple point. The calculation is analytical apart from one-dimensional Gaussian quadratures and the solution of a single scalar equation. The specific-heat difference is found to be positive, meaning reentrance, for all integers $3\le p\le12$ and throughout the continuous range of $p$ examined numerically. The same quantity is also proven analytically at leading asymptotic order to be positive for $p$ sufficiently close to $2$ and for sufficiently large $p$. Thus, the FM--SG boundary is now established to be reentrant in the vicinity of the triple point.

Combining this result with the $p$-body extension of the NOO proposition, we conclude that there is at least one pair of distinct temperatures in the SG phase for which the distribution function of the two-temperature spin-state overlap is the delta function at the origin. The logical reasoning itself guarantees only the existence of at least one such temperature pair. Nevertheless, it would be physically quite unusual for the overlap to vanish for some temperature pairs but remain finite for others within a single SG phase. The present result therefore strongly suggests temperature chaos in the usual broader sense that the overlap distribution is the delta function at the origin for all temperature pairs in the spin glass phase. To support this conclusion, we carried out a two-temperature replica calculation, unrelated to the analyses around the triple point, and found it to be consistent with the above-mentioned conclusion.

The rest of this paper is organized as follows. In Sec.~\ref{sec:model}, we state the NOO proposition and define the two-temperature overlap distribution, introduce the model, summarize the exact constraints from gauge symmetry, and characterize the triple point. In Sec.~\ref{sec:expansion}, we expand the FM--SG boundary around the triple point, derive its curvature, and determine its sign, including the limits $p\to2^+$ and $p\to\infty$. Section~\ref{sec:chaos} discusses the consequence for temperature chaos and the independent two-temperature replica calculation. Section~\ref{sec:conclusion} summarizes the results and their limitations. The list of symbols follows, and details of the calculations are given in the Appendices.

\section{Formulation of the problem and the triple point}\label{sec:model}

In this section, we formulate the problem that underlies the subsequent analyses, and then derive the properties of the triple point, which will serve as the reference point for the expansions in the following sections.

\subsection{The NOO proposition and the overlap distribution}\label{sec:noo}

Before introducing the details of the model, let us state the proposition of NOO \cite{Nishimori2025a} that forms the starting point of the present paper. Consider an Ising spin-glass model with a gauge-invariant Hamiltonian $H$, and prepare two real replicas $\{S_i^{(1)}(=\pm 1)\}$ and $\{S_i^{(2)}(=\pm 1)\}$ at inverse temperatures $\beta_1$ and $\beta_2$, respectively, under the same realization of disorder. The two-temperature overlap distribution is defined by
\begin{equation}
\begin{split}
P_2(x\mid\beta_1,\beta_2)
&=\left[\frac{\displaystyle\sum_{S^{(1)},S^{(2)}}
\delta\Bigl(x-\frac1N\sum_{i=1}^N S_i^{(1)}S_i^{(2)}\Bigr)\,
e^{-\beta_1H(S^{(1)})}\,e^{-\beta_2H(S^{(2)})}}
{Z(\beta_1)\,Z(\beta_2)}\right], \\
Z(\beta)&=\sum_{S}e^{-\beta H(S)}
\end{split}
\label{eq:overlapx}
\end{equation}
where $N$ is the total number of spins in each system and $[\cdots]$ denotes the configurational average over the disorder in the couplings. The relation 
\begin{equation}
    P_2(x\mid\beta_1,\beta_2)\to\delta(x)\quad (N\to \infty)
\end{equation}
for $\beta_1\ne\beta_2$ means that the spin configurations at the two different temperatures become completely uncorrelated in the thermodynamic limit. If this relation holds for all temperature pairs within a spin glass phase, it means that temperature chaos exists in that phase.

In terms of this distribution function, the proposition of NOO and its contrapositive can be stated as follows.

\begin{quote}
\textbf{Proposition.} If a spin-glass phase exists at finite temperature and the spin-glass phase does not have temperature chaos, then the FM--SG phase boundary is not reentrant.\\
\textbf{Contrapositive.} If the FM--SG phase boundary is reentrant, then either there exists at least one pair of distinct temperatures in the SG phase for which $P_2(x\mid\beta_1,\beta_2)=\delta(x)$, or no SG phase exists at finite temperatures.
\end{quote}

The assumption of ``no temperature chaos'' in NOO is a uniform assumption that $P_2(x\mid\beta_1,\beta_2)\ne\delta(x)$ for every pair of distinct temperatures in the SG phase. Consequently, its contrapositive yields the non-constructive conclusion that there exists at least one temperature pair for which $P_2(x\mid\beta_1,\beta_2)=\delta(x)$; it does not by itself make a statement about all temperature pairs as required in the definition of temperature chaos\footnote{In the NOO paper, the contrapositive is stated as follows: if the FM--SG phase boundary is reentrant, then either temperature chaos exists in the SG phase or no SG phase exists at finite temperature. Here we state the content in a more precise logical form.}. This is, however, a limitation of the logical implication rather than a physical argument. Physically, it would be highly unnatural if $P_2(x\mid\beta_1,\beta_2)= \delta(x)$ for some temperature pairs but $P_2(x\mid\beta_1,\beta_2)\ne\delta(x)$ for others within the same SG phase. It is therefore natural to expect that if $P_2(x\mid\beta_1,\beta_2)=\delta(x)$ for even one pair, the same property holds for arbitrary distinct temperature pairs in the same SG phase. We return to this distinction in Sec.~\ref{sec:chaos}. We note that the gauge-transformation argument used in the proof of the NOO proposition extends directly to the $p$-body model discussed in the present paper by replacing a bond $ij$ with a $p$-body hyperedge as elaborated in Sec.~\ref{sec:chaoslogic}.

It is also important to note that the proposition does not determine the shape of the FM--SG phase boundary in the reverse direction: temperature chaos may exist even in the absence of reentrance. The SK model \cite{Sherrington1975} provides such an example: Temperature chaos exists \cite{Rizzo2003}, whereas the phase boundary is not reentrant \cite{Toulouse1980,GabayToulouse1981}. The present work provides an example in the opposite direction, for which the contrapositive has nontrivial content: we establish reentrance, which yields the existence statement of at least one temperature pair described above and strongly suggests temperature chaos.

\subsection{Model}\label{sec:constraints}

We consider a fully connected system of $N$ Ising spins $\sigma_i=\pm1$ with $p$-body interactions,
\begin{equation}
H=-\sum_{i_1<\cdots<i_p}J_{i_1\cdots i_p}\,\sigma_{i_1}\cdots\sigma_{i_p}.
\label{eq:H}
\end{equation}
We assume $p>2$ in this paper.
The couplings are independent Gaussian variables and, following the convention of Ref.~\cite{Nishimori1999}, we take the average and variance as
\begin{equation}
\bigl[J_{i_1\cdots i_p}\bigr]=\frac{j_0\,p!}{N^{p-1}},\qquad
\bigl[(\delta J_{i_1\cdots i_p})^2\bigr]=\frac{J^2\,p!}{2N^{p-1}}.
\label{eq:dist}
\end{equation}
We set $J=1$ below. The NL is given by $\beta=2j_0$, or equivalently $T=1/(2j_0)$. Under the gauge transformation
$\sigma_i\to\sigma_i\xi_i$,
$J_{i_1\cdots i_p}\to J_{i_1\cdots i_p}\xi_{i_1}\cdots\xi_{i_p}~(\xi_i=\pm 1)$
the Hamiltonian is invariant. Moreover, because the Gaussian coupling distribution has the standard gauge-covariant form, the various identities and correlation inequalities known for the $p=2$ model extend to the present model without modification \cite{Nishimori1981,Nishimori2001,Nishimori1999,Gillin2001}.

A schematic phase diagram of the present model derived in Ref.~\cite{Nishimori1999} is reproduced in Fig.~\ref{fig:schematic}. In what follows, we abbreviate the paramagnetic phase as PM and the replica-symmetric solution as RS. It is useful to summarize three constraints on this phase diagram: (i) the transition line between the PM and 1RSB-SG phases is independent of $j_0$ and is the horizontal line $T=T_c$; (ii) the NL does not enter the SG phase \cite{Nishimori2001,Nishimori2001b}; and (iii) the phase boundary between the RS-FM and 1RSB-SG phases below the triple point\footnote{Following the terminology of Nishimori and Wong \cite{Nishimori1999}, we call the point at which the first-order PM--FM and FM--SG transition lines and the second-order PM--SG transition line meet the triple point.} $\Mpt$ is either vertical or reentrant in such a way that the 1RSB-SG phase bends into the lower-temperature side of the RS-FM phase, whereas the opposite shape, in which the RS-FM phase extends into the lower-temperature side of the 1RSB-SG phase, is excluded by the inequality derived in Refs.~\cite{Nishimori1981,Nishimori2001}.

The FM phase continues as a metastable state beyond the equilibrium phase boundaries to the spinodal line shown as a thin dashed line in the figure. Since we are concerned only with thermodynamic equilibrium, we do not analyze the metastable state. As will also be shown below, in the vicinity of the triple point $\Mpt$, the RS-FM solution is stable against replica-symmetry-breaking perturbations (positive replicon eigenvalue; Sec.~\ref{sec:mpoint} and Appendix~\ref{app:replicon}).

\begin{figure}[!htb]
\centering
\begin{tikzpicture}[x=7.6cm,y=5.6cm]
\draw[thick] (0,0.3) rectangle (1.02,1.2);
\foreach \xx/\lb in {0/0, 0.2/0.2, 0.4/0.4, 0.6/0.6, 0.8/0.8, 1.0/1.0}
  {\draw[thick] (\xx,0.3)--(\xx,0.318); \node[below,font=\small] at (\xx,0.3) {\lb};}
\foreach \yy/\lb in {0.3/0.3, 0.4/0.4, 0.6/0.6, 0.8/0.8, 1.0/1.0, 1.2/1.2}
  {\draw[thick] (0,\yy)--(0.012,\yy); \node[left,font=\small] at (0,\yy) {\lb};}
\node[below,font=\small] at (0.51,0.235) {$j_0$};
\node[above,font=\small] at (0.0,1.235) {$T$};
\draw[black!75,dash dot,semithick]
  plot[smooth] coordinates {(0.723,0.692) (0.734,0.62) (0.748,0.54)
                            (0.7676,0.44) (0.786,0.36) (0.800,0.30)};
\draw[black!75,dashed,semithick]
  plot[smooth] coordinates {(0.723,0.692) (0.736,0.781) (0.752,0.881)
                            (0.771,0.981) (0.793,1.081) (0.817,1.181) (0.8225,1.20)};
\draw[black!75,dashed,semithick] (0.723,0.692)--(0.723,0.30);
\draw[dashed,very thick,domain=0.4167:1.02,samples=140,smooth,variable=\x]
  plot (\x,{1/(2*\x)});
\draw[very thick] (0,0.6514)--(0.7676,0.6514);
\draw[very thick] plot[smooth] coordinates
  {(0.7676,0.6514) (0.781,0.74) (0.797,0.84) (0.816,0.94)
   (0.838,1.04) (0.862,1.14) (0.876,1.20)};
\draw[very thick] (0.7676,0.6514)--(0.7676,0.30);
\fill (0.7676,0.6514) circle (1.5pt);
\node[font=\small] at (0.806,0.575) {$\Mpt$};
\node[font=\small] at (0.33,1.00) {PM};
\node[font=\small] at (0.33,0.48) {1RSB-SG};
\node[font=\small] at (0.900,0.68) {RS-FM};
\node[font=\small] at (0.555,1.03) {NL};
\end{tikzpicture}
\caption{Schematic phase diagram of the present model for $p=3$ \cite{Nishimori1999}. Thick solid lines denote equilibrium phase boundaries. The horizontal line is the PM--1RSB-SG transition, which is independent of $j_0$; the gently curved line extending upward to the right from the triple point $\Mpt$ is the first-order PM--RS-FM transition; and the almost vertical line extending downward from $\Mpt$ is the RS-FM--1RSB-SG phase boundary. The thick dashed line is the Nishimori line (NL), $T=1/(2j_0)$, which passes through the triple point $\Mpt$ and does not enter the SG phase. The thin dashed line is the metastability limit (spinodal line) of the FM phase. Since only equilibrium states are considered in the present paper, the spinodal line will not be analyzed; it is included to indicate correspondence with Fig.~7 of Ref.~\cite{Nishimori1999}. The thin dash-dotted line is the de Almeida--Thouless (AT) line, below which replica symmetry of the RS-FM solution is broken and the RS-FM phase (or the metastable FM state to the left of the equilibrium phase boundary) becomes a mixed phase (RSB ferromagnetic phase). Its crossing with the equilibrium phase boundary is at $T_{\rm AT}\simeq0.435$ (Appendix~\ref{app:replicon}). The reentrance of the RS-FM--1RSB-SG phase boundary to be derived later in this paper is indistinguishable from a vertical line on this scale; an enlarged view is shown in Fig.~\ref{fig:phase}(b). Below $T_G\simeq0.24$ the SG phase becomes full RSB. Since that region is outside the scope of the present analysis, only $T>0.3$ is shown. The code used to draw this schematic figure was produced with assistance from the generative-AI tools described in the Methods section.}
\label{fig:schematic}
\end{figure}

\subsection{Characterization of the triple point}\label{sec:mpoint}

In this subsection, we write the free-energy functional and derive the conditions characterizing the triple point. We write $f$ for the free energy per spin and define $\varphi\equiv-\beta f$, together with the abbreviation
\begin{equation}
\int Dz\,(\cdots)\equiv\int_{-\infty}^{\infty}\frac{dz}{\sqrt{2\pi}}e^{-z^2/2}(\cdots).
\label{eq:Dz}
\end{equation}

\subsubsection{General 1RSB functional}
Nishimori and Wong \cite{Nishimori1999} derived the 1RSB free energy of the present model. In our convention ($J=1$), it reads, in terms of the variational parameters $(m,q_0,q_1,x)$, 
\begin{align}
\varphi_{\rm 1RSB}&=\frac{\beta^2}{4}+\frac{\beta^2}{4}(p-1)\bigl(xq_0^p+(1-x)q_1^p\bigr)
-\frac{\beta^2}{4}p\,q_1^{p-1}-(p-1)\beta j_0m^p\\
&\qquad+\frac{1}{x}\int Du\,\ln\!\int Dv\,\bigl(2\cosh\Xi\bigr)^{x},
\label{eq:NWfunc}\\
\Xi&=\sqrt{\Lam_0}\,u+\sqrt{\Lam_1-\Lam_0}\,v+p\beta j_0m^{p-1},
\qquad
\Lam_a=\frac{p}{2}\beta^2q_a^{p-1},
\label{eq:NWfield}
\end{align}
where $m$ and $q_0, q_1$ are the order parameters for magnetization and spin-glass ordering, respectively, and $x$ is the breakpoint between $q_0$ and $q_1$. The stationarity conditions are
\begin{equation}
m=\int Du\,\langle\tanh\Xi\rangle,\qquad
q_0=\int Du\,\langle\tanh\Xi\rangle^2,\qquad
q_1=\int Du\,\langle\tanh^2\Xi\rangle,
\label{eq:NWstat}
\end{equation}
together with $\partial\varphi_{\rm 1RSB}/\partial x=0$. Here $\langle\cdot\rangle$ denotes the average over $v$ with weight $(2\cosh\Xi)^x$. We now specialize this common functional to the paramagnetic, ferromagnetic, and spin glass phases.

\subsubsection{Paramagnetic phase}
The solution $m=q_0=q_1=0$ (and hence $\Lam_a=0$ and $\Xi=0$) is always stationary, and the dependence on $x$ disappears, giving
\begin{equation}
\varphi_{\PM}(\beta)=\frac{\beta^2}{4}+\ln2.
\label{eq:phiPM}
\end{equation}
This is the paramagnetic free energy.

\subsubsection{Ferromagnetic phase}
The RS solution is realized by setting $q_0=q_1\equiv q$ (and hence $\Lam_0=\Lam_1\equiv\Lam$). In this limit, the $v$ dependence of $\Xi$ disappears and the dependence on $x$ also drops out, yielding
\begin{equation}
\varphi_{\FM}(m,q;\beta,j_0)
=\frac{\beta^2}{4}+\frac{(p-1)\beta^2}{4}q^p-(p-1)\beta j_0m^p
-\frac{\Lam}{2}+\int Dz\,\lc\bigl(H+\sqrt{\Lam}\,z\bigr).
\label{eq:phiFM}
\end{equation}
The effective field and its mean and variance are
\begin{equation}
\eta\equiv H+\sqrt{\Lam}\,z,\qquad H=p\beta j_0m^{p-1},\qquad
\Lam=\frac{p}{2}\beta^2q^{p-1}.
\label{eq:FMfield}
\end{equation}
Variation with respect to $m$ and $q$ gives the self-consistency equations
\begin{equation}
m=\int Dz\,\tanh\eta,\qquad q=\int Dz\,\tanh^2\eta,
\label{eq:FMstat}
\end{equation}
respectively.

\subsubsection{Spin glass phase}
In the spin glass phase, the magnetization vanishes, $m=0$, so that the ferromagnetic bias $j_0$ does not contribute to the effective field ($H=0$). In addition, the 1RSB solution for zero external field has $q_0=0$ \cite{Gardner1985,Gross1984}. Substituting $m=q_0=0$ into Eq.~\eqref{eq:NWfunc}, for which $\Xi=\sqrt{\Lam_1}\,v$, gives
\begin{equation}
\begin{split}
\varphi_{\SG}(q_1,x;\beta)
&=\frac{\beta^2}{4}
+\frac1x\Bigl(-\frac{(p-1)\beta^2}{4}\,x(x-1)\,q_1^p-\frac{\Lam_1 x}{2}
+\ln \int Dz\,\bigl(2\cosh\sqrt{\Lam_1}\,z\bigr)^{x}\Bigr), \\
\Lam_1&=\frac{p}{2}\beta^2q_1^{p-1}
\end{split}
\label{eq:phiSG}
\end{equation}
for the 1RSB-SG free energy. Variation with respect to $q_1$ gives the self-consistency equation
\begin{equation}
q_1=\frac{\int Dz\,\bigl(2\cosh\sqrt{\Lam_1}\,z\bigr)^{x}\tanh^2\!\bigl(\sqrt{\Lam_1}\,z\bigr)}
{\int Dz\,\bigl(2\cosh\sqrt{\Lam_1}\,z\bigr)^{x}},
\label{eq:SGstat}
\end{equation}
which is the last relation in Eq.~\eqref{eq:NWstat} specialized to $m=q_0=0$. The remaining stationarity condition, $\partial\varphi_{\SG}/\partial x=0$, is needed only at the triple point and is derived below.

\subsubsection{Conditions at the triple point}
We next derive the conditions that characterize the triple point $\Mpt=(\jM,T_c)$\footnote{We attach the subscript $\Mpt$ to the $j_0$ coordinate and write it as $\jM$ to indicate the triple point, whereas the temperature coordinate is denoted by $T_c$, rather than $T_{\Mpt}$. As shown below, the FM transition temperature on the NL coincides with the PM--SG transition temperature $T_c$, which is defined independently in Sec.~\ref{sec:constraints} and is independent of $j_0$. We therefore use the independently defined $T_c$ (and $\bc\equiv1/T_c$) throughout.}. At the triple point, $m=q$ and $j_0=\beta/2$ \cite{Nishimori1999}. Equation~\eqref{eq:FMfield} then gives $H=\Lam$. We therefore introduce the Gaussian field $h\equiv\Lam+\sqrt{\Lam}\,z$, whose mean and variance are both $\Lam$. Writing $\Lam_c$ for the value of $\Lam$ at the triple point, we denote the field there by $h_c\equiv\Lam_c+\sqrt{\Lam_c}\,z$. For this distribution, the identity
\begin{equation}
\int Dz\,\tanh^{2k-1}h_c=\int Dz\,\tanh^{2k}h_c\qquad(k=1,2,\dots)
\label{eq:tilted}
\end{equation}
holds (the NL identity; Appendix~\ref{app:A1}(c)).

Let us first consider variation with respect to $q_1$. At the triple point, the breakpoint reaches $x^*(T_c)=1$ \cite{Gardner1985}. In the limit $x\to1^-$, the stationarity condition \eqref{eq:SGstat} for $q_1$ becomes
\begin{equation}
q_{1c}=\int Dz\,\tanh^2h_c.
\label{eq:q1c}
\end{equation}
By the NL identity \eqref{eq:tilted} with $k=1$, the right-hand side is equal to $\int Dz\,\tanh h_c$. We denote this common value by $\mu$:
\begin{equation}
q_{1c}=\int Dz\,\tanh^2h_c=\int Dz\,\tanh h_c\equiv\mu.
\label{eq:mudef}
\end{equation}
We note that $\mu\ne0$ because the order parameter is discontinuous at the transition for $p>2$. For a second-order transition ($p=2$), the order parameters vanish at the transition point, including the multicritical point.

We next consider variation with respect to $x$. Evaluating the breakpoint condition $\partial\varphi_{\SG}/\partial x|_{x=1}=0$ at $\beta=\bc$, $q_1=q_{1c}=\mu$, and $\Lam_1=\Lam_c$ (Appendix~\ref{app:A2}(b), Eq.~\eqref{eq:phix}) gives
\begin{equation}
\int Dz\,\lc h_c=\ln2+\frac{\Lam_c}{2}+\frac{(p-1)\bc^2}{4}\mu^p.
\label{eq:bpcond}
\end{equation}
Equations~\eqref{eq:q1c} and \eqref{eq:bpcond} are the two conditions characterizing the triple point, and show that $\Mpt$ can be described by a single scalar $\mu\in(0,1)$. Here $\Lam_c=(p/2)\bc^2\mu^{p-1}$, $T_c=1/\bc$, and $\jM=\bc/2$. The fact that $\Mpt$ lies on the NL will be discussed after Table~\ref{tab:mpoint}. The quantity $\mu$ is the value of the order parameter at the triple point and should be distinguished from $m$ and $q$, which vary with general $(\beta,j_0)$.

Furthermore, defining
\begin{equation}
W(\Lam)\equiv\int Dz\,\lc h-\ln2-\frac{\Lam}{2}
\label{eq:Wdefmain}
\end{equation}
and using $(p-1)\bc^2\mu^p/4=(p-1)\Lam_c\mu/(2p)$, the triple-point conditions \eqref{eq:q1c} and \eqref{eq:bpcond} reduce to the following single equation for $\Lam_c$:
\begin{equation}
W(\Lam_c)=\frac{p-1}{2p}\Lam_c\,\mu(\Lam_c).
\label{eq:onevar}
\end{equation}
Here $\mu(\Lam)\equiv\int Dz\,\tanh h$, with $h=\Lam+\sqrt{\Lam}\,z$, is a known function of $\Lam$. Thus, for a given $p$, Eq.~\eqref{eq:onevar} determines $\Lam_c>0$. Substitution of this root gives $\mu=\mu(\Lam_c)$, followed by $\bc^2=(2/p)\Lam_c\mu^{1-p}$, $T_c=1/\bc$, and $\jM=\bc/2$, thereby determining the triple point completely.

Table~\ref{tab:mpoint} also lists the replicon eigenvalue $\lambda(\Mpt)$, which measures stability at $\Mpt$. The replicon eigenvalue is an eigenvalue of the Hessian of the free energy in a replica-symmetry-breaking direction; a positive value means local stability in that direction. For an RS solution this is the AT eigenvalue \cite{deAlmeida1978}, whereas for a 1RSB solution it is the innermost-block replicon eigenvalue. For the two solutions compared in this paper, they are given by (Appendix~\ref{app:replicon})
\begin{subequations}\label{eq:repdef}
\begin{equation}
\lambda_{\rm RS}=1-\frac{p(p-1)}{2}\,\beta^2q^{p-2}\int Dz\,\mathrm{sech}^4\eta,
\label{eq:repdefRS}
\end{equation}
\begin{equation}
\lambda_1=1-\frac{p(p-1)}{2}\,\beta^2q_1^{p-2}\,
\frac{\int Dz\,\bigl(2\cosh\sqrt{\Lam_1}\,z\bigr)^{x}\,\mathrm{sech}^4\bigl(\sqrt{\Lam_1}\,z\bigr)}
{\int Dz\,\bigl(2\cosh\sqrt{\Lam_1}\,z\bigr)^{x}},
\label{eq:repdefSG}
\end{equation}
\end{subequations}
for the RS-FM and 1RSB-SG solutions, respectively, where $\eta=H+\sqrt{\Lam}\,z$ is the FM effective field of Eq.~\eqref{eq:FMfield}, and the average on the SG side carries the same weight as in Eq.~\eqref{eq:SGstat}. At the triple point, $H=\Lam=\Lam_c$, so that $\eta$ reduces to $h_c$. On the SG side, $x=1$ and $\Lam_1=\Lam_c$ there, and the weighted average in Eq.~\eqref{eq:repdefSG} becomes the Gaussian average over $h_c$ by the identity~\eqref{eq:tiltshift}. Expanding $\mathrm{sech}^4=(1-\tanh^2)^2$ and applying the $k=1$ and $k=2$ cases of the NL identity \eqref{eq:tilted} gives $\int Dz\,\mathrm{sech}^4h_c=1-2\mu+\tau$ (Appendix~\ref{app:B2}). We therefore find
\begin{equation}
\lambda(\Mpt)=1-\Omega\,(1-2\mu+\tau),\qquad
\Omega=\frac{p(p-1)}{2}\bc^2\mu^{p-2},\qquad
\tau=\int Dz\,\tanh^3h_c.
\label{eq:lamMmain}
\end{equation}
Representative values are shown in Table~\ref{tab:mpoint}.

\begin{table}[htb]
\centering
\caption{Quantities at the triple point. Here $\mu=m_{\Mpt}=q_{\Mpt}=q_{1c}$, and $\lambda(\Mpt)$ is the replicon eigenvalue common to the two branches.}
\label{tab:mpoint}
\begin{tabular}{cccccc}
\toprule
$p$ & $T_c$ & $\jM$ & $\mu$ & $\Lam_c$ & $\lambda(\Mpt)$ \\
\midrule
3 & 0.651385 & 0.767595 & 0.813518 & 2.3396 & $+0.3305$ \\
4 & 0.616883 & 0.810526 & 0.948088 & 4.4789 & $+0.5743$ \\
5 & 0.606952 & 0.823789 & 0.981568 & 6.2996 & $+0.7348$ \\
6 & 0.603296 & 0.828781 & 0.992590 & 7.9416 & $+0.8369$ \\
\bottomrule
\end{tabular}
\end{table}

It follows that, at the triple point, the FM stationarity condition \eqref{eq:FMstat} reduces to the same equation as \eqref{eq:q1c}, and hence $m_{\Mpt}=q_{\Mpt}=q_{1c}=\mu$. Here $q_{1c}$ is the innermost 1RSB overlap $q_1$ of the SG solution evaluated at $T_c$, whereas $m_{\Mpt}$ and $q_{\Mpt}$ are the RS order parameters of the FM solution at the triple point. They belong to different solutions and coincide only at $\Mpt$; below $T_c$ the two branches separate (Appendix~\ref{app:B2}). Moreover, the breakpoint condition \eqref{eq:bpcond} is identical to the condition that the FM and PM free energies coincide on the NL, so that the SG transition temperature and the FM transition temperature on the NL coincide identically (Appendix~\ref{app:A2}(e)). This is the reason why we use $T_c$, rather than $T_{\Mpt}$, for the temperature coordinate of the triple point.

The same coincidence holds for the replicon eigenvalues mentioned above: at $\Mpt$, the AT eigenvalue on the FM side and the innermost replicon eigenvalue on the SG side take the common value given in Eq.~\eqref{eq:lamMmain} [see Eq.~\eqref{eq:lamM}]. Over the range evaluated numerically, $\lambda(\Mpt)>0$, and the two branches compared in the vicinity of the triple point are replicon stable.

\section{Expansion around the triple point and criterion for reentrance}\label{sec:expansion}

The purpose of this section is to determine analytically the shape of the FM--SG phase boundary in the vicinity of the triple point $\Mpt$ and thereby establish whether the transition is reentrant. A slight bending of this boundary for $p=3$ was noticed in the numerical 1RSB solution of Ref.~\cite{ZdeborovaKrzakala2010PRB}; the aim here is to derive the local shape analytically as a function of $p$. For $p\ge3$, the FM--SG transition is first order \cite{Nishimori1999}, and the phase boundary is determined by equality of the free energies of the two phases. We introduce
\begin{equation}
b\equiv\beta-\bc,\qquad t\equiv T_c-T,\qquad \delta\equiv j_0^*-\jM
\label{eq:btdelta}
\end{equation}
and expand the free energies of the two phases in the two deviations $b$ and $\delta$ from the triple point. The phase boundary $j_0^*(T)$ is obtained by equating the two free energies. As will be shown below, $\delta$ is of order $b^2$, so it is sufficient to retain terms through second order in $b$ and first order in $\delta$. We present the structure of the derivation in the main text and give the details in Appendix~\ref{app:expansion}.

\subsection{Vanishing of the first-order term}\label{sec:firstorder}

Let us first fix the notation. The quantity $\varphi_{\FM}(m,q;\beta,j_0)$ in Eq.~\eqref{eq:phiFM} is a functional, namely, a function of the four variables consisting of the variational variables $(m,q)$ and the external variables $(\beta,j_0)$. Substituting the stationary solution gives
\begin{equation}
\hat\varphi_{\FM}(\beta,j_0)\equiv
\varphi_{\FM}\bigl(m^*(\beta,j_0),\,q^*(\beta,j_0);\,\beta,j_0\bigr),
\label{eq:onshell}
\end{equation}
which is a function of only the two variables $(\beta,j_0)$. We call this the on-shell function and distinguish it by a hat. Similarly, we write the SG on-shell function as $\hat\varphi_{\SG}(\beta)$. Since $m=0$ in the SG phase, its dependence on $j_0$ disappears and it is a one-variable function. For first derivatives, the stationarity conditions $\partial\varphi_{\FM}/\partial m=\partial\varphi_{\FM}/\partial q=0$ eliminate contributions from the implicit dependence of the variational variables, so derivatives of the on-shell function with respect to $\beta$ or $j_0$ coincide with partial derivatives of the functional. The same observation applies to $\varphi_{\SG}$.

We first consider the derivative with respect to $j_0$. Differentiating Eq.~\eqref{eq:phiFM} with respect to $j_0$ at fixed $(m,q,\beta)$ gives
\begin{equation}
\frac{\partial\hat\varphi_{\FM}}{\partial j_0}=\partial_{j_0}\varphi_{\FM}
=-(p-1)\beta m^p+\frac{\partial H}{\partial j_0}\int Dz\,\tanh\eta
=-(p-1)\beta m^p+p\beta m^{p-1}\cdot m=\beta m^p.
\label{eq:dphidj0}
\end{equation}
At the triple point, where $m=\mu$, this becomes
\begin{equation}
\frac{\partial\hat\varphi_{\FM}}{\partial j_0}\bigg|_{\Mpt}=\bc\mu^p>0.
\label{eq:c1}
\end{equation}

We next consider the derivative with respect to $\beta$. On the FM side, for the same reason, $\partial\hat\varphi_{\FM}/\partial\beta=\partial_\beta\varphi_{\FM}|_{m,q}$, and this quantity gives the internal energy $u=-\partial\hat\varphi/\partial\beta$. Immediately below the triple point the NL passes through the FM phase. We can therefore use the energy identity on the NL, namely, the gauge-symmetry result that the equilibrium internal energy on the NL is $u=-\beta/2$, to obtain
\begin{equation}
\frac{\partial\hat\varphi_{\FM}}{\partial\beta}\bigg|_{\Mpt}=\frac{\bc}{2}
\label{eq:c2}
\end{equation}
at the triple point.

On the SG side, this result does not follow from the same identity because the NL does not enter the SG phase [constraint (ii) in Sec.~\ref{sec:constraints}]. Instead, we use the fact that the breakpoint reaches $x^*(T_c)=1$ at the triple point (Sec.~\ref{sec:mpoint}) \cite{Gardner1985}. Setting $x=1$ in Eq.~\eqref{eq:phiSG}, the identity $\int Dz\,2\cosh(\sqrt{\Lam_1}z)=2e^{\Lam_1/2}$ gives
\begin{equation}
\varphi_{\SG}(q_1,1;\beta)=\frac{\beta^2}{4}+\ln2=\varphi_{\PM}(\beta)
\label{eq:slicemain}
\end{equation}
identically, independently of the values of $q_1$ and $\beta$ [Appendix~\ref{app:A2}(a)]. Differentiating this identity with respect to $\beta$ yields $\partial\hat\varphi_{\SG}/\partial\beta|_{\Mpt}=\varphi_{\PM}'(\bc)=\bc/2$, the same value as in Eq.~\eqref{eq:c2}. This expresses the absence of latent heat at the PM--SG transition.

Collecting these results, we have
\begin{equation}
\frac{\partial\hat\varphi_{\FM}}{\partial j_0}\Big|_{\Mpt}=\bc\,\mu^p\;(>0),\qquad
\frac{\partial\hat\varphi_{\FM}}{\partial\beta}\Big|_{\Mpt}
=\frac{\partial\hat\varphi_{\SG}}{\partial\beta}\Big|_{\Mpt}=\frac{\bc}{2},
\label{eq:firstder}
\end{equation}
where the second relation states that the internal energies of the two phases coincide at the triple point (i.e., the latent heat vanishes). As seen above, however, the origins of this equality are different on the two sides.

To examine the behavior of the phase boundary, let us introduce the function
\begin{equation}
F(b,\delta)\equiv\hat\varphi_{\FM}(\bc+b,\,\jM+\delta)-\hat\varphi_{\SG}(\bc+b).
\label{eq:Fdef}
\end{equation}
The implicit function theorem states that if $F(b,\delta)$ is of class $C^1$ in a neighborhood of the origin and satisfies
\begin{equation*}
F(0,0)=0,\qquad
\frac{\partial F}{\partial\delta}\bigg|_{(0,0)}\ne0,
\end{equation*}
then, in some neighborhood $|b|<\varepsilon$ of the origin, the equation $F(b,\delta)=0$ determines $\delta$ uniquely as a function of $b$. In other words, there exists a $C^1$ function $\delta(b)$ with $\delta(0)=0$ such that $F(b,\delta(b))=0$ identically. Differentiating with respect to $b$ gives
\begin{equation*}
\delta'(b)=-\,\frac{\partial F/\partial b}{\partial F/\partial\delta}\bigg|_{(b,\,\delta(b))}.
\end{equation*}
If $F$ is of class $C^k$, then $\delta$ is also of class $C^k$. Since we use the expansion through second order with a remainder $O(b^3)$, we assume $k\ge3$. Smoothness of $\hat\varphi$ follows from smoothness of the stationary branches in the vicinity of the triple point. For the FM branch, the stationarity conditions $\partial\varphi_{\FM}/\partial m=\partial\varphi_{\FM}/\partial q=0$ for the functional \eqref{eq:phiFM} determine $m$ and $q$ as smooth functions of $\beta$ and $j_0$, because the $2\times2$ matrix of the second derivatives of $\varphi_{\FM}$ with respect to $m$ and $q$ has a nonzero determinant at the triple point for the values of $p$ studied (Appendix~\ref{app:A5}); this matrix is the Hessian $\mathsf H$ defined in Eq.~\eqref{eq:Hvdef}. For the SG branch, smoothness follows from the implicit function theorem applied to regularized stationarity conditions in Appendix~\ref{app:A4}(b${}^{\prime}$).

Let us verify the hypotheses. As discussed in Sec.~\ref{sec:mpoint}, the free energies of the two phases coincide at the triple point, so that $F(0,0)=0$. The first relation in Eq.~\eqref{eq:firstder} gives $\partial F/\partial\delta|_{(0,0)}=\bc\mu^p\ne0$. Hence there exists a unique smooth phase-boundary curve $\delta(b)$ through the triple point, whose slope is
\begin{equation}
\delta'(0)
=-\,\frac{\partial F/\partial b}{\partial F/\partial\delta}\bigg|_{(0,0)}
=-\,\frac{1}{\bc\,\mu^p}\Bigl(\frac{\partial\hat\varphi_{\FM}}{\partial\beta}-\frac{\partial\hat\varphi_{\SG}}{\partial\beta}\Bigr)_{\Mpt}
=0,
\label{eq:IFT}
\end{equation}
and is therefore quadratic near the triple point,
\begin{equation}
\delta=O(b^2),
\label{eq:Ob2}
\end{equation}
or, equivalently,
\begin{equation}
\frac{dj_0^*}{dT}\bigg|_{\Mpt}=0.
\label{eq:vertical}
\end{equation}
Thus the FM--SG phase boundary is tangent to the temperature axis, i.e., to the vertical direction, at the triple point.

\subsection{Formula for the phase boundary}\label{sec:secondorder}

\subsubsection{Expansion to second order and determination of the boundary}\label{sec:deltadet}

Since the first-order term vanishes, the phase boundary is determined by the second-order coefficients. We first fix $j_0=\jM$ and Taylor-expand the on-shell functions of the two phases with respect to $\beta$ through second order around the triple point.

The first-order coefficient is $\bc/2$ for both phases [Eq.~\eqref{eq:c2} and the second relation in Eq.~\eqref{eq:firstder}], and hence we can write
\begin{equation}
\begin{split}
\hat\varphi_{\SG}(\beta)&=\hat\varphi_{\SG}(\bc)+\frac{\bc}{2}\,b
+\frac12\frac{\partial^2\hat\varphi_{\SG}}{\partial\beta^2}\bigg|_{\Mpt} b^2+O(b^3), \\
\hat\varphi_{\FM}(\beta,\jM)&=\hat\varphi_{\FM}(\bc,\jM)+\frac{\bc}{2}\,b
+\frac12\frac{\partial^2\hat\varphi_{\FM}}{\partial\beta^2}\bigg|_{\Mpt} b^2+O(b^3).
\end{split}
\label{eq:PMtangent}
\end{equation}
We define the following quantities:
\begin{equation}
A_{\SG}\equiv\frac12-\frac{\partial^2\hat\varphi_{\SG}}{\partial\beta^2}\bigg|_{\Mpt},\qquad
A_{\FM}\equiv\frac12-\frac{\partial^2\hat\varphi_{\FM}}{\partial\beta^2}\bigg|_{\Mpt}.
\label{eq:Acoef}
\end{equation}
The common constant $1/2$ is the curvature $\varphi_{\PM}''=1/2$ of the PM free energy [Eq.~\eqref{eq:phiPM}], and therefore $A_{\SG}$ and $A_{\FM}$ measure the deviations of the curvatures in the two phases from the PM value. This common term drops out of the difference $A_{\FM}-A_{\SG}$ which appears in the formula derived later.

It is useful to clarify the role of these two expansions. Both are expansions with respect to $\beta$ alone around the triple point, with $j_0$ fixed at $j_0=\jM$ rather than varied along the phase boundary. The quantity $\delta=j_0^*-\jM$ therefore does not appear here. On the SG side, $\hat\varphi_{\SG}$ is independent of $j_0$ in the first place and is thus a one-variable function unrelated to $\delta$. On the FM side, $\hat\varphi_{\FM}(\beta,\jM)$ is likewise evaluated at fixed $j_0$ and contains no $\delta$.

We next subtract the two relations in Eq.~\eqref{eq:PMtangent}. The zeroth-order terms cancel because the free energies coincide at the triple point. The first-order terms also cancel because the coefficient $\bc/2$ is common to the two phases [the second relation in Eq.~\eqref{eq:firstder}]. Consequently, at fixed $j_0=\jM$, the free-energy difference starts at second order. In the notation of Eq.~\eqref{eq:Acoef},
\begin{equation}
\hat\varphi_{\FM}(\beta,\jM)-\hat\varphi_{\SG}(\beta)
=-\frac{A_{\FM}-A_{\SG}}{2}\,b^2+O(b^3),
\label{eq:phidiff}
\end{equation}
which will be used below.

The deviation $\delta$ enters only at the stage where this difference is inserted into the phase-boundary condition. The boundary is determined by $\hat\varphi_{\FM}(\beta,j_0^*)=\hat\varphi_{\SG}(\beta)$. Let us write both sides explicitly to order $b^2$. For the left-hand side $\hat\varphi_{\FM}(\beta,j_0^*)$, with $j_0^*=\jM+\delta$ on the boundary, we use
\begin{equation}
\hat\varphi_{\FM}(\beta,\jM+\delta)
=\hat\varphi_{\FM}(\beta,\jM)
+\frac{\partial\hat\varphi_{\FM}}{\partial j_0}\Big|_{\Mpt}\,\delta
+O(b\delta,\delta^2)
\label{eq:j0shift}
\end{equation}
for the difference from the value at fixed $j_0=\jM$. The term of $O(b\delta)$ comes from the difference between $\partial\hat\varphi_{\FM}/\partial j_0$ evaluated at a general $\beta$ and at the triple point ($\beta=\bc$) as will be explained later. One point concerning Eq.~\eqref{eq:j0shift} is worth emphasizing. Although the left-hand side appears to be a two-variable function, in the present application $\delta$ is the phase-boundary displacement $\delta(b)$ determined by the boundary condition as a function of $\beta$, so the left-hand side is a one-variable function restricted to the phase boundary. Equation~\eqref{eq:j0shift} is a Taylor expansion with respect to the second variable performed before substituting $\delta=\delta(b)$ in order to evaluate this one-variable function.
As shown in Sec.~\ref{sec:firstorder}, $\delta=O(b^2)$, so the correction $\bc\mu^p\delta$ [Eq.~\eqref{eq:c1}], the second term on the right-hand side, is of order $b^2$. We now verify in turn that the remaining second-order Taylor terms do not contribute at this accuracy.

First consider the $\delta^2$ term. Its coefficient is the second derivative $\partial^2\hat\varphi_{\FM}/\partial j_0^2|_{\Mpt}$, which is finite by the smoothness of the stationary branch discussed in Sec.~\ref{sec:firstorder}. Since $\delta=O(b^2)$, we have $\delta^2=O(b^4)$, and this term can be omitted at order $b^2$.

Next consider the cross term. In Eq.~\eqref{eq:j0shift}, the coefficient of the second term was replaced by its value $\bc\mu^p$ at the triple point [Eq.~\eqref{eq:c1}], whereas strictly speaking it should be evaluated at a general $\beta$. The difference is
\begin{equation}
\frac{\partial\hat\varphi_{\FM}}{\partial j_0}\bigg|_{(\beta,\jM)}
=\bc\mu^p
+\frac{\partial^2\hat\varphi_{\FM}}{\partial\beta\,\partial j_0}\bigg|_{\Mpt}\,b
+O(b^2).
\label{eq:crosscoef}
\end{equation}
The second term on the right-hand side, multiplied by $\delta$, is the cross term. It is proportional to $b\delta$ with coefficient $\partial^2\hat\varphi_{\FM}/(\partial\beta\,\partial j_0)|_{\Mpt}$. Since $\delta=O(b^2)$, we have $b\delta=O(b^3)$, and this term also drops out at order $b^2$.

Thus, to order $b^2$, the only term containing $\delta$ that remains is $\bc\mu^p\delta$. Using Eq.~\eqref{eq:phidiff} for the difference between $\hat\varphi_{\FM}(\beta,\jM)$ and $\hat\varphi_{\SG}(\beta)$, the phase-boundary condition therefore becomes
\begin{equation}
-\frac{A_{\FM}-A_{\SG}}{2}\,b^2+\bc\mu^p\,\delta
=O(b^3),
\label{eq:bcond2}
\end{equation}
which is solved for $\delta$ to give
\begin{equation}
\delta=\frac{A_{\FM}-A_{\SG}}{2\bc\mu^p}\,b^2+O(b^3).
\label{eq:deltasolve}
\end{equation}
Hence the coefficient of $b^2$ in $\delta$ is determined by the second-order coefficient of the free-energy difference at fixed $\jM$ [Eq.~\eqref{eq:phidiff}], namely by the difference $A_{\FM}-A_{\SG}$ between the deviations of the two curvatures from the PM value.

\subsubsection{Second-order coefficients in closed form}\label{sec:coeffs}

As shown in Appendixes~\ref{app:A4}--\ref{app:A5}, both coefficients $A_{\SG}$ and $A_{\FM}$ can be expressed entirely in terms of moments of the Gaussian distribution $h_c=\Lam_c+\sqrt{\Lam_c}\,z$ at the triple point:
\begin{align}
A_{\SG}&=\frac{\varphi_{x\beta}^2}{\varphi_{xx}},\qquad
\varphi_{x\beta}=\frac{\bc\,\mu^p}{2},\qquad
\varphi_{xx}=V-\frac{(p-1)\bc^2}{2}\mu^p,
\label{eq:ASGa}\\
&\hspace{28mm}
V=\int Dz\,(\lc h_c)^2-\Bigl(\int Dz\,\lc h_c\Bigr)^{\!2},
\label{eq:ASG}\\[2pt]
A_{\FM}&=\frac{\mu^p}{2}+\frac{\Lam_c^2}{\bc^2}\,B
+\bm v^{\!\top}\mathsf H^{-1}\bm v,\qquad
B=1-3\mu+2\tau,\qquad \tau=\int Dz\,\tanh^3h_c,
\label{eq:AFMmain}\\
&\hspace{20mm}
\bm v^{\!\top}\mathsf H^{-1}\bm v
=\varphi_{m\beta}^2\,\frac{\varphi_{mm}+2\varphi_{mq}+\varphi_{qq}}{\det\mathsf H}.
\label{eq:AFM}
\end{align}
Here $\varphi_{x\beta}$ and $\varphi_{xx}$ on the SG side are the second partial derivatives of the 1RSB functional \eqref{eq:phiSG} with respect to $x$ and $\beta$, evaluated at the triple point:
\begin{equation}
\varphi_{x\beta}=\frac{\partial^2\varphi_{\SG}}{\partial x\,\partial\beta}\bigg|_{\Mpt},\qquad
\varphi_{xx}=\frac{\partial^2\varphi_{\SG}}{\partial x^2}\bigg|_{\Mpt}.
\label{eq:SGderivdef}
\end{equation}
On the FM side, $\mathsf H$ is the $2\times2$ Hessian with respect to $(m,q)$ and $\bm v$ is the vector of mixed second derivatives in the $\beta$ direction, defined by
\begin{equation}
\mathsf H_{ab}=\frac{\partial^2\varphi_{\FM}}{\partial a\,\partial b}\bigg|_{\Mpt}
\quad(a,b\in\{m,q\}),\qquad
\bm v=(\varphi_{m\beta},\varphi_{q\beta})^{\!\top},\qquad
\varphi_{a\beta}\equiv\frac{\partial^2\varphi_{\FM}}{\partial a\,\partial\beta}\bigg|_{\Mpt},
\label{eq:Hvdef}
\end{equation}
respectively.

As shown in Eq.~\eqref{eq:vsym} of Appendix~\ref{app:A5}(b), $\varphi_{q\beta}=-\varphi_{m\beta}$ at the triple point. Hence $\bm v=\varphi_{m\beta}(1,-1)^{\!\top}$, with component
\begin{equation}
\varphi_{m\beta}=\frac{\Omega\Lam_c}{\bc}\,B.
\label{eq:vmbmain}
\end{equation}
Here $\Omega=p(p-1)\bc^2\mu^{p-2}/2$ has already appeared in Eq.~\eqref{eq:lamMmain}. In addition, as shown in Eqs.~\eqref{eq:hessian}--\eqref{eq:hessianqq} of Appendix~\ref{app:A5}(a), the elements of $\mathsf H$ are also given in closed form in terms of the same small set of quantities $(\Omega,\mu,\tau)$:
\begin{equation}
\varphi_{mm}=\Omega\bigl((1-\mu)\Omega-1\bigr),\qquad
\varphi_{mq}=-\Omega^2(\mu-\tau),\qquad
\varphi_{qq}=\frac{\Omega}{2}\bigl(1-\Omega(1-4\mu+3\tau)\bigr),
\label{eq:hessmain}
\end{equation}
with $\det\mathsf H=\varphi_{mm}\varphi_{qq}-\varphi_{mq}^2$.

The coefficient $A_{\FM}$ can therefore be evaluated from Eqs.~\eqref{eq:AFMmain}, \eqref{eq:AFM}, \eqref{eq:hessmain}, and \eqref{eq:vmbmain}. The quadratic-form expression in Eq.~\eqref{eq:AFM} follows by inserting $\bm v=\varphi_{m\beta}(1,-1)^{\!\top}$ and the $2\times2$ inverse matrix
\begin{equation}
\mathsf H^{-1}=\frac{1}{\det\mathsf H}
\begin{pmatrix}\varphi_{qq}&-\varphi_{mq}\\ -\varphi_{mq}&\varphi_{mm}\end{pmatrix}
\label{eq:Hinvmain}
\end{equation}
as detailed in Appendix~\ref{app:A5}(d).

The phase boundary is therefore
\begin{equation}
j_0^*(T)-\jM
=K_b\,b^2+O(b^3)
=K\,t^2+O(t^3),\qquad
K_b=\frac{A_{\FM}-A_{\SG}}{2\bc\,\mu^p},\qquad
K=K_b\,\bc^4 .
\label{eq:boundary}
\end{equation}
The difference of the coefficients has a direct thermodynamic interpretation. The common $1/2$ in Eq.~\eqref{eq:Acoef} cancels, giving $A_{\FM}-A_{\SG}=\bigl(\partial^2\hat\varphi_{\SG}/\partial\beta^2-\partial^2\hat\varphi_{\FM}/\partial\beta^2\bigr)|_{\Mpt}$. Using the specific heat $C=\beta^2\,\partial^2\hat\varphi/\partial\beta^2$, we obtain
\begin{equation}
K=\frac{\bc\,\bigl(C_{\SG}-C_{\FM}\bigr)\big|_{\Mpt}}{2\,\mu^p}.
\label{eq:CC}
\end{equation}
Thus the condition that the SG specific heat immediately below the triple point exceeds the FM specific heat, $C_{\SG}>C_{\FM}$, is precisely the criterion for a reentrant transition, $K>0$. The gauge-symmetry constraint [constraint (iii) in Sec.~\ref{sec:constraints}] guarantees $K\ge0$ in advance, provided that the phase boundary defined by equality of the free energies of the two solutions is the physical equilibrium boundary. It is therefore sufficient to establish that $C_{\SG}-C_{\FM}$ is strictly positive.

It is worth noting that the factor $\bc\mu^p$ appearing in the denominator of $K_b$ in Eq.~\eqref{eq:boundary} is the discontinuity in $\partial\hat\varphi/\partial j_0$ across the phase boundary [Eq.~\eqref{eq:c1}], or equivalently $\bc$ times the discontinuity $\Delta(m^p)=\mu^p$, and thus measures the strength of the first-order transition. The factor $2$ originates from the factor $1/2$ in the second-order Taylor coefficient.

\subsubsection{Determination of the parameters}\label{sec:params}

For completeness, we summarize how the parameters appearing in the above formulas are determined. Although Eqs.~\eqref{eq:ASGa}--\eqref{eq:CC} contain several quantities, $\mu,\tau,V,\Lam_c,\bc,\Omega,$ and $B$, the only independent input is $p$. All quantities are determined by the numerical solution of a single transcendental equation and a small number of Gaussian integrals. We collect the defining equations here, at the expense of some repetition. With $h=\Lam+\sqrt{\Lam}\,z$, the one-variable condition [Eq.~\eqref{eq:onevar}] is
\begin{equation}
W(\Lam)=\frac{p-1}{2p}\,\Lam\,\mu(\Lam),
\qquad
W(\Lam)=\int Dz\,\lc h-\ln2-\frac{\Lam}{2},
\qquad
\mu(\Lam)=\int Dz\,\tanh h
\label{eq:paramroot}
\end{equation}
and its positive numerical root gives $\Lam_c$ ($\Lam_c=2.339642$ for $p=3$). The Gaussian averages over the local field $h_c=\Lam_c+\sqrt{\Lam_c}\,z$ at the triple point then determine
\begin{equation}
\begin{split}
&\mu=\int Dz\,\tanh h_c,\qquad
\tau=\int Dz\,\tanh^3h_c, \\
&V=\int Dz\,(\lc h_c)^2-\Bigl(\int Dz\,\lc h_c\Bigr)^{\!2}.
\end{split}
\label{eq:paramgauss}
\end{equation}
The quantity $\mu$ is itself the order parameter at the triple point: $m_{\Mpt}=q_{\Mpt}=q_{1c}=\mu$; see Sec.~\ref{sec:mpoint}. All remaining quantities follow from the algebraic relations
\begin{equation}
\bc^2=\frac{2\Lam_c}{p}\,\mu^{1-p},\qquad
T_c=\frac{1}{\bc},\qquad
\jM=\frac{\bc}{2},\qquad
\Omega=\frac{p(p-1)}{2}\,\bc^2\mu^{p-2},\qquad
B=1-3\mu+2\tau.
\label{eq:paramalg}
\end{equation}
The first relation is obtained by solving the defining equation $\Lam_c=(p/2)\bc^2\mu^{p-1}$ at the triple point for $\bc$. Substitution into Eqs.~\eqref{eq:ASGa}--\eqref{eq:hessmain} gives $A_{\SG}$ and $A_{\FM}$, and Eq.~\eqref{eq:boundary} then yields $K_b$ and $K$. In short, the required numerical work consists only of a single one-dimensional root search and four Gaussian quadratures ($\mu,\tau,W,V$); no large-scale numerical calculation such as Monte Carlo simulation is involved. Below, the term ``numerical'' refers to the evaluation of these closed-form quadratures.

\subsection{Positivity of $K$}\label{sec:evaluation}

\begin{figure}[!htb]
\centering
\includegraphics[scale=1.0]{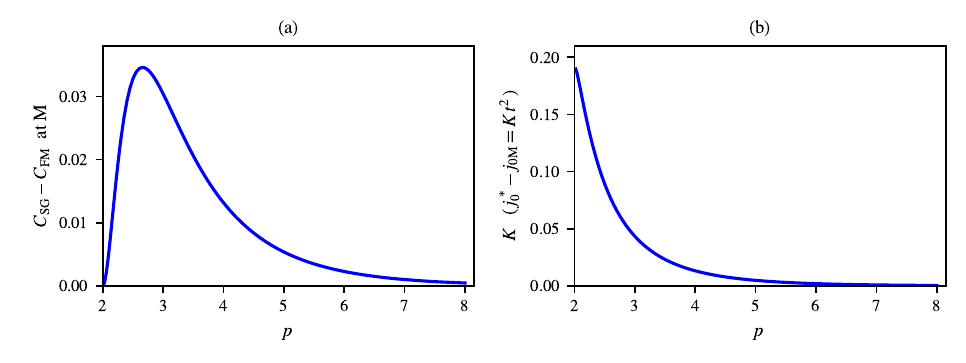}
\caption{$p$ dependence of the driving quantity for reentrance and of the curvature. (a) Specific-heat difference $C_{\SG}-C_{\FM}$ at the triple point. It approaches zero at both endpoints, $p\to2^+$ and $p\to\infty$. As shown in Sec.~\ref{sec:endpoints}, the behavior is $(27/32)(p-2)^2$ for $p\to2^+$ and $2^{-p}/\sqrt p$ for $p\to\infty$ (Appendixes~\ref{app:p2} and \ref{app:largep}). The maximum is $\approx0.035$ at $p\approx2.66$. (b) Curvature $K$ of the phase boundary. It is positive throughout the range scanned; as shown in Sec.~\ref{sec:endpoints}, it approaches the finite value $3/16$ (with logarithmic corrections) as $p\to2^+$ and tends to zero as $p\to\infty$. The numerical data and plotting code used to generate this figure were produced with assistance from the generative-AI tools described in the Methods section.}
\label{fig:pdep}
\end{figure}

The results obtained by evaluating Eqs.~\eqref{eq:ASGa}--\eqref{eq:CC} are shown in Table~\ref{tab:coeffs}. We find $C_{\SG}-C_{\FM}>0$, and hence $K>0$, to all quoted digits for all integers $3\le p\le12$. The decay is approximately geometric. The ratio upon increasing $p$ by one decreases slowly from about $2.3$ to $2.1$; for $p=6\to12$ the successive ratios are $2.31,\,2.22,\,2.17,\,2.13,\,2.11,\,2.10$.
Furthermore, since the reduction to the one-variable equation \eqref{eq:onevar} allows $p$ to be treated as a continuous variable, we can scan $C_{\SG}-C_{\FM}$ continuously in $p$. It remains positive throughout the range examined and has a broad maximum of approximately $0.035$ near $p\simeq2.66$. The result is shown in Fig.~\ref{fig:pdep}.

The smooth behavior over the scanned range, together with the analytically established positivity of the endpoint asymptotic expansions (Sec.~\ref{sec:endpoints}), strongly suggests that $K>0$ for every finite $p>2$.

\begin{table}[htb]
\centering
\caption{Second-order coefficients of the expansion around the triple point. Here $C_{\SG}-C_{\FM}=\bc^2(A_{\FM}-A_{\SG})$ is the specific-heat difference at $\Mpt$, $K$ is the curvature of the phase boundary ($j_0^*-\jM=Kt^2$), and $\lambda(\Mpt)$ is the replicon eigenvalue common to the two branches at the triple point; a positive value means that both branches being compared are replicon stable.}
\label{tab:coeffs}
\begin{tabular}{cccccc}
\toprule
$p$ & $A_{\SG}$ & $A_{\FM}$ & $C_{\SG}-C_{\FM}$ & $K$ & $\lambda(\Mpt)$ \\
\midrule
3 & 0.402977 & 0.415872 & $+0.030392$ & 0.0433296 & $+0.3305$ \\
4 & 0.469474 & 0.474478 & $+0.013149$ & 0.0131910 & $+0.5743$ \\
5 & 0.488151 & 0.490126 & $+0.005364$ & 0.0048492 & $+0.7348$ \\
6 & 0.494915 & 0.495745 & $+0.002280$ & 0.0019756 & $+0.8369$ \\
\midrule
7 & 0.497698 & 0.498062 & $+0.001007$ & 0.0008556 & $+0.9009$ \\
8 & 0.498925 & 0.499090 & $+0.000457$ & 0.0003849 & $+0.9406$ \\
9 & 0.499489 & 0.499565 & $+0.000212$ & 0.0001774 & $+0.9648$ \\
10 & 0.499754 & 0.499790 & $+0.0000995$ & 0.0000831 & $+0.9794$ \\
11 & 0.499881 & 0.499898 & $+0.0000472$ & 0.0000394 & $+0.9881$ \\
12 & 0.499942 & 0.499950 & $+0.0000225$ & 0.0000188 & $+0.9932$ \\
\bottomrule
\end{tabular}
\end{table}

\begin{figure}[!htb]
\centering
\includegraphics[scale=1.0]{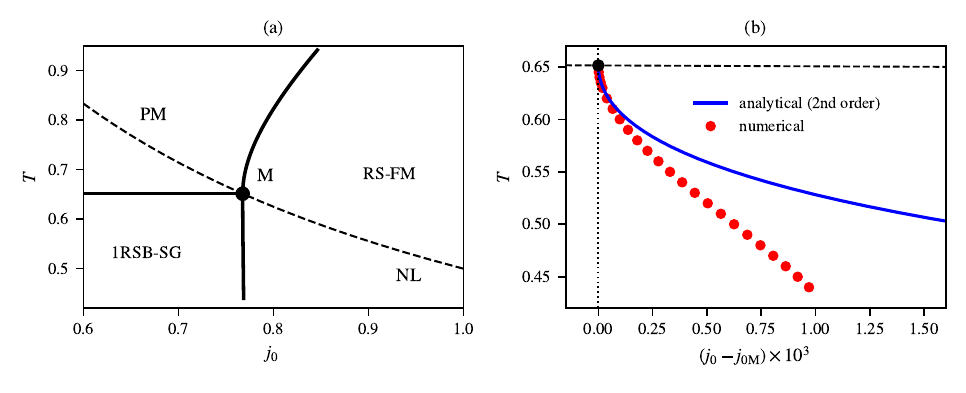}
\caption{Phase diagram for $p=3$. (a) Overall phase diagram. All solid lines are phase boundaries. This is a quantitatively calculated version of the schematic diagram in Fig.~\ref{fig:schematic}, with the spinodal line (thin dashed line) and the AT line (dash-dotted line) omitted. Horizontal line: PM--SG transition ($T_c=0.651385$). Almost vertical line extending downward from $\Mpt$: FM--SG phase boundary. Line extending upward to the right from $\Mpt$: first-order PM--FM transition line. Dashed line: Nishimori line $T=1/(2j_0)$. Black dot: triple point $\Mpt$. On this scale the FM--SG boundary is indistinguishable from a vertical line. (b) Enlarged view near the phase boundary. The horizontal axis is $10^3$ times the deviation from $\jM$. ``numerical'' (red dots): direct determination from equality of the numerically evaluated free energies of the AT-stable RS-FM solution and the replicon-stable 1RSB-SG solution (stability window $0.435\lesssim T<T_c$; Appendix~\ref{app:numeric}). ``analytical (2nd order)'' (solid line): analytical formula $\jM+K_b(\beta-\bc)^2$. The curvature extrapolated from the numerical phase boundary near the triple point agrees with the analytical value of $K_b$ (Sec.~\ref{sec:evaluation} and Appendix~\ref{app:E3}). Vertical dotted line: $j_0=\jM$. Dashed line: NL. The numerical data and plotting code used to generate this figure were produced with assistance from the generative-AI tools described in the Methods section.}
\label{fig:phase}
\end{figure}

We therefore conclude, within the standard RS-FM/1RSB-SG replica mean-field description, that the FM--SG phase boundary is reentrant immediately below the triple point for all integers $3\le p\le12$. The scan over noninteger $p$ strongly suggests the same behavior throughout the range covered. For a fixed small $\epsilon=j_0-\jM>0$, the local boundary $\epsilon=Kt^2+O(t^3)$ is crossed at $t=\sqrt{\epsilon/K}+O(\epsilon)$, so cooling along this vertical line produces the sequence PM $\to$ FM $\to$ SG.
As discussed in Sec.~\ref{sec:mpoint}, this conclusion is based only on the free-energy expansion in the vicinity of the triple point along the stationary branches for which both solutions entering the comparison, namely the AT-stable RS-FM solution and the replicon-stable 1RSB-SG solution, are replicon stable.
In principle, comparison of two replicon-stable solutions leaves open the possibility that a third solution intervenes. As regards a continuous instability, Fig.~7 of Ref.~\cite{Nishimori1999} shows that the AT instability of the RS-FM solution occurs only at temperatures a finite distance below the triple point, so the RS-FM solution is stable in the vicinity of $\Mpt$ (Fig.~\ref{fig:schematic} of the present paper, $T_{\rm AT}\simeq0.435$; Appendix~\ref{app:replicon}). An intervening discontinuous 1RSB-FM solution also appears unlikely. According to the classification of the $p$-spin model with an $r$-body ferromagnetic term by Gillin, Nishimori, and Sherrington \cite{Gillin2001}, the present model with $p=r$ (for $p=3$, $r_1\simeq2.92<r$) lies in the region where no discontinuous 1RSB transition line appears inside the ferromagnetic phase.

A numerical continuation of the phase boundary beyond the immediate neighborhood in which the second-order expansion around the triple point is valid is presented in Appendix~\ref{app:numeric}. For $p=3$, the phase boundary can be determined directly from equality of the free energies of the AT-stable RS-FM solution and the replicon-stable 1RSB-SG solution. Throughout the stability window $0.435\lesssim T<T_c$, the boundary remains monotonically reentrant and joins smoothly onto Eq.~\eqref{eq:boundary} (Fig.~\ref{fig:phase}). This is consistent with the slight bending toward larger bias noticed in the 1RSB phase diagram of Ref.~\cite{ZdeborovaKrzakala2010PRB}. In addition, the curvature (second-order coefficient) near the triple point extrapolated from the numerical boundary agrees well with the analytical result in Table~\ref{tab:coeffs}: for $p=3$, $K=0.04333$ numerically versus $0.043330$ analytically, and for $p=4$, $0.013192$ versus $0.013191$. See Appendix~\ref{app:E3} for details.

\subsection{Limiting cases: The REM and SK models}\label{sec:endpoints}

The purpose of this subsection is to show that the second-order coefficient, or equivalently the specific-heat difference $C_{\SG}-C_{\FM}$, can be obtained analytically at the two endpoints $p\to\infty$ (REM) and $p\to2^+$ (SK), and to clarify the special nature of the two limits. Detailed derivations are given in Appendixes~\ref{app:largep} and \ref{app:p2}; here we state the results and discuss their implications.

\subsubsection{$p\to\infty$ (REM)}\label{sec:REM}
In the REM, the on-shell free energy in the frozen phase is exactly $\hat\varphi_{\SG}(\beta)=\beta\bc/2$ \cite{Derrida1981,Gross1984}. Since this is linear in $\beta$, $\partial^2\hat\varphi_{\SG}/\partial\beta^2=0$, and Eq.~\eqref{eq:Acoef} gives $A_{\SG}=1/2$. On the FM side, the limit $A_{\FM}\to1/2$ is not immediate, but the large-$p$ expansion in Appendix~\ref{app:largep} gives $A_{\FM}=1/2-\pi^3\bar\rho/(16\ln2)+\cdots$, where $\bar\rho=e^{-\Lam_c/2}/\sqrt{2\pi\Lam_c}$, and shows exponential convergence to the same limiting value because $\Lam_c\to\infty$ as $p\to\infty$ [Eq.~\eqref{eq:Lamcasym}]. The numerical sequences in Table~\ref{tab:coeffs} show that both $A_{\SG}$ and $A_{\FM}$ approach $1/2$ monotonically (both are $0.4999$ at $p=12$). Consequently, $C_{\SG}-C_{\FM}=\bc^2(A_{\FM}-A_{\SG})\to0$ and $K\to0$, so the second-order curvature vanishes in the REM limit. The local expansion establishes only this vanishing of the second-order curvature and does not determine the higher-order terms. The verticality of the boundary itself is known from the exact solution of the REM with a ferromagnetic bias \cite{Nishimori2001}: below $T_c$, the free energies per spin of the two phases are $f_{\FM}=-j_0$ and $f_{\SG}=-\sqrt{\ln2}$, so that their equality gives $j_0^*(T)=\sqrt{\ln2}$ at all temperatures in this range, independently of the second-order expansion. The present result is consistent with it. Verticality to all orders also follows from the absence of temperature chaos in the REM \cite{Derrida2021} through the NOO proposition, together with the gauge-symmetry constraint excluding the opposite bending of the FM--SG boundary. The same connection between the absence of chaos and the straight boundary in the REM limit was pointed out in Ref.~\cite{ZdeborovaKrzakala2010PRB}.

The rate of approach can also be determined. The large-$p$ expansion in Appendix~\ref{app:largep} gives
\begin{equation}
C_{\SG}-C_{\FM}\simeq\Bigl(4\pi\ln2-\frac{\pi^3}{4}\Bigr)\frac{e^{-\Lam_c/2}}{\sqrt{2\pi\Lam_c}},
\qquad e^{-\Lam_c/2}\simeq2^{-p}
\end{equation}
so the magnitude, including the algebraic prefactor, scales as $2^{-p}/\sqrt p$. The approximately geometric decay in Table~\ref{tab:coeffs}, with a ratio approaching $2$, reflects this exponential factor $2^{-p}$ [Fig.~\ref{fig:pdep}(a)]. In particular, the positive coefficient of the leading term ($4\pi\ln2>\pi^3/4$) establishes reentrance analytically for sufficiently large $p$.

\subsubsection{$p\to2^+$ (SK)}\label{sec:SK}
For the SK model ($p=2$), the FM (mixed)--SG phase boundary is known to be vertical within the full-RSB (fRSB) solution \cite{Toulouse1980,Parisi1979,Parisi1980a}. On the other hand, the numerical scan in Fig.~\ref{fig:pdep} shows that, as $p\to2^+$, $C_{\SG}-C_{\FM}$ approaches zero [Fig.~\ref{fig:pdep}(a)] whereas $K$ approaches a finite value [Fig.~\ref{fig:pdep}(b)]. There is no contradiction in these behaviors of $C_{\SG}-C_{\FM}$ and $K$: the factor $\mu^p$ in the denominator of Eq.~\eqref{eq:CC} also tends to zero as $p\to2^+$ and compensates the vanishing numerator.

The endpoint behavior can be obtained in closed form from the small-$\Lam$ expansion in Appendix~\ref{app:p2}. As $p\to2^+$, the PM--SG transition approaches the continuous (second-order) SK transition, and the jump of the order parameter at the triple point vanishes linearly as $\mu=(3/2)(p-2)+O((p-2)^2)$. The specific-heat difference and curvature then behave as
\begin{equation}
C_{\SG}-C_{\FM}=\frac{27}{32}(p-2)^2\,\bigl(1+O(p-2)\bigr),
\qquad K\to\frac{3}{16}
\end{equation}
with $\bc\to1$; the logarithmic correction to $K$ is $O(\mu|\ln\mu|)$. Since the coefficient $27/32$ of the leading term is positive, reentrance for $p$ sufficiently close to $2$ follows analytically at leading asymptotic order, just as on the REM side. The numerical scan agrees quantitatively with these closed forms: $\mu/(p-2)\to1.50$ and $(C_{\SG}-C_{\FM})/\mu^p\to0.375$, corresponding to $3/2$ and $3/8$, respectively (Table~\ref{tab:p2} in Appendix~\ref{app:p2}(l)).

The finite limiting value of $K$ implies that the reentrant behavior persists all the way to the limit $p\to2^+$. This value, however, cannot be extrapolated to $p=2$. As $p\to2^+$, the common replicon eigenvalue approaches zero from the positive side as $\lambda(\Mpt)=\mu/3+O(\mu^2)$ and the 1RSB solution becomes marginally stable. Exactly at $p=2$, the appropriate description is fRSB, to which the present analysis does not directly apply. The vertical boundary of the SK model is obtained by the fRSB treatment made at exactly $p=2$, and is not the limit $p\to2^+$ of the present 1RSB curvature. For the same reason, the implicit-function-theorem argument in Appendix~\ref{app:A4}(b${}^{\prime}$) is not uniform in $p$.

In summary, $K(p)>0$ is established analytically by the asymptotic expansions at both endpoints and numerically from values close to $p=2$ through approximately $p=12$. This strongly suggests that the boundary is reentrant for every finite $p>2$. The vertical boundaries at the two endpoints themselves are not derived here; they are known independently, and the present results are consistent with them in different ways. In the REM, the curvature itself smoothly tends to zero, and the model indeed has no temperature chaos. The absence of both reentrance and temperature chaos is consistent with both the NOO proposition and its contrapositive. On the SK side, the 1RSB description ceases to be applicable while the curvature remains finite ($K\to3/16$), with $\lambda(\Mpt)\to0^+$, and at exactly $p=2$ the phase boundary is vertical within fRSB. The contrapositive used in Sec.~\ref{sec:chaos} therefore has no implication in that limiting case. It provides a sufficient condition, whereas the (weak) temperature chaos of the SK model \cite{Rizzo2003} is established by other means.

\section{Temperature chaos}\label{sec:chaos}

In this section, we connect the reentrant transition established in the previous section with the proposition of NOO described in Sec.~\ref{sec:noo} and derive the consequence that there exists a temperature pair in the SG phase for which the distribution of the overlap of spin configurations becomes a delta function at the origin, $P_2(x\mid\beta_1,\beta_2)=\delta(x)$, a necessary condition for temperature chaos. We then test this consequence independently using a two-temperature replica calculation that is logically independent of the proposition and its contrapositive.

\subsection{Existence statement from the contrapositive and its scope}\label{sec:chaoslogic}

In Sec.~\ref{sec:expansion}, we established that, for all integers $3\le p\le12$, the FM--SG phase boundary of the present model is reentrant in a neighborhood immediately below the triple point, where both solutions are replicon stable. The analysis for noninteger $p$ from values close to $2$ through approximately $12$ leads to the same conclusion.

The proposition of NOO was formulated for two-body interactions, but the gauge-transformation argument used in its proof extends directly to the present model by replacing a bond $ij$ with a $p$-body hyperedge $A=(i_1,\ldots,i_p)$. Under $S_i\to S_i\xi_i$ and $J_A\to J_A\xi_A$, with $\xi_A=\prod_{i\in A}\xi_i$, the interaction term $J_A\prod_{i\in A}S_i$ is invariant. The interaction order $p$ enters through the definition of $\xi_A$, but it does not alter the algebraic structure of the gauge-transformation argument.
The contrapositive of the extended proposition then states that, if the phase boundary is reentrant, either there exists at least one temperature pair in the SG phase for which $P_2(x\mid\beta_1,\beta_2)=\delta(x)$ in the thermodynamic limit, or no SG phase exists at finite temperature. In the present model, within the standard replica mean-field analysis, a replicon-stable 1RSB-SG phase exists in the window $T_G<T<T_c$ between the Gardner temperature $T_G$, where the innermost replicon eigenvalue of the 1RSB solution vanishes, and the equilibrium SG transition temperature $T_c$ \cite{Nishimori1999,Gardner1985} ($T_G\simeq0.24$ and $T_c\simeq0.6514$ for $p=3$). It is widely accepted that this replica solution represents the physical equilibrium state, and hence the second alternative in the contrapositive, namely the absence of a finite-temperature SG phase, is excluded. Subject to the additional condition that the RS-FM/1RSB-SG free-energy crossing identified above is the physical equilibrium phase boundary, we obtain the following observation.

\begin{quote}
\textbf{Observation.} Within the standard replica mean-field description, for each $p$ for which reentrance has been established (all integers $3\le p\le12$ and sufficiently large $p$), there must exist at least one pair of distinct temperatures $(T_1,T_2)$ in the finite-temperature SG phase such that $P_2(x\mid\beta_1,\beta_2)=\delta(x)$ in the thermodynamic limit, provided that the RS-FM/1RSB-SG free-energy crossing is the physical equilibrium boundary and that the finite-temperature 1RSB-SG solution represents the physical equilibrium SG phase.
\end{quote}

The existence of the replicon-stable 1RSB interval $T_G<T<T_c$ is used only to guarantee the existence of a finite-temperature SG phase and thereby exclude the second alternative in the contrapositive; it does not imply that both temperatures of the pair whose existence is established lie within this interval.  
Note that the stronger statement of temperature chaos in the strict sense, that the delta-function collapse occurs for every pair of distinct temperatures in the SG phase, does not follow automatically.

The contrapositive gives only this nonconstructive existence statement and does not identify which temperature pair in the SG phase exhibits the delta-function collapse. Nor does it specify at which disorder condition (bias $j_0$) the SG phase containing that pair is located; in particular, temperature chaos in the symmetric model ($j_0=0$) does not follow directly from the contrapositive. In the present mean-field model, however, $j_0$ drops out exactly at the SG saddle point ($m=0$), so at the level of the replica description $P_2$ in the SG phase is independent of the bias. Consequently, if a temperature pair satisfying $P_2(x\mid\beta_1,\beta_2)=\delta(x)$ exists in the SG phase for one value of $j_0$, the same statement carries over to any $j_0$ for which both temperatures lie in the equilibrium SG phase.

As emphasized above, the contrapositive does not specify which temperature pairs exhibit the delta-function collapse. This is again a limitation of the logical implication, and physically it is natural to expect a stronger statement. It would be unusual for temperature pairs with $P_2(x\mid\beta_1,\beta_2)=\delta(x)$ and those without this property to coexist within the same SG phase; temperature chaos is naturally regarded as a property of the phase rather than of individual temperature pairs. Numerical studies of the finite-dimensional Edwards--Anderson model support this view by mutually independent approaches \cite{Katzgraber2007b,Fernandez-2013,Wang2015,Baity-Jesi2021}. Thus, although a mathematical proof is difficult, it is physically plausible that $P_2(x\mid\beta_1,\beta_2)=\delta(x)$ holds for arbitrary distinct temperature pairs in the SG phase. For $p=3$, the state-following calculation of Ref.~\cite{ZdeborovaKrzakala2010PRB} exhibits free-energy level crossings for states in equilibrium at several temperatures spread over the interval $T_G<T<T_c$, which supports this uniform picture at the 1RSB level. The two-temperature replica calculation in the next subsection is also consistent with this uniform picture.

The same logic is not restricted to mean-field models. Since the proof of NOO relies only on gauge symmetry and not on details of the lattice, the proposition also applies to hierarchical lattices, on which a reentrant FM--SG boundary \cite{Guven2008,Hinczewski2005} and temperature chaos at the symmetric point \cite{Banavar1987,Ney-Nifle-1997,Aspelmeier2002,Sasaki2003} have been established in independent lines of work. That case is discussed in Sec.~\ref{sec:conclusion}.

\subsection{Independent test by a two-temperature replica calculation}\label{sec:chaostest}

To test the logical consequence described in the previous section independently, as in Refs.~\cite{Rizzo2001,Rizzo2003}, we performed a two-temperature replica calculation in which two replica species at different inverse temperatures ($n$ replicas at $\beta_1$ and $m$ replicas at $\beta_2$) are placed under the same disorder realization and $[Z_1^nZ_2^m]$ is evaluated by the saddle-point method. No coupling between the two species is added to the Hamiltonian. In Refs.~\cite{Rizzo2001,Rizzo2003}, a term proportional to $\sum_i\sigma_i\tau_i$ is added, and its coefficient acts as a Lagrange multiplier that fixes the overlap between the configurations at the two temperatures; the free energy at fixed overlap is then studied. Here, for a fixed disorder realization the two species are statistically independent, and the cross overlap is treated as a variational parameter determined by stationarity. An effective coupling between the two species, the last term of Eq.~\eqref{eq:Xi2Tmain}, appears only after the disorder average, and its strength is fixed self-consistently by the cross overlap rather than imposed from outside. Here and in Appendix~\ref{app:twotemp}, $m$ denotes the number of replicas at $\beta_2$ and is distinct from the magnetization $m$ used elsewhere in this paper. The full formulation is given in Appendix~\ref{app:twotemp}. Here we summarize the main results, reproducing the principal equations from Appendix~\ref{app:twotemp} so that the logic of the calculation can be followed.

\subsubsection{Two-temperature replica action and overlap distribution}\label{sec:2Taction}
For the replicas $\sigma^a$ at inverse temperature $\beta_1$ and $\tau^\alpha$ at inverse temperature $\beta_2$, we take as variational variables the within-temperature overlaps $q_{ab}$ for $\beta_1$ and $r_{\alpha\beta}$ for $\beta_2$ and the cross overlap $s_{a\alpha}=(1/N)\sum_i\sigma_i^a\tau_i^\alpha$.

It is useful to display the structure used below before writing the formulas. The overlap matrix is of size $(n+m)\times(n+m)$ and consists of $q_{ab}$ ($n\times n$), $r_{\alpha\beta}$ ($m\times m$), and the cross overlap $s_{a\alpha}$ ($n\times m$) together with its transpose. An example of the overlap matrix, with the replicas at the two inverse temperatures arranged consecutively ($n=m=6$, all block sizes equal to 3, and one matched pair with $s_1=s_{a\alpha}$), is
{\setlength{\arraycolsep}{2.2pt}
\begin{equation}
\left(\begin{array}{ccc|ccc||ccc|ccc}
0&\tilde q_1&\tilde q_1&&&&s_1&s_1&s_1&&&\\
\tilde q_1&0&\tilde q_1&&0&&s_1&s_1&s_1&&0&\\
\tilde q_1&\tilde q_1&0&&&&s_1&s_1&s_1&&&\\ \hline
&&&0&q_1&q_1&&&&&&\\
&0&&q_1&0&q_1&&0&&&0&\\
&&&q_1&q_1&0&&&&&&\\ \hline\hline
s_1&s_1&s_1&&&&0&\tilde r_1&\tilde r_1&&&\\
s_1&s_1&s_1&&0&&\tilde r_1&0&\tilde r_1&&0&\\
s_1&s_1&s_1&&&&\tilde r_1&\tilde r_1&0&&&\\ \hline
&&&&&&&&&0&r_1&r_1\\
&0&&&0&&&0&&r_1&0&r_1\\
&&&&&&&&&r_1&r_1&0
\end{array}\right)
\label{eq:examplematmain}
\end{equation}}
where all blank entries are zero. The upper-left $6\times6$ block is $q_{ab}$, the lower-right $6\times6$ block is $r_{\alpha\beta}$, and the upper-right block separated by the double line (together with the lower-left block) is the cross overlap $s_{a\alpha}$. The $3\times3$ block of $s_1$ has no zeros on its diagonal, because $\sigma^a$ and $\tau^\alpha$ belong to different replica sets. Each diagonal block is a 1RSB cluster, and only the first cluster at inverse temperature $\beta_1$ and the first cluster at inverse temperature $\beta_2$ are connected by $s_1\ne0$. This connected pair is referred to as a matched pair; it is the replica representation of a one-to-one correspondence between pure states at the two temperatures. The values inside the pair are denoted by tildes because they generally differ from the equilibrium values $q_1$ and $r_1$ of unpaired clusters. The equal block sizes and the single pair are chosen only for illustration. In the calculation, the unpaired clusters keep the block sizes of the single-temperature 1RSB solutions, whereas the sizes $\tilde x_1$ and $\tilde x_2$ of the two blocks forming a pair are variational parameters that may differ from them and from each other. At fixed $n$ and $m$, a change of $\tilde x_i$ changes the number of unpaired clusters rather than their size, and the stationarity condition for the size of an unpaired cluster coincides with the single-temperature one. The number $k$ of matched pairs is not fixed either: it is not a variational parameter, but all admissible values of $k$ are summed over in Eq.~\eqref{eq:ksummain}. If $s_1$ remains finite, configurations at the two temperatures retain a correlation; if the weight associated with $s_1$ vanishes, the necessary condition for temperature chaos is satisfied.

We restrict below to the SG sector, in which the magnetizations vanish at both temperatures ($m_a=\tilde m_\alpha=0$). The terms containing $j_0$ then drop out (the replica representation of constraint (i) in Sec.~\ref{sec:constraints}), and the replica action becomes
\begin{equation}
G=\frac14\bigl(n\beta_1^2+m\beta_2^2\bigr)
-\frac{p-1}{4}\Bigl(
 \beta_1^2\sum_{a\ne b}q_{ab}^p
 +\beta_2^2\sum_{\alpha\ne\beta}r_{\alpha\beta}^p
 +2\beta_1\beta_2\sum_{a,\alpha}s_{a\alpha}^p
 \Bigr)
 +\ln\operatorname{Tr}e^{\Xi'},
\label{eq:G2Tmain}
\end{equation}
where $\Xi'$ is the one-site exponent containing the conjugate fields,
\begin{equation}
\Xi'=\frac{p}{4}\Bigl(
 \beta_1^2\sum_{a\ne b}q_{ab}^{p-1}\sigma^a\sigma^b
 +\beta_2^2\sum_{\alpha\ne\beta}r_{\alpha\beta}^{p-1}\tau^\alpha\tau^\beta
 \Bigr)
 +\frac{p}{2}\beta_1\beta_2
 \sum_{a,\alpha}s_{a\alpha}^{p-1}\sigma^a\tau^\alpha.
\label{eq:Xi2Tmain}
\end{equation}
In this representation, the two-temperature overlap distribution to be diagnosed is the histogram of cross overlaps,
\begin{equation}
P_2(\zeta)=\lim_{n,m\to0}\frac1{nm}\sum_{a,\alpha}\delta(\zeta-s_{a\alpha}).
\label{eq:P2repmain}
\end{equation}
The condition $P_2(\zeta)=\delta(\zeta)$ means that the overlap of a pair of configurations independently drawn from the Gibbs measures at $\beta_1$ and $\beta_2$ is zero with probability one in the thermodynamic limit, or equivalently that the weight of the normalized overlap distribution is concentrated at $\zeta=0$. Thus typical spin configurations at the two temperatures are uncorrelated, which is a condition for temperature chaos. Since this is a statement about the weight of a distribution, the mere existence of a stationary sector with nonzero overlap whose weight vanishes in the thermodynamic limit does not contradict $P_2(\zeta)=\delta(\zeta)$. The calculation below is designed precisely to diagnose the weight of such a sector.

If only the replicas at inverse temperature $\beta_1$ are retained and the 1RSB ansatz is substituted, the action separates into a term proportional to $n$ and a term proportional to the number of blocks $n/x$, the coefficient of the latter being the contribution per cluster,
\begin{equation}
g_1(x)=-\frac{(p-1)\beta_1^2}{4}x(x-1)q_1^p+\ln I_1(x),
\qquad
I_1(x)=e^{-\Lambda_1x/2}\int Dz\,\bigl(2\cosh\sqrt{\Lambda_1}z\bigr)^x,
\label{eq:g1main}
\end{equation}
where
\begin{equation}
    \Lambda_1=\frac{p\beta_1^2q_1^{p-1}}{2}.
\end{equation}
We call a solution $(q_i,x_i)~(i=1,2)$ of the single-temperature 1RSB equations at inverse temperature $\beta_i$ the equilibrium values, and a block containing them an equilibrium cluster; $g_i(x_i)$ denotes $g_i$ evaluated at these values.

\subsubsection{Matched pairs and the diagnostic quantity}\label{sec:2Tcriterion}
A matched pair is written generally as a pair of blocks $B_1$ and $B_2$, consisting of $\tilde x_1$ replicas at inverse temperature $\beta_1$ and $\tilde x_2$ replicas at inverse temperature $\beta_2$, respectively. Within the pair, we set
\begin{equation}
q_{ab}=\tilde q_1\ (a\ne b\in B_1),\qquad
r_{\alpha\beta}=\tilde r_1\ (\alpha\ne\beta\in B_2),\qquad
s_{a\alpha}=s_1\ (a\in B_1,\ \alpha\in B_2)
\label{eq:pairansatzmain}
\end{equation}
and all overlaps with replicas outside the pair are set to zero. Tildes indicate variational values defined only inside the pair. All five quantities $(\tilde q_1,\tilde x_1;\,\tilde r_1,\tilde x_2;\,s_1>0)$, including the block sizes, are varied. The action of one such pair is
\begin{equation}
g_{12}=-\frac{p-1}{4}\Bigl(
 \beta_1^2\tilde x_1(\tilde x_1-1)\tilde q_1^p
 +\beta_2^2\tilde x_2(\tilde x_2-1)\tilde r_1^p
 +2\beta_1\beta_2\tilde x_1\tilde x_2s_1^p
 \Bigr)+\ln\tilde I,
\label{eq:g12main}
\end{equation}
where $\ln\tilde I$ is the contribution of the paired trace with a two-dimensional Gaussian measure,
\begin{equation}
\tilde I=e^{-(\tilde\Lambda_1\tilde x_1+\tilde\Lambda_2\tilde x_2)/2}
 \int D[h_1,h_2]\,
 (2\cosh h_1)^{\tilde x_1}(2\cosh h_2)^{\tilde x_2}
\label{eq:Itildemain}
\end{equation}
and $D[h_1,h_2]$ is the zero-mean two-dimensional Gaussian measure whose covariance matrix is given by the conjugate fields inside the pair,
\begin{equation}
\tilde\Lambda_1=\frac{p\beta_1^2\tilde q_1^{p-1}}{2},\qquad
\tilde\Lambda_2=\frac{p\beta_2^2\tilde r_1^{p-1}}{2},\qquad
\tilde\Lambda_{12}=\frac{p\beta_1\beta_2s_1^{p-1}}{2}.
\label{eq:Lamtildemain}
\end{equation}
Thus the one-dimensional Gaussian integral in $g_1$ is replaced by a two-dimensional Gaussian integral coupled through the cross overlap, with correlation coefficient
\begin{equation}
\rho=\frac{\tilde\Lambda_{12}}{\sqrt{\tilde\Lambda_1\tilde\Lambda_2}}=\left(\frac{s_1}{\sqrt{\tilde q_1\tilde r_1}}\right)^{p-1}.
\end{equation}

Taking the difference of the actions before and after forming the pair cancels all terms proportional to $n$ and $m$, leaving the action difference per matched pair,
\begin{equation}
\Delta=g_{12}-\frac{\tilde x_1}{x_1}\,g_1(x_1)-\frac{\tilde x_2}{x_2}\,g_2(x_2).
\label{eq:Deltamain}
\end{equation}
The subtraction terms on the right-hand side are the contributions from the equilibrium clusters to which the replicas used in the pair originally belonged. Thus $\Delta$ is the cost relative to the reference state in which the two temperatures are treated independently, without forming a pair.

The way in which $\Delta$ enters the overlap distribution is as follows. In the matched-pair ansatz, every cross-overlap component is either $s_1$ or $0$, so Eq.~\eqref{eq:P2repmain} reduces to two delta functions,
\begin{equation}
P_2(\zeta)=\bigl(1-w_{\rm pair}\bigr)\delta(\zeta)+w_{\rm pair}\,\delta(\zeta-s_1),
\qquad
w_{\rm pair}=\lim_{n,m\to0}\frac{\tilde x_1\tilde x_2}{nm}\,\langle k\rangle,
\label{eq:P2ansatzmain}
\end{equation}
where $k$ is the number of matched pairs, $\langle k\rangle$ is its average, and $w_{\rm pair}$ is the weight of the nonzero-overlap peak. Within the present matched-pair ansatz, the action is linear in $k$. If $M(k)$ denotes the number of ways of choosing $k$ pairs, we can write
\begin{equation}
\bigl[Z_1^nZ_2^m\bigr]\simeq\sum_kM(k)\,e^{N(G(0)+k\Delta)},
\label{eq:ksummain}
\end{equation}
and the distribution of $k$ is weighted by $e^{Nk\Delta}$. Within the present ansatz and the usual saddle-point interpretation, if the stationary value of $\Delta$ (to be denoted as $\Delta^*$) is negative, we identify the leading exponential dependence as $w_{\rm pair}\propto e^{N\Delta^*}$. The contribution from $k\ge2$ is then exponentially small compared with that from $k=1$, and an exponential suppression of $w_{\rm pair}$ with rate $\Delta^*$ follows. The diagnostic criterion is then as follows. If a stationary point with $s_1^*>0$ and $\Delta^*=0$ exists, $w_{\rm pair}$ is not exponentially suppressed. If $\Delta^*<0$, the weight of the nonzero-overlap peak vanishes exponentially in the thermodynamic limit; in the latter case, within the present ansatz, the weight of $P_2(\zeta)$ becomes concentrated at $\zeta=0$, and spin configurations at the two temperatures become uncorrelated.

The results are as follows. For unequal-temperature pairs, a high-overlap stationary point with $s_1^*>0$ exists, but its cost is systematically negative, $\Delta^*\sim-10^{-5}$--$10^{-4}$, increases in magnitude with $|T_1-T_2|$, and connects continuously to zero in the equal-temperature limit (Table~\ref{tab:deltamain}; $p=3$). As $p$ increases, $\Delta^*$ decreases in magnitude toward $0^-$ (Table~\ref{tab:deltapmain}).

\begin{table}[htb]
\centering
\caption{Stationary points in the matched-pair sector for $p=3$ (reproduced from Table~\ref{tab:delta}). For unequal temperatures, $\Delta^*<0$ is obtained systematically.}
\label{tab:deltamain}
\begin{tabular}{cccc}
\toprule
$T_1$ & $T_2$ & $s_1^*$ & $\Delta^*$ \\
\midrule
0.45 & 0.40 & 0.92029 & $-6.79\times10^{-6}$ \\
0.45 & 0.35 & 0.92569 & $-2.20\times10^{-5}$ \\
0.50 & 0.35 & 0.91356 & $-5.71\times10^{-5}$ \\
0.55 & 0.35 & 0.89932 & $-1.15\times10^{-4}$ \\
0.60 & 0.40 & 0.87972 & $-1.59\times10^{-4}$ \\
0.60 & 0.35 & 0.88273 & $-2.03\times10^{-4}$ \\
\bottomrule
\end{tabular}
\end{table}

\begin{table}[htb]
\centering
\caption{$p$ dependence of $\Delta^*$ at fixed reduced temperatures $(T_1/T_c,T_2/T_c)=(0.70,0.50)$ (reproduced from Table~\ref{tab:deltap}).}
\label{tab:deltapmain}
\begin{tabular}{cc}
\toprule
$p$ & $\Delta^*$ \\
\midrule
3 & $-3.35\times10^{-5}$ \\
4 & $-2.13\times10^{-5}$ \\
5 & $-9.83\times10^{-6}$ \\
6 & $-4.42\times10^{-6}$ \\
8 & $-9.29\times10^{-7}$ \\
10 & $-2.05\times10^{-7}$ \\
12 & $-4.65\times10^{-8}$ \\
\bottomrule
\end{tabular}
\end{table}

Thus, for every pair of temperatures in the SG phase examined here, the weight of the most natural matched-pair peak representing the absence of chaos is exponentially suppressed as $e^{N\Delta^*}\to0$. This provides evidence consistent with the delta-function collapse required as an existence statement by the contrapositive of the NOO proposition. It is also in line with the explicit level crossings found for $p=3$ by the state-following method \cite{KrzakalaZdeborova2010EPL,ZdeborovaKrzakala2010PRB}, which computes the free energy of the state in equilibrium at one temperature when the temperature is changed and finds it to lie above the equilibrium free energy; the negative $\Delta^*$ is qualitatively consistent with this, as it expresses, within a different formalism, a thermodynamic suppression of a one-to-one correspondence of states at the two temperatures. The suppression is, however, extremely weak. A rough estimate for the disappearance of the peak is $N\gtrsim1/|\Delta^*|\sim10^4$--$10^5$ for $p=3$.

As $p\to\infty$, the present model reduces to the REM, which is known to have no temperature chaos; equivalently, the nonzero peak of the two-temperature overlap retains an $O(1)$ weight \cite{Derrida2021}. For finite $p$, $\Delta^*$ approaches $0^-$ monotonically as $p$ increases and $s_1^*\to1$ (Table~\ref{tab:deltapmain}), continuously connecting to this nonchaotic REM limit.

If the present ansatz had produced, at finite $p$, a legitimate zero-cost $\Delta^*=0$ structure of finite weight for every pair of distinct temperatures in the SG phase, the no-chaos assumption would have held and, through the forward direction of the proposition, would have contradicted the reentrant transition found in Sec.~\ref{sec:expansion}. No such structure was found. The limitations of the ansatz are discussed in Appendix~\ref{app:F5}, in particular the absence of hierarchical cross-block structures, splitting and merging of states, and fluctuations of block sizes.

\section{Discussion and conclusion}\label{sec:conclusion}

In the present paper, we have determined analytically the shape of the FM--SG phase boundary in the vicinity of the triple point $\Mpt$ for the fully connected Ising $p$-spin glass with a ferromagnetic bias, and have connected the resulting reentrance to the question of temperature chaos.

We first expanded the location of the phase boundary as a function of temperature around the triple point. The first-order term vanishes exactly. The phase boundary is therefore tangent to the temperature axis, i.e., vertical, at the triple point. Its local shape is consequently determined by the second-order coefficient, the curvature. We obtained a closed-form expression showing that this curvature is proportional to the difference between the SG and FM specific heats at the triple point. The geometrical question of whether the phase boundary is reentrant or vertical is thus reduced to the sign of a single thermodynamic quantity, the specific-heat difference at the triple point. It therefore remains only to establish that the specific-heat difference is positive.

We found that the specific-heat difference is positive for all integers $3\le p\le12$, and a numerical evaluation treating $p$ as a continuous variable is consistent with positivity from values close to $2$ through approximately $12$. Also, analytical results are available at both endpoints. In both the REM limit $p\to\infty$ and the SK limit $p\to2^+$, the leading term of the specific-heat difference is positive; it approaches zero as $2^{-p}/\sqrt p$, with exponential factor $2^{-p}$ and algebraic prefactor $p^{-1/2}$, in the former limit and as $(p-2)^2$ in the latter. These results strongly suggest that reentrance occurs for every finite $p>2$. For $p=3$, this reentrance is the slight bending of the boundary noticed in the numerical 1RSB solution of Ref.~\cite{ZdeborovaKrzakala2010PRB}, now characterized by a curvature in closed form; the vanishing of the curvature as $2^{-p}/\sqrt p$ also makes quantitative the remark in that reference that the boundary becomes strictly vertical in the REM limit. A uniform analytical proof over the entire range would require error bounds on the two endpoint asymptotic expansions together with analytical control of the sign over the intervening finite-$p$ interval, and remains an open problem. The two endpoints have different meanings. In the REM limit, the curvature itself tends to zero, consistently with the vertical boundary known from the exact solution of the REM. In the SK limit, by contrast, the specific-heat difference and the order-parameter discontinuity both vanish, leaving a finite limiting curvature. At the same time the replicon eigenvalue approaches zero and the appropriate description changes from 1RSB to fRSB at $p=2$, so our finite-curvature result cannot be extrapolated to the SK model itself.

The second result is obtained by connecting this reentrant transition with the contrapositive of the NOO proposition. The proposition states that the phase boundary cannot be reentrant in the absence of temperature chaos. If the RS-FM/1RSB-SG free-energy crossing identified above is the physical equilibrium boundary and the finite-temperature 1RSB-SG phase of the present model is accepted as a physical equilibrium solution, the second alternative in the contrapositive, namely the absence of a finite-temperature SG phase, is excluded. There must then exist at least one pair of temperatures within the finite-temperature SG phase for which the two-temperature overlap distribution collapses to a delta function. Temperature chaos has long been a difficult problem for numerical investigations; the present results provide a concrete example showing that a nontrivial constraint on two-temperature overlaps can instead be inferred from the thermodynamic shape of a phase boundary. This reverses the direction of the argument in Ref.~\cite{ZdeborovaKrzakala2010PRB}, where the level crossings of Gibbs states obtained by the state-following method for $p=3$ were used to interpret the bending of the boundary. Both approaches exploit gauge symmetry, but the present route does not require following individual Gibbs states or the mapping of state following onto the biased model, which is limited to states taken from equilibrium in the paramagnetic phase between the dynamical transition temperature and $T_c$, and it applies to every $p$ for which reentrance is established.

This consequence is nevertheless nonconstructive. The contrapositive of NOO alone determines neither which temperature pair exhibits the collapse nor at which disorder bias $j_0$ the corresponding SG phase is located, and it does not imply temperature chaos for every pair of distinct temperatures in the SG phase.
Physically, it is difficult to imagine that the presence or absence of chaos depends on the choice of temperature pair within the same SG phase, and it is natural to interpret the result as applying to arbitrary distinct temperature pairs. This interpretation, however, is not a logical consequence of the contrapositive. The independent two-temperature replica calculation performed to test this picture finds that, for every unequal-temperature pair examined, the stationary matched-pair structure corresponding to the absence of temperature chaos has a negative action difference $\Delta^*$ relative to the independent sector. This means that the matched-pair sector with nonzero overlap is thermodynamically disfavored and provides evidence consistent with the existence of temperature chaos. For $p=3$, this is in line with the explicit level crossings of Ref.~\cite{ZdeborovaKrzakala2010PRB}. The magnitude of the cost is small. Using only the exponential factor $e^{N\Delta^*}$ as a guide, the suppression is very weak for system sizes accessible to ordinary numerical calculations, so finite systems may appear nonchaotic. 

Several qualifications accompany these conclusions. The evidence for a positive coefficient of the curvature $K>0$ is analytical at the two endpoints and numerical in the intervening range. The numerical part, however, involves only a finite set of quadratures rather than a large-scale computation such as Monte Carlo simulation, and there is no ambiguity at the level of the numerical errors. The present analysis also does not independently prove that no additional global equilibrium branch intervenes immediately below the triple point between the RS-FM and 1RSB-SG branches compared here (Appendix~\ref{app:numeric}). The contrapositive gives only an existence statement and does not locate the temperature pair. The replicon-stable 1RSB interval used to establish the existence of a finite-temperature SG phase is restricted to $T_G<T<T_c$; at lower temperatures the SG phase itself becomes fRSB, and a different framework is required to follow the phase boundary. A similar qualification applies on the FM side. For $p=3$, the 1RSB mixed phase becomes replicon unstable at low temperatures $T\lesssim0.435$, so a more general RSB mixed phase would have to be constructed to determine the boundary in that region. Finally, the two-temperature replica calculation is based on a restricted matched-pair ansatz and does not include hierarchical cross structures or splitting and merging of states. Our analysis should therefore be regarded as supporting evidence for the main conclusion rather than as a definitive proof.

Even with these reservations, the two main results of the present work are separately well defined. The closed-form reduction of the phase-boundary curvature to the specific-heat difference at the triple point completely characterizes the local FM--SG boundary through second order near the triple point. The logical implication from reentrance to the existence of a temperature pair with a collapsed overlap distribution connects the phase diagram of a mean-field model with a general proposition based on gauge symmetry. Hierarchical lattices provide another setting in which this connection can be tested beyond mean-field theory. The Migdal--Kadanoff recursion is exact there, and reentrant FM--SG boundaries have been obtained in models with a finite-temperature SG phase. In particular, G\"uven et al.\ showed for an anisotropic $d=3$ model that this boundary is always reentrant when the multicritical point terminating it lies on the Nishimori line \cite{Guven2008}; a related multicritical-point relation was obtained in Ref.~\cite{Hinczewski2005}. Temperature chaos at the symmetric point has also been established independently \cite{Banavar1987,Sasaki2003}. Since the NOO proposition does not depend on details of the lattice, whenever a finite-temperature SG phase and a reentrant phase boundary coexist in the same hierarchical model, the same logic applies and provides another arena in which the two phenomena are linked.

Another point of interest concerns finite-dimensional models in which the FM--SG boundary is expected to be a first-order line, such as short-range spin glasses with $p$-body ($p\ge3$) interactions. The ingredients of Sec.~\ref{sec:firstorder} are then essentially the Clausius--Clapeyron relation combined with the exact internal energy on the NL, both of which remain valid in finite dimensions for the corresponding gauge-compatible disorder. Thus, provided that the coexisting free-energy branches are sufficiently smooth and that the discontinuity of the $p$-body correlation conjugate to $j_0$ does not vanish at the multicritical point, the vertical tangency and the relation between the local curvature of the phase boundary and the specific-heat difference have direct finite-dimensional analogues. For continuous transitions, the comparison of two free-energy branches is no longer available, and the local shape of the boundary near the multicritical point would have to be addressed by different methods.

A further implication concerns dynamics. Temperature chaos is conventionally defined as an equilibrium property, but it has long been discussed in connection with the rejuvenation and memory effects observed when a spin glass is aged under temperature cycling \cite{Jonason1998,Bouchaud2001}. More recently, dynamic temperature chaos was identified directly in the off-equilibrium dynamics of the three-dimensional Edwards--Anderson model \cite{Baity-Jesi2021}, and subsequent simulations reproduced memory and rejuvenation effects and found that strong temperature chaos correlates with full rejuvenation \cite{Baity-Jesi2023}. For fully connected $p$-spin models, a classic analytic treatment of off-equilibrium dynamics was developed for the spherical $p$-spin model \cite{Cugliandolo1993}. For the Ising case, the adiabatic evolution of an individual Gibbs state under slow temperature changes has been described by the state-following method of Refs.~\cite{KrzakalaZdeborova2010EPL,ZdeborovaKrzakala2010PRB}; for $p=3$, it gives 1RSB lower bounds on the limiting energy of a slow annealing and the temperatures at which a followed state melts upon heating or undergoes a further replica-symmetry-breaking transition upon cooling. Also for $p=3$, Billoire et al.\ studied equilibrium dynamics numerically and found that the correlation times grow as $\ln\tau\propto N$ \cite{Billoire2005}. Their simulations used an unbiased binary coupling distribution, corresponding to zero ferromagnetic bias (the analogue of $j_0=0$ here), and the correlation-time measurements reported at $T=0.48$ and $0.54$ both lie in the replicon-stable interval $T_G<T<T_c$. The fully connected $p$-spin glass therefore offers a setting in which equilibrium statics is analytically controlled and the response of individual Gibbs states to temperature changes can be computed. It would be interesting to investigate temperature-shift protocols in this model in light of the present results and of the state-following description.

\section*{Methods: Use of generative-AI tools}

Generative-AI tools were used extensively as research assistants in carrying out the analytical and numerical calculations reported in this work. The author determined the physical questions to be addressed, the detailed strategy of the analysis, and the calculations to be performed. Claude Opus 4.8, Claude Opus 5, and Claude Fable 5 (Anthropic) were used primarily to carry out analytical derivations and numerical calculations under detailed instructions from the author.

For the analytical calculations, the author examined and independently verified each step of the AI-generated derivations. GPT-5.5 and GPT-5.6 Sol (OpenAI) were additionally used to independently examine the derivations and identify possible errors or inconsistencies. Issues found either by the author or through this cross-checking process were fed back to the Claude models for correction, after which the revised calculations were again examined by the author. This iterative procedure was repeated throughout the analytical development.

Numerical calculations were cross-checked by having the Claude and GPT models independently generate implementations of the relevant calculations and by comparing their results. The numerical results were tested by the author and the AI tools against independently established analytical results where applicable. All decisions concerning the physical approach, the acceptance or rejection of AI-generated results, and their interpretation were made by the author, who takes full responsibility for the scientific content of this work.

\begin{acknowledgments}
Generative-AI tools, Claude Opus 4.8, Claude Opus 5, and Claude Fable 5 (Anthropic), with additional use of GPT-5.5 and GPT-5.6 Sol (OpenAI) and Gemini 3.1 Pro (Google), were used extensively in preparing the manuscript through iterative interaction with the author (their use in the calculations is described in the Methods section). The author provided detailed and repeated instructions for revision, correction, and restructuring, and reviewed and edited the resulting text. The author takes full responsibility for the final wording and presentation of the manuscript.
\end{acknowledgments}

\section*{Data availability}
The data that support the findings of this article and the source code used to generate them are publicly available \cite{NishimoriData2026}.

\section*{List of symbols}
The symbols used repeatedly in this paper are summarized below together with the places where they are defined.

{\small
\begin{tabular}{@{}lp{95mm}l@{}}
\toprule
\multicolumn{3}{@{}l}{\textbf{Model and conventions (Sec.~\ref{sec:model})}}\\
\midrule
Symbol & Meaning and definition & Defined in\\
\midrule
$p$ & Order of the interaction & Sec.~\ref{sec:constraints}\\
$j_0$ & Ferromagnetic bias specifying the mean coupling & Eq.~\eqref{eq:dist}\\
NL & Nishimori line $j_0=\beta/2$, i.e., $T=1/(2j_0)$ & Sec.~\ref{sec:constraints}\\
$\varphi$ & $-\beta f$, where $f$ is the free energy per spin & Sec.~\ref{sec:mpoint}\\
$\int Dz$ & Standard Gaussian average & Eq.~\eqref{eq:Dz}\\
\bottomrule
\end{tabular}

\vspace{3mm}

\begin{tabular}{@{}lp{95mm}l@{}}
\toprule
\multicolumn{3}{@{}l}{\textbf{Functionals and stationary solutions (Sec.~\ref{sec:mpoint})}}\\
\midrule
$\varphi_{\rm 1RSB}$ & 1RSB functional of Ref.~\cite{Nishimori1999} (variational variables $m,q_0,q_1,x$) & Eq.~\eqref{eq:NWfunc}\\
$\Xi$, $\Lam_a$ & 1RSB effective field and variance, $\Lam_a=(p/2)\beta^2q_a^{p-1}$ & Eq.~\eqref{eq:NWfield}\\
$x$ & 1RSB breakpoint; $x^*(T_c)=1$ at $\Mpt$ & Below Eq.~\eqref{eq:NWfield}; above Eq.~\eqref{eq:q1c}\\
$\varphi_{\SG}(q_1,x;\beta)$ & 1RSB functional in the SG phase & Eq.~\eqref{eq:phiSG}\\
$\Lam_1$ & $(p/2)\beta^2q_1^{p-1}$ (field variance on the SG side) & Eq.~\eqref{eq:phiSG}\\
$g$ & Numerator in $\varphi_{\SG}=\beta^2/4+g/x$ & Eq.~\eqref{eq:gdef}\\
$\varphi_{\FM}(m,q;\beta,j_0)$ & RS functional in the FM phase (variational variables $m,q$) & Eq.~\eqref{eq:phiFM}\\
$\eta$, $H$, $\Lam$ & FM effective field $\eta=H+\sqrt{\Lam}z$, $H=p\beta j_0m^{p-1}$, $\Lam=(p/2)\beta^2q^{p-1}$ & Eq.~\eqref{eq:FMfield}\\
$\varphi_{\PM}$ & Functional in the PM phase $\beta^2/4+\ln2$ & Eq.~\eqref{eq:phiPM}\\
\bottomrule
\end{tabular}

\vspace{3mm}

\begin{tabular}{@{}lp{95mm}l@{}}
\toprule
\multicolumn{3}{@{}l}{\textbf{Triple point $\Mpt$ (Sec.~\ref{sec:mpoint})}}\\
\midrule
$\Mpt=(\jM,T_c)$ & Triple point; $\bc=1/T_c$ and $\jM=\bc/2$ (on the NL) & Sec.~\ref{sec:mpoint}\\
$\mu$ & Scalar characterizing $\Mpt$; $m_{\Mpt}=q_{\Mpt}=q_{1c}=\mu$& Eq.~\eqref{eq:mudef}; below Table~\ref{tab:mpoint}\\
$q_{1c}$ & Innermost 1RSB overlap $q_1$ of the SG solution at $T_c$ & Eq.~\eqref{eq:q1c}\\
$m_{\Mpt}$, $q_{\Mpt}$ & RS order parameters of the FM solution at $\Mpt$ & Below Table~\ref{tab:mpoint}\\
$\Lam_c$ & $(p/2)\bc^2\mu^{p-1}$ & Below Eq.~\eqref{eq:bpcond}\\
$h_c$ & Local field at $\Mpt$, $\Lam_c+\sqrt{\Lam_c}\,z$ & Above Eq.~\eqref{eq:tilted}\\
$\tau$ & $\int Dz\,\tanh^3h_c=\int Dz\,\tanh^4h_c$ & Eqs.~\eqref{eq:lamMmain} and \eqref{eq:moments}\\
$\Omega$ & $p(p-1)\bc^2\mu^{p-2}/2$ ($\Lam_q$ at $\Mpt$) & Eq.~\eqref{eq:lamMmain}\\
$\lambda(\Mpt)$ & Common replicon eigenvalue at $\Mpt$ (general forms in Eqs.~\eqref{eq:repdefRS} and \eqref{eq:repdefSG}) & Eq.~\eqref{eq:lamMmain}\\
$W(\Lam)$ & $\int Dz\,\lc h-\ln2-\Lam/2$, with $h=\Lam+\sqrt{\Lam}z$ & Eq.~\eqref{eq:Wdefmain}\\
\bottomrule
\end{tabular}

\vspace{3mm}

\begin{tabular}{@{}lp{95mm}l@{}}
\toprule
\multicolumn{3}{@{}l}{\textbf{Expansion variables and coefficients (Sec.~\ref{sec:expansion})}}\\
\midrule
$b$, $t$, $\delta$ & $b=\beta-\bc$, $t=T_c-T$, $\delta=j_0^*-\jM$ & Eq.~\eqref{eq:btdelta}\\
$\hat\varphi_{\FM}(\beta,j_0)$, $\hat\varphi_{\SG}(\beta)$ & On-shell free energies, obtained by substituting the stationary solutions into the functionals $\varphi_{\FM}$ and $\varphi_{\SG}$ & Eq.~\eqref{eq:onshell} and below\\
$F(b,\delta)$ & Function defining the phase-boundary condition, $\hat\varphi_{\FM}-\hat\varphi_{\SG}$ & Eq.~\eqref{eq:Fdef}\\
$A_{\SG}$, $A_{\FM}$ & Quadratic coefficients of the deviations from $\varphi_{\PM}$ & Eq.~\eqref{eq:Acoef}\\
$V$ & Variance of $\lc h_c$ & Eq.~\eqref{eq:ASG}\\
$B$ & $1-3\mu+2\tau$ & Eq.~\eqref{eq:AFMmain}\\
$\mathsf H$, $\bm v$ & Hessian in $(m,q)$ and vector of mixed derivatives in the $\beta$ direction & Eq.~\eqref{eq:Hvdef}; Eq.~\eqref{eq:hessmain}\\
$K_b$, $K$ & Phase-boundary curvatures in the $\beta$ and temperature variables, with $K=K_b\bc^4$ & Eq.~\eqref{eq:boundary}\\
$C$ & Thermodynamic specific heat $\beta^2\,\partial^2\hat\varphi/\partial\beta^2$ & Above Eq.~\eqref{eq:CC}\\
\bottomrule
\end{tabular}

\vspace{3mm}

\begin{tabular}{@{}lp{95mm}l@{}}
\toprule
\multicolumn{3}{@{}l}{\textbf{Auxiliary quantities in Appendix~\ref{app:expansion}}}\\
\midrule
$h$ & General NL-type local field $\Lam+\sqrt{\Lam}\,z$ & Beginning of Appendix~\ref{app:expansion}\\
$\theta$, $\lambda$ & Generic notation in Appendix~\ref{app:A1}(d) (variational-variable vector and external variable) & Appendix~\ref{app:A1}(d)\\
$\Lam_q$ & $\partial_q\Lam_1=p(p-1)\beta^2q^{p-2}/2$ & Above Eq.~\eqref{eq:polyderivq}\\
$\langle\cdot\rangle_x$ & Normalized Gaussian average with weight $(2\cosh\sqrt{\Lam_1}z)^x$ & Appendix~\ref{app:A2}(c)\\
$Q(q,x,\beta)$ & $\langle\tanh^2(\sqrt{\Lam_1}z)\rangle_x$ (right-hand side of the $q$ equation) & Eq.~\eqref{eq:Qdef}\\
$\Psi$ & $\varphi_q/(x-1)$ (regularized $q$ stationarity condition) & Eq.~\eqref{eq:Psidef}\\
$\mathsf J$ & Jacobian with respect to $(q,x)$ of the regularized system $(\Psi,\varphi_x)$ & Eq.~\eqref{eq:Jdef}\\
$\dot q$, $\dot x$ & Responses to $\beta$ along the stationary branch & Appendix~\ref{app:A4}(b),(b${}^{\prime}$)\\
\bottomrule
\end{tabular}

\vspace{3mm}

\begin{tabular}{@{}lp{95mm}l@{}}
\toprule
\multicolumn{3}{@{}l}{\textbf{$p\to2^+$ asymptotic expansion (Appendix~\ref{app:p2})}}\\
\midrule
$s$ & $p-2$ & Beginning of Appendix~\ref{app:p2}\\
$L$ & $\mu^{p-2}$ ($\to1$, with logarithmic corrections) & Beginning of Appendix~\ref{app:p2}\\
$G(\mu)$ & $(A_{\FM}-A_{\SG})/L=(3/8)\mu^2+O(\mu^3)$ & Eq.~\eqref{eq:Gmu}\\
\bottomrule
\end{tabular}

\vspace{3mm}

\begin{tabular}{@{}lp{95mm}l@{}}
\toprule
\multicolumn{3}{@{}l}{\textbf{Large-$p$ asymptotic expansion (Appendix~\ref{app:largep})}}\\
\midrule
$\bar\rho$ & $e^{-\Lam_c/2}/\sqrt{2\pi\Lam_c}$  & Beginning of Appendix~\ref{app:largep}\\
$r(h)$ & $\ln(1+e^{-2h})$ ($\lc h=h+r(h)$) & Appendix~\ref{app:largep}(a)\\
$\delta_1$, $\delta_3$ & $1-\mu$, $1-\tau$ & Appendix~\ref{app:largep}(a),(c)\\
$I_m[G]$ & $\int e^h h^m G(h)\,dh$  & Eq.~\eqref{eq:master}\\
\bottomrule
\end{tabular}

\vspace{3mm}

\begin{tabular}{@{}lp{95mm}l@{}}
\toprule
\multicolumn{3}{@{}l}{\textbf{Stability (Appendix~\ref{app:replicon})}}\\
\midrule
$\lambda_{\rm RS}$ & AT (replicon) eigenvalue of the RS-FM solution & Eqs.~\eqref{eq:repdefRS} and \eqref{eq:ATRS}\\
$\lambda_1$ & Innermost replicon eigenvalue of the 1RSB solution & Eqs.~\eqref{eq:NW41} and \eqref{eq:repdefSG}\\
\bottomrule
\end{tabular}

\vspace{3mm}

\begin{tabular}{@{}lp{95mm}l@{}}
\toprule
\multicolumn{3}{@{}l}{\textbf{Two-temperature calculation (Appendix~\ref{app:twotemp})}}\\
\midrule
$P_2$ & Two-temperature overlap distribution \cite{Nishimori2025a} & Eq.~\eqref{eq:overlapx}\\
$\tilde x_1,\tilde x_2,\tilde q_1,\tilde r_1,s_1$ & Internal parameters of a matched pair (all variational) & Eq.~\eqref{eq:pairansatzmain} and below\\
$\Delta$, $\Delta^*$ & Cost per matched pair and its stationary value & Eq.~\eqref{eq:Deltamain}; below Eq.~\eqref{eq:ksummain}\\
\bottomrule
\end{tabular}}

\appendix

\section{Expansion around the triple point}\label{app:expansion}

This appendix gives the details of the expansion presented in Sec.~\ref{sec:expansion}. The main symbols used throughout the paper and the places where they are defined are summarized in the List of symbols above. We set $J=1$ and $\varphi=-\beta f$, with $f$ the free energy per spin. For a general $\Lam>0$, we write an NL-type local field as $h=\Lam+\sqrt{\Lam}\,z$. Under $\int Dz$, $h$ is Gaussian with both mean and variance equal to $\Lam$. The field at the triple point ($\Lam=\Lam_c$) is denoted separately by $h_c=\Lam_c+\sqrt{\Lam_c}\,z$. We also write the general FM effective field as $\eta=H+\sqrt{\Lam}\,z$ [Eq.~\eqref{eq:FMfield}] and distinguish it from $h$.

\subsection{Functionals, stationarity conditions, and the NL identity}\label{app:A1}

We first collect the formulas that will be used repeatedly below. We write the free-energy functionals and stationarity conditions on the SG and FM sides, derive the NL identity for a Gaussian field of the $\Lam$ type, and define the moments $\mu$, $\tau$, and $V$ at the triple point. We then summarize the derivatives with respect to an external variable along a stationary branch.

\subsubsection*{(a) SG side}
Setting $m=q_0=0$ in the 1RSB mixed-phase functional of Ref.~\cite{Nishimori1999} [Eq.~\eqref{eq:NWfunc}] gives Eq.~\eqref{eq:phiSG} for the variational variables $(q_1,x)$, with $0\le q_1\le1$ and $0\le x\le1$:
\begin{equation}
\begin{split}
\varphi_{\SG}(q_1,x;\beta)&=\frac{\beta^2}{4}+\frac{g(q_1,x;\beta)}{x}, \\
g&=-\frac{(p-1)\beta^2}{4}x(x-1)q_1^p-\frac{\Lam_1x}{2}
+\ln\int Dz\,\bigl(2\cosh\sqrt{\Lam_1}z\bigr)^x,
\end{split}
\label{eq:gdef}
\end{equation}
where $\Lam_1=(p/2)\beta^2q_1^{p-1}$. The stationarity conditions are $\partial_{q_1}\varphi=\partial_x\varphi=0$.

\subsubsection*{(b) FM side}
The FM functional is Eq.~\eqref{eq:phiFM}:
\begin{equation}
\varphi_{\FM}(m,q;\beta,j_0)
=\frac{\beta^2}{4}+\frac{(p-1)\beta^2}{4}q^p-(p-1)\beta j_0m^p
-\frac{\Lam}{2}+\int Dz\,\lc\bigl(H+\sqrt{\Lam}\,z\bigr).
\label{eq:phiFMapp}
\end{equation}
We derive its stationarity conditions for later use. Let $\eta=H+\sqrt{\Lam}z$, $H=p\beta j_0m^{p-1}$, and $\Lam=(p/2)\beta^2q^{p-1}$. We use the standard differentiation formulas for a Gaussian average, where $c$ is the mean and $\Lam$ the variance,
\begin{equation}
\frac{\partial}{\partial c}\,\int Dz\,f(c+\sqrt{\Lam}z)=\int Dz\,f',\qquad
\frac{\partial}{\partial \Lam}\,\int Dz\,f(c+\sqrt{\Lam}z)=\frac12\,\int Dz\,f'',
\label{eq:gaussdiff}
\end{equation}
which give
\begin{align}
\frac{\partial\varphi_{\FM}}{\partial m}
&=-p(p-1)\beta j_0m^{p-1}+\frac{\partial H}{\partial m}\,\int Dz\,\tanh \eta
\nonumber\\
&=p(p-1)\beta j_0m^{p-2}\bigl(\int Dz\,\tanh \eta-m\bigr),
\label{eq:statm}\\
\frac{\partial\varphi_{\FM}}{\partial q}
&=\frac{p(p-1)\beta^2}{4}q^{p-1}-\frac{1}{2}\frac{\partial\Lam}{\partial q}
+\frac{\partial\Lam}{\partial q}\cdot\frac12\,\int Dz\,\mathrm{sech}^2\eta
\nonumber\\
&=\frac{p(p-1)\beta^2}{4}q^{p-2}\bigl(q-\int Dz\,\tanh^2\eta\bigr).
\label{eq:statq}
\end{align}
Thus the stationarity conditions are $m=\int Dz\,\tanh\eta$ and $q=\int Dz\,\tanh^2\eta$, as in Eq.~\eqref{eq:FMstat}.

\subsubsection*{(c) NL identity for the local field}
\begin{quote}
\textbf{NL identity}\; (Eq.~\eqref{eq:tilted} in the main text)\label{lem:tilted}\quad
For $h=\Lam+\sqrt{\Lam}\,z$ and every $k\ge1$,
\begin{equation}
   \int Dz\,\tanh^{2k-1}h=\int Dz\,\tanh^{2k}h.
\end{equation}
\end{quote}
\textbf{Derivation.}\quad With the change of variables $h=\sqrt{\Lam}\,y$, where $y=z+\sqrt{\Lam}$,
we have, for any function $f$,
\begin{equation}
\int Dz\,f(h)=e^{-\Lam/2}\int Dy\,e^{\sqrt{\Lam}\,y}\,f(\sqrt{\Lam}\,y).
\end{equation}
Splitting $e^{\sqrt{\Lam}y}=\cosh(\sqrt{\Lam}y)+\sinh(\sqrt{\Lam}y)$ and using parity gives
\begin{align}
\int Dy\,e^{\sqrt{\Lam}y}\,\tanh^{2k-1}(\sqrt{\Lam}y)
&=\int Dy\,\sinh(\sqrt{\Lam}y)\,\tanh^{2k-1}(\sqrt{\Lam}y),\\
\int Dy\,e^{\sqrt{\Lam}y}\,\tanh^{2k}(\sqrt{\Lam}y)
&=\int Dy\,\cosh(\sqrt{\Lam}y)\,\tanh^{2k}(\sqrt{\Lam}y).
\end{align}
Since $\cosh x\,\tanh^{2k}x=\sinh x\,\tanh^{2k-1}x$ identically, the right-hand sides are equal.\hfill$\square$

We use the following notation for averages over the local field $h_c=\Lam_c+\sqrt{\Lam_c}\,z$ at the triple point:
\begin{equation}
\begin{gathered}
\mu=\int Dz\,\tanh h_c=\int Dz\,\tanh^2 h_c,\qquad
\tau=\int Dz\,\tanh^3 h_c=\int Dz\,\tanh^4 h_c,\\
V=\int Dz\,(\lc h_c)^2-\Bigl(\int Dz\,\lc h_c\Bigr)^{\!2}.
\end{gathered}
\label{eq:moments}
\end{equation}
This defines $\mu$, $\tau$, and $V$.

\subsubsection*{(d) Parameter derivatives at a stationary point}
We next prepare the first- and second-derivative formulas for the free energy along a stationary branch when an external variable ($\beta$ or $j_0$) is changed. To keep the structure transparent, we first write the formulas component by component for the FM side, with variational variables $m,q$ and external variable $\beta$ at fixed $j_0$, and then give the general form in Eq.~\eqref{eq:envelope2}. On the SG side one simply replaces $(m,q)$ by $(q_1,x)$.

We follow the notation introduced in Sec.~\ref{sec:firstorder}, Eq.~\eqref{eq:onshell}. Thus $\varphi$ denotes the functional before imposing stationarity, whereas $\hat\varphi$ denotes the on-shell function obtained after substituting the stationary solution. The derivative $\partial\hat\varphi/\partial\beta$ is taken at fixed value of the other external variable ($j_0$ here). By contrast, $\partial_\beta\varphi$ and subscripts such as $\varphi_m\equiv\partial\varphi/\partial m$ denote derivatives of the functional with the variational variables and the other external variable held fixed. In particular, $\partial\hat\varphi/\partial\beta$ is not a derivative along the phase boundary.
\vspace{1mm}\\
\textbf{First derivative.} The stationary solution $(m^*(\beta),q^*(\beta))$ satisfies $\varphi_m=\varphi_q=0$. Differentiating the on-shell function $\hat\varphi(\beta)=\varphi(m^*(\beta),q^*(\beta);\beta)$ gives
\begin{equation}
\frac{\partial\hat\varphi}{\partial\beta}
=\partial_\beta\varphi+\varphi_m\,\dot m^*+\varphi_q\,\dot q^*
=\partial_\beta\varphi
\qquad\Bigl(\dot m^*\equiv\frac{\partial m^*}{\partial\beta},\ \dot q^*\equiv\frac{\partial q^*}{\partial\beta}\Bigr).
\label{eq:first1}
\end{equation}
The last two terms vanish by stationarity. Hence the implicit $\beta$ dependence of the variational variables does not contribute to the first derivative, and $\partial\hat\varphi/\partial\beta=\partial_\beta\varphi$.
\vspace{1mm}\\
\textbf{Response of the stationary point.} At second order this cancellation no longer occurs, and the responses $\dot m^*$ and $\dot q^*$ are needed. The stationarity conditions hold identically along the branch $(m^*(\beta),q^*(\beta))$, so differentiating them with respect to $\beta$ gives
\begin{equation}
0=\frac{\partial}{\partial\beta}\,\varphi_m\bigl(m^*(\beta),q^*(\beta);\beta\bigr)
=\varphi_{mm}\,\dot m^*+\varphi_{mq}\,\dot q^*+\varphi_{m\beta},
\qquad
0=\frac{\partial}{\partial\beta}\,\varphi_q
=\varphi_{qm}\,\dot m^*+\varphi_{qq}\,\dot q^*+\varphi_{q\beta}.
\end{equation}
This is a pair of linear equations for $(\dot m^*,\dot q^*)$, whose solution is
\begin{equation}
\begin{pmatrix}\dot m^*\\ \dot q^*\end{pmatrix}
=-\,\mathsf H^{-1}
\begin{pmatrix}\varphi_{m\beta}\\ \varphi_{q\beta}\end{pmatrix},
\qquad
\mathsf H=
\begin{pmatrix}\varphi_{mm}&\varphi_{mq}\\ \varphi_{mq}&\varphi_{qq}\end{pmatrix}.
\label{eq:response}
\end{equation}
Here $\mathsf H$ is the $2\times2$ Hessian with respect to the variational variables, and we assume $\det\mathsf H\ne0$ so that it can be inverted. Equation~\eqref{eq:response} gives the displacement of the stationary point under a change in $\beta$.
\vspace{1mm}\\
\textbf{Second derivative.} We now differentiate $\partial\hat\varphi/\partial\beta=\partial_\beta\varphi(m^*(\beta),q^*(\beta);\beta)$ once more along the stationary branch. Since $\partial_\beta\varphi$ itself depends on $(m^*,q^*)$,
\begin{equation}
\frac{\partial^2\hat\varphi}{\partial\beta^2}
=\varphi_{\beta\beta}+\varphi_{\beta m}\,\dot m^*+\varphi_{\beta q}\,\dot q^*
\end{equation}
and substitution of Eq.~\eqref{eq:response} gives
\begin{equation}
\frac{\partial^2\hat\varphi}{\partial\beta^2}
=\varphi_{\beta\beta}
-\sum_{a,b\,\in\{m,q\}}\varphi_{a\beta}\,(\mathsf H^{-1})_{ab}\,\varphi_{b\beta}.
\label{eq:second1}
\end{equation}
The first term is the explicit second derivative at fixed variational variables. The second term is the contribution from the response of the stationary point and is a quadratic form involving $\mathsf H^{-1}$.
\vspace{1mm}\\
\textbf{General form.} The preceding calculation does not depend on the names of the variational variables. Let $\theta$ be the two-component vector of variational variables [$(m,q)$ on the FM side and $(q_1,x)$ on the SG side], and let $\lambda$ be an external variable ($\beta$ or $j_0$; this generic symbol is unrelated to the replicon eigenvalue $\lambda(\Mpt)$). Denote the stationary branch by $\theta^*(\lambda)$, its response by $\dot\theta^*\equiv\partial\theta^*/\partial\lambda$, the vector of mixed derivatives by $\varphi_{\theta\lambda}$, and the Hessian by $\mathsf H\equiv\varphi_{\theta\theta}$. Equations~\eqref{eq:first1}, \eqref{eq:response}, and \eqref{eq:second1} then become
\begin{equation}
\frac{\partial\hat\varphi}{\partial\lambda}
=\partial_\lambda\varphi+(\partial_\theta\varphi)^{\!\top}\dot\theta^*
=\partial_\lambda\varphi,
\qquad
\dot\theta^*=-\mathsf H^{-1}\varphi_{\theta\lambda},
\end{equation}
\begin{equation}
\frac{\partial^2\hat\varphi}{\partial\lambda^2}
=\varphi_{\lambda\lambda}-\varphi_{\theta\lambda}^{\!\top}\,\mathsf H^{-1}\,\varphi_{\theta\lambda}.
\label{eq:envelope2}
\end{equation}
Here $(\partial_\theta\varphi)^{\!\top}\dot\theta^*$ is the scalar product $\varphi_m\dot m^*+\varphi_q\dot q^*$, while $\varphi_{\theta\lambda}^{\!\top}\mathsf H^{-1}\varphi_{\theta\lambda}$ is the quadratic form
\begin{equation}
\varphi_{\theta\lambda}^{\!\top}\,\mathsf H^{-1}\,\varphi_{\theta\lambda}
=\sum_{i,j=1}^{2}
\frac{\partial^2\varphi}{\partial\theta_i\,\partial\lambda}\,
(\mathsf H^{-1})_{ij}\,
\frac{\partial^2\varphi}{\partial\theta_j\,\partial\lambda}.
\end{equation}
Since $\lambda$ is a scalar, $\varphi_{\lambda\lambda}$ is also a scalar. Equation~\eqref{eq:envelope2} will be used explicitly in Appendix~\ref{app:A5} on the FM side, with $\theta=(m,q)$ and $\lambda=\beta$, where the three components of $\mathsf H$ and the vector $\varphi_{\theta\lambda}=\bm v$ are evaluated. On the SG side, however, setting $x=1$ allows the Gaussian integral in Eq.~\eqref{eq:phiSG} to be performed exactly and gives $\varphi_{\SG}(q_1,1;\beta)=\varphi_{\PM}(\beta)$, independent of $q_1$. Since the triple point lies on this $x=1$ slice, $\varphi_{q_1q_1}=0$ there. At the triple point, $\varphi_{q_1x}=0$ also holds, as shown in Appendix~\ref{app:A4}(a); hence the Hessian is singular ($\det\mathsf H=0$). Equation~\eqref{eq:envelope2} therefore cannot be applied directly. The SG side is treated separately in Appendix~\ref{app:A4}.

\subsection{Exact identities at the triple point}\label{app:A2}

We derive the two conditions that characterize the triple point $\Mpt$ in the main text, namely the $q_1$ stationarity condition, Eq.~\eqref{eq:q1c}, and the breakpoint condition, Eq.~\eqref{eq:bpcond}, and establish several exact relations that hold there. We start from the degeneracy identity at fixed $x=1$, show that the $x$ derivative gives the breakpoint condition, and then show that the regularized $q$ equation reduces to $q_{1c}=\mu$ as $x\to1^-$. We also verify that the FM solution on the NL obeys the same equation and that $T_{\Mpt}=T_c$ is an identity.

\subsubsection*{(a) Degeneracy at fixed $x=1$}
Setting $x=1$ in Eq.~\eqref{eq:gdef} and using $\int Dz\,2\cosh(\sqrt{\Lam_1}z)=2e^{\Lam_1/2}$ gives
\begin{equation}
\varphi_{\SG}(q_1,1;\beta)=\frac{\beta^2}{4}+\ln2=\varphi_{\PM}(\beta).
\label{eq:slice}
\end{equation}
This identity holds for arbitrary $q_1$ and $\beta$. Differentiating it with respect to $q_1$ and $\beta$ while keeping $x=1$ therefore gives, at the triple point,
\begin{equation}
\varphi_q\big|_{x=1}=\varphi_{qq}\big|_{x=1}=\varphi_{q\beta}\big|_{x=1}=0,\qquad
\varphi_{\beta\beta}\big|_{x=1}=\varphi_{\PM}''=\frac12,
\label{eq:slicederiv}
\end{equation}
where the subscripts denote partial derivatives. We henceforth write $q\equiv q_1$.

\subsubsection*{(b) $x$ derivative and the breakpoint condition}
From $\varphi_{\SG}=\beta^2/4+g/x$, we have $\varphi_x=(xg'-g)/x^2$, where the prime denotes differentiation with respect to $x$. At $x=1$, $\varphi_x=g'(1)-g(1)$. Differentiating $g$ with respect to $x$ gives
\begin{equation}
g'(x)=-\frac{(p-1)\beta^2}{4}(2x-1)q^p-\frac{\Lam_1}{2}
+\frac{\int Dz\,\bigl[(2\cosh\sqrt{\Lam_1}z)^x\,\lc(\sqrt{\Lam_1}z)\bigr]}{\int Dz\,(2\cosh\sqrt{\Lam_1}z)^x}.
\end{equation}
At $x=1$, the Gaussian average with weight $2\cosh(\sqrt{\Lam_1}z)\propto e^{\sqrt{\Lam_1}z}+e^{-\sqrt{\Lam_1}z}$ can be converted into a shifted Gaussian average. For an even function $f$,
\begin{equation}
\frac{\int Dz\,\bigl[2\cosh(\sqrt{\Lam_1}z)\,f(\sqrt{\Lam_1}z)\bigr]}{\int Dz\,2\cosh(\sqrt{\Lam_1}z)}
=\int Dz\,f\bigl(\Lam_1+\sqrt{\Lam_1}\,z\bigr).
\label{eq:tiltshift}
\end{equation}
Therefore, at $x=1$,
\begin{equation}
\varphi_x\big|_{x=1}
=\int Dz\,\lc h-\frac{\Lam_1}{2}-\frac{(p-1)\beta^2}{4}q^p-\ln2
=W(\Lam_1)-\frac{(p-1)\beta^2}{4}q^p.
\label{eq:phix}
\end{equation}
Here,
\begin{equation}
    W(\Lam)=\int Dz\,\lc(h)-\ln2-\Lam/2, \quad h=\Lam+\sqrt{\Lam}\,z.
\end{equation}
The breakpoint condition $\varphi_x|_{x=1}=0$ is Eq.~\eqref{eq:bpcond} of the main text. Using $(p-1)\beta^2q^p/4=(p-1)\Lam_1q/(2p)$ reduces it to the one-variable form Eq.~\eqref{eq:onevar}.

\subsubsection*{(c) The $q$ equation as $x\to1^-$}
We first clarify the logical role of this step. At the triple point, $x=1$ and $q_0=0$, so the thermodynamic overlap in the SG state, $[\langle S_i\rangle^2]=xq_0+(1-x)q_1$ in 1RSB notation, vanishes. Thus the gauge-symmetry identity $m=q$ is trivially satisfied as $0=0$ in the SG state and does not constrain the intracluster order parameter $q_1$.

We now derive the $q$ stationarity condition for general $x$. The $q$ dependence of $g$ in Eq.~\eqref{eq:gdef} occurs in two polynomial terms and in the logarithmic term $\ln I$, with $I(q,x)\equiv\int Dz\,(2\cosh\sqrt{\Lam_1}z)^x$ and $\Lam_1=(p/2)\beta^2q^{p-1}$. Writing $\Lam_q\equiv\partial_q\Lam_1=p(p-1)\beta^2q^{p-2}/2$, and using $\Lam_qq=p(p-1)\beta^2q^{p-1}/2$, the derivatives of the two polynomial terms are
\begin{equation}
\partial_q\Bigl[-\frac{(p-1)\beta^2}{4}\,x(x-1)\,q^p\Bigr]=-\frac{\Lam_q}{2}\,x(x-1)\,q,
\qquad
\partial_q\Bigl[-\frac{\Lam_1 x}{2}\Bigr]=-\frac{x}{2}\,\Lam_q.
\label{eq:polyderivq}
\end{equation}
The derivative of the integrand in the logarithmic term with respect to $\Lam_1$ is
\begin{equation}
\partial_{\Lam_1}\bigl(2\cosh\sqrt{\Lam_1}z\bigr)^{x}
=\frac{x\,z}{2\sqrt{\Lam_1}}\,\bigl(2\cosh\sqrt{\Lam_1}z\bigr)^{x}\tanh\bigl(\sqrt{\Lam_1}z\bigr),
\end{equation}
which brings down a factor $z$. We remove it by Gaussian integration by parts. For convenience, define
\[
F=(2\cosh\sqrt{\Lam_1}z)^x\tanh(\sqrt{\Lam_1}z).
\]
Then
\begin{equation}
F'(z)=\sqrt{\Lam_1}\,\bigl(2\cosh\sqrt{\Lam_1}z\bigr)^{x}\Bigl(x\tanh^2\bigl(\sqrt{\Lam_1}z\bigr)+\mathrm{sech}^2\bigl(\sqrt{\Lam_1}z\bigr)\Bigr).
\end{equation}
If $\langle\cdot\rangle_x$ denotes the normalized average with weight $(2\cosh\sqrt{\Lam_1}z)^x$, this gives
\begin{equation}
\partial_q\ln I=\frac{\Lam_q\,x}{2}\Bigl(1+(x-1)\,\bigl\langle\tanh^2\bigl(\sqrt{\Lam_1}z\bigr)\bigr\rangle_x\Bigr).
\label{eq:dlnIq}
\end{equation}
We use the abbreviation
\begin{equation}
Q(q,x,\beta)\equiv\bigl\langle\tanh^2\bigl(\sqrt{\Lam_1}z\bigr)\bigr\rangle_x
\label{eq:Qdef}
\end{equation}
below.

Adding the three contributions, i.e., the derivatives of the two polynomial terms in Eq.~\eqref{eq:polyderivq} and the logarithmic contribution in Eq.~\eqref{eq:dlnIq}, the first term $\Lam_qx/2$ on the right-hand side of Eq.~\eqref{eq:dlnIq} cancels exactly against the second polynomial derivative $-x\Lam_q/2$. The remaining terms factorize as
\begin{equation}
\partial_q g=\frac{\Lam_q\,x}{2}\Bigl(-(x-1)\,q-1+1+(x-1)\,Q\Bigr)=\frac{\Lam_q\,x}{2}\,(x-1)\,\bigl(Q-q\bigr).
\end{equation}
Since $\varphi_q=(\partial_qg)/x$, we obtain
\begin{equation}
\varphi_q=\frac{\Lam_q}{2}\,(x-1)\,\bigl(Q(q,x,\beta)-q\bigr),\qquad
Q(q,x,\beta)=\frac{\int Dz\,\bigl(2\cosh\sqrt{\Lam_1}z\bigr)^{x}\tanh^2\!\bigl(\sqrt{\Lam_1}z\bigr)}{\int Dz\,\bigl(2\cosh\sqrt{\Lam_1}z\bigr)^{x}}.
\label{eq:phiqfactor}
\end{equation}
For $x\ne1$, the 1RSB stationarity condition $\varphi_q=0$ is therefore the self-consistency equation $q=Q$, i.e., $q=\langle\tanh^2\rangle_x$ with the $(2\cosh)^x$ weight. At $x=1$, Eq.~\eqref{eq:tiltshift} gives $Q(q,1,\beta)=\int Dz\,\tanh^2h$, with $h=\Lam_1+\sqrt{\Lam_1}\,z$. Hence the $q$ equation becomes $q=\int Dz\,\tanh^2h$. By the NL identity, Eq.~\eqref{eq:tilted}, this is identical to the FM equation on the NL, $\mu=\int Dz\,\tanh h$. It follows that $q_{1c}=\mu$.

\subsubsection*{(d) FM solution on the NL}
In part (c), the gauge-symmetry identity $m=q$ was trivial in the SG state and did not determine $q_1$. Here we instead verify that the RS ansatz on the FM side satisfies the same requirement $m=q$ for an equilibrium state on the NL. On the NL, $j_0=\beta/2$, so $H=(p/2)\beta^2m^{p-1}$. Setting $m=q$ gives $H=\Lam$, and the FM effective field $\eta$ has the same distribution as the NL-type field $h=\Lam+\sqrt{\Lam}\,z$. The stationarity conditions, Eq.~\eqref{eq:FMstat}, become $m=\int Dz\,\tanh h$ and $q=\int Dz\,\tanh^2h$, which are identical by Eq.~\eqref{eq:tilted}. Thus the FM solution on the NL forms a one-parameter family $m=q=\mu(\Lam)$, and at $T_c$ it reduces to the same equation as in part (c): $m_{\Mpt}=q_{\Mpt}=q_{1c}=\mu$, as stated in Sec.~\ref{sec:mpoint}.

\subsubsection*{(e) $T_{\Mpt}\equiv T_c$}
We next evaluate the difference between the FM and PM free energies on the NL for the solution $m=q=\mu$. Substituting $j_0=\beta/2$ into Eq.~\eqref{eq:phiFM}, the polynomial terms become
\begin{equation}
\frac{(p-1)\beta^2}{4}\mu^p-(p-1)\beta\cdot\frac{\beta}{2}\mu^p
=-\frac{(p-1)\beta^2}{4}\mu^p
\end{equation}
so that
\begin{equation}
\varphi_{\FM}\big|_{\rm NL}-\varphi_{\PM}
=\int Dz\,\lc h-\ln2-\frac{\Lam}{2}-\frac{(p-1)\beta^2}{4}\mu^p
=W(\Lam)-\frac{(p-1)\beta^2}{4}\mu^p .
\label{eq:NLcross}
\end{equation}
The right-hand side is identical to $\varphi_x|_{x=1}$ in Eq.~\eqref{eq:phix}, with $q=\mu$. Thus the condition for equality of the FM and PM free energies along the NL is the same equation as the 1RSB breakpoint condition $x^*(T_c)=1$. It follows identically that $T_{\Mpt}=T_c$, as stated in Sec.~\ref{sec:mpoint}.

\subsection{Quadratic coefficient $A_{\SG}$ on the SG side}\label{app:A4}

We evaluate the second derivative along the SG branch $(q^*(\beta),x^*(\beta))$, with $x^*(\bc)=1$ and $x^*<1$ for $\beta>\bc$.

\subsubsection*{(a) Structure of the degeneracy}
Equation~\eqref{eq:slicederiv} gives $\varphi_{qq}=\varphi_{q\beta}=0$ at the triple point. We first show that the mixed derivative $\varphi_{qx}$ also vanishes there. The quantity $\varphi_x|_{x=1}$ in Eq.~\eqref{eq:phix} is a function of $q$, with $\Lam_1(q)=(p/2)\beta^2q^{p-1}$ and $\partial_q\Lam_1=\Omega=p(p-1)\beta^2q^{p-2}/2$. Since the Gaussian field has both mean and variance equal to $\Lam_1$, both parts of the differentiation formula~\eqref{eq:gaussdiff} contribute:
\begin{equation}
\frac{d}{d\Lam}\int Dz\,\lc\bigl(\Lam+\sqrt{\Lam}\,z\bigr)
=\int Dz\,\tanh h+\frac12\,\int Dz\,\mathrm{sech}^2h .
\label{eq:dW}
\end{equation}
Using $\int Dz\,\tanh h_c=\mu$, $\int Dz\,\mathrm{sech}^2h_c=1-\mu$, and $p(p-1)\beta^2q^{p-1}/4=(p-1)\Lam_1/2=\Omega\mu/2$, we obtain at the triple point
\begin{equation}
\varphi_{qx}\Big|_{\Mpt}
=\Omega\Bigl(\mu+\frac{1-\mu}{2}-\frac12\Bigr)-\frac{(p-1)\Lam_c}{2}
=\frac{\Omega\mu}{2}-\frac{\Omega\mu}{2}=0 .
\label{eq:phiqx}
\end{equation}
Again the NL identity, Eq.~\eqref{eq:tilted}, is essential for this cancellation.

\subsubsection*{(b) Second derivative of the on-shell function}
Differentiating the stationarity conditions $\varphi_q=\varphi_x=0$ with respect to $\beta$ gives
\begin{equation}
\begin{pmatrix}\varphi_{qq}&\varphi_{qx}\\ \varphi_{xq}&\varphi_{xx}\end{pmatrix}
\begin{pmatrix}\dot q\\ \dot x\end{pmatrix}
=-\begin{pmatrix}\varphi_{q\beta}\\ \varphi_{x\beta}\end{pmatrix}.
\end{equation}
At the triple point, the first row degenerates to $0=0$ in the $q$ direction. This is not a singularity of the 1RSB branch itself. Rather, setting $x=1$ in Eq.~\eqref{eq:phiSG} allows the Gaussian integral to be carried out exactly and gives $\varphi(q,1;\beta)=\beta^2/4+\ln2=\varphi_{\PM}(\beta)$, independent of $q$. Thus the degeneracy is a degeneracy associated with the flat $q$ direction on the $x=1$ slice. As shown in part (b${}^{\prime}$), removing the known factor $x-1$ from $\varphi_q$ regularizes the stationarity equations; the implicit-function theorem then guarantees a unique smooth branch through the triple point and a finite $\dot q$. Moreover, $\dot q$ drops out of the second derivative below, and the expression for $A_{\SG}$ does not depend on the detailed motion of the branch. The second row gives
\begin{equation}
\dot x=-\frac{\varphi_{x\beta}}{\varphi_{xx}}.
\end{equation}
The second derivative is then
\begin{equation}
\frac{\partial^2\hat\varphi_{\SG}}{\partial\beta^2}
=\varphi_{\beta\beta}+\varphi_{\beta q}\dot q+\varphi_{\beta x}\dot x
=\frac12-\frac{\varphi_{x\beta}^2}{\varphi_{xx}}.
\label{eq:d2SG}
\end{equation}
Here $\varphi_{\beta\beta}=1/2$ and $\varphi_{\beta q}=0$ follow from Eq.~\eqref{eq:slicederiv}; the $\dot q$ term vanishes because its coefficient $\varphi_{\beta q}$ is zero. Therefore, from the definition~\eqref{eq:Acoef}, $A_{\SG}=1/2-\partial^2\hat\varphi_{\SG}/\partial\beta^2=\varphi_{x\beta}^2/\varphi_{xx}$, with $1/2=\partial^2\varphi_{\PM}/\partial\beta^2$.

\subsubsection*{(b${}^{\prime}$) Regularization of the degeneracy and the implicit-function theorem}
The purpose of this step is to justify the smooth branch and the finiteness of $\dot q$ assumed in part (b). As shown in Appendix~\ref{app:A2}(c), Eq.~\eqref{eq:phiqfactor}, the $q$ stationarity condition factorizes exactly as $\varphi_q=(\Lam_q/2)(x-1)\bigl[Q(q,x,\beta)-q\bigr]$. Removing the known factor $x-1$, define
\begin{equation}
\Psi(q,x,\beta)\equiv\frac{\varphi_q}{x-1}=\frac{\Lam_q}{2}\,\bigl(Q(q,x,\beta)-q\bigr).
\label{eq:Psidef}
\end{equation}
This function extends smoothly to $x=1$. The SG branch can therefore be defined by the regularized system $\Psi=0$ and $\varphi_x=0$, which removes the apparent degeneracy at $x=1$. Only the $q$ equation is replaced by its regularized form: $\varphi_q$ vanishes identically at $x=1$ because of the factor $x-1$, whereas the $x$ equation remains nondegenerate and is kept unchanged. Its value at $x=1$ is precisely the breakpoint equation~\eqref{eq:phix}.

For fixed $\beta$, regard these two equations as a map $(q,x)\mapsto(\Psi,\varphi_x)$. Its Jacobian is
\begin{equation}
\mathsf J\equiv
\begin{pmatrix}
\partial_q\Psi & \partial_x\Psi\\[2pt]
\partial_q\varphi_x & \partial_x\varphi_x
\end{pmatrix}
=
\begin{pmatrix}
\Psi_q & \Psi_x\\[2pt]
\varphi_{xq} & \varphi_{xx}
\end{pmatrix}.
\label{eq:Jdef}
\end{equation}
The first row contains the derivatives of the regularized $q$ equation $\Psi$, and the second row those of the $x$ equation $\varphi_x$, both with respect to $(q,x)$. We evaluate the matrix at the triple point, where $q=\mu$, $x=1$, and $\beta=\bc$. At $x=1$, the average with weight $(2\cosh\sqrt{\Lam_1}z)^x$ becomes the Gaussian NL average over $h_c=\Lam_c+\sqrt{\Lam_c}\,z$ by Eq.~\eqref{eq:tiltshift}. Hence $Q(\mu,1,\bc)=\mu$, so $\Psi(\Mpt)=0$ and the triple point is indeed a solution of the regularized system. Expanding the $2\times2$ determinant in Eq.~\eqref{eq:Jdef} and using $\varphi_{xq}(\Mpt)=0$ from part (a), Eq.~\eqref{eq:phiqx}, gives
\begin{equation}
\det\mathsf J_{\Mpt}=\bigl(\Psi_q\varphi_{xx}-\Psi_x\varphi_{xq}\bigr)\Big|_{\Mpt}=\Psi_q(\Mpt)\,\varphi_{xx}(\Mpt),
\end{equation}
which is the desired simplification.

The derivative $\Psi_q(\Mpt)$ is directly related to the replicon eigenvalue. Differentiating Eq.~\eqref{eq:Psidef} with respect to $q$ gives $\Psi_q=(\Lam_q/2)(\partial_qQ-1)+(\partial_q\Lam_q/2)(Q-q)$. At the triple point, the second term vanishes because $Q-q=0$. With $\Lam_q|_{\Mpt}=\Omega$, as defined in Sec.~\ref{sec:mpoint}, we obtain $\Psi_q(\Mpt)=(\Omega/2)(\partial_qQ|_{\Mpt}-1)$. It remains to evaluate $\partial_qQ|_{\Mpt}$.

Since $x$ is held fixed at 1, we may first use Eq.~\eqref{eq:tiltshift} to write $Q(q,1,\beta)=\int Dz\,\tanh^2h$, where $h=\Lam_1+\sqrt{\Lam_1}\,z$, and then differentiate with respect to $q$. The $q$ dependence enters only through $\Lam_1$, so $\partial_qQ=\Lam_q\,\partial_{\Lam_1}Q$. Because both the mean and variance of $h$ are $\Lam_1$, differentiation with respect to $\Lam_1$ changes both simultaneously. Applying Eq.~\eqref{eq:gaussdiff} to $f=\tanh^2$ therefore gives
\begin{equation}
\partial_{\Lam_1}\int Dz\,\tanh^2h=\int Dz\,(\tanh^2)'(h)+\frac12\int Dz\,(\tanh^2)''(h).
\label{eq:dQdL}
\end{equation}
The derivatives are $(\tanh^2)'=2\tanh\,\mathrm{sech}^2=2(\tanh-\tanh^3)$ and $(\tanh^2)''=2(1-4\tanh^2+3\tanh^4)$. At the triple point, $h$ has the NL distribution $h_c$. Using the $k=1,2$ cases of Eq.~\eqref{eq:tilted}, namely $\int Dz\,\tanh h_c=\int Dz\,\tanh^2h_c=\mu$ and $\int Dz\,\tanh^3h_c=\int Dz\,\tanh^4h_c=\tau$, we find
\begin{equation}
\partial_{\Lam_1}Q\Big|_{\Mpt}=2\,(\mu-\tau)+\bigl(1-4\mu+3\tau\bigr)=1-2\mu+\tau.
\label{eq:dQeval}
\end{equation}
Together with $\partial_qQ|_{\Mpt}=\Lam_q|_{\Mpt}\,\partial_{\Lam_1}Q|_{\Mpt}$, this gives
\begin{equation}
\frac{\partial Q}{\partial q}\Big|_{\Mpt}=\Omega\,(1-2\mu+\tau)
\quad\Longrightarrow\quad
\Psi_q(\Mpt)=\frac{\Omega}{2}\bigl(\Omega(1-2\mu+\tau)-1\bigr)=-\frac{\Omega}{2}\,\lambda(\Mpt).
\label{eq:PsiqM}
\end{equation}
Thus the replicon eigenvalue in Eq.~\eqref{eq:lamM} appears directly, and
\begin{equation}
\det\mathsf J_{\Mpt}=-\frac{\Omega}{2}\,\lambda(\Mpt)\,\varphi_{xx}(\Mpt).
\label{eq:detJM}
\end{equation}
If $\lambda(\Mpt)\ne0$ and $\varphi_{xx}(\Mpt)\ne0$, the implicit-function theorem gives a unique smooth local branch $(q(\beta),x(\beta))$ through the triple point. For the integers $3\le p\le12$ evaluated numerically, $\lambda(\Mpt)>0$ (Tables~\ref{tab:mpoint} and \ref{tab:coeffs}) and $\varphi_{xx}(\Mpt)>0$ [Eq.~\eqref{eq:phixx}, consistent with $A_{\SG}>0$], and hence the nondegeneracy condition is satisfied.

To make this conclusion explicit, differentiate the regularized conditions $\Psi=0$ and $\varphi_x=0$ with respect to $\beta$. In addition to $\dot x=-\varphi_{x\beta}/\varphi_{xx}$ from part (b), one obtains
\begin{equation}
\dot q=-\frac{\Psi_\beta+\Psi_x\,\dot x}{\Psi_q}.
\label{eq:qdot}
\end{equation}
The denominator in the first expression is $\Psi_q(\Mpt)=-(\Omega/2)\lambda(\Mpt)\ne0$ by Eq.~\eqref{eq:PsiqM}, and the denominator of $\dot x$ is $\varphi_{xx}(\Mpt)\ne0$. Both responses are therefore finite. These conditions are precisely equivalent to $\det\mathsf J_{\Mpt}\ne0$. This justifies the omission of the $\dot q$ contribution in part (b), where its coefficient $\varphi_{\beta q}(\Mpt)$ vanishes in Eq.~\eqref{eq:d2SG}, and establishes $A_{\SG}=\varphi_{x\beta}^2/\varphi_{xx}$. The argument is not uniform as $p\to2^+$, because both $\lambda(\Mpt)$ and $\varphi_{xx}(\Mpt)$ tend to zero. It is nevertheless valid for every fixed $p>2$ satisfying the nondegeneracy conditions, in particular for all integers $3\le p\le12$ considered in the main text.

\subsubsection*{(c) Evaluation of $\varphi_{x\beta}$}
Differentiate Eq.~\eqref{eq:phix} with respect to $\beta$ at fixed $q=\mu$. Using $\partial_\beta\Lam_1=2\Lam_1/\beta$ and Eq.~\eqref{eq:dW},
\begin{equation}
\begin{split}
\varphi_{x\beta}
&=\frac{2\Lam_c}{\bc}\Bigl(\mu+\frac{1-\mu}{2}-\frac12\Bigr)-\frac{(p-1)\bc}{2}\mu^p \\
&=\frac{\Lam_c\,\mu}{\bc}-\frac{(p-1)\bc}{2}\mu^p
=\frac{p\bc}{2}\mu^p-\frac{(p-1)\bc}{2}\mu^p
=\frac{\bc\,\mu^p}{2},
\end{split}
\label{eq:phixb}
\end{equation}
where we used $\Lam_c\mu/\bc=(p/2)\bc\mu^p$.

\subsubsection*{(d) Evaluation of $\varphi_{xx}$}
For $\varphi=\beta^2/4+g/x$, the second derivative with respect to $x$ is $(g/x)''=g''/x-2g'/x^2+2g/x^3$, which at $x=1$ becomes $g''(1)-2g'(1)+2g(1)$. The identity~\eqref{eq:slice} at fixed $x=1$ [Appendix~\ref{app:A2}(a)] gives $g(1)=\ln2$. In addition, $\varphi_x|_{x=1}=g'(1)-g(1)$ [Appendix~\ref{app:A2}(b)] and the breakpoint condition $\varphi_x|_{x=1}=0$ at the triple point, Eq.~\eqref{eq:bpcond}, imply $g'(1)=g(1)$. Hence $g'(1)=g(1)=\ln2$, the last two terms cancel, and $\varphi_{xx}|_{\Mpt}=g''(1)$. The second derivative of the logarithmic term $\ln\int Dz\,e^{xL}$, with $L\equiv\lc(\sqrt{\Lam_1}z)$, is the variance of $L$ under the $(2\cosh)^x$ weight. At $x=1$, Eq.~\eqref{eq:tiltshift} converts this to the Gaussian variance $V$. The first term in $g$ contributes $-(p-1)\beta^2q^p/2$. Therefore
\begin{equation}
\varphi_{xx}\Big|_{\Mpt}=V-\frac{(p-1)\bc^2}{2}\mu^p .
\label{eq:phixx}
\end{equation}
Using the breakpoint condition~\eqref{eq:onevar}, this may also be written as $\varphi_{xx}=V(\Lam_c)-2W(\Lam_c)$. Thus, on the triple-point manifold, $\varphi_{xx}$ is a function of $\Lam_c$ alone.

Combining Eqs.~\eqref{eq:phixb} and \eqref{eq:phixx} gives
\begin{equation}
A_{\SG}=\frac{(\bc\mu^p/2)^2}{V-\frac{(p-1)\bc^2}{2}\mu^p} .
\label{eq:ASGcoef}
\end{equation}

\subsection{Quadratic coefficient $A_{\FM}$ on the FM side}\label{app:A5}

The equilibrium FM solution is obtained by variation with respect to the two variables $(m,q)$. The Hessian at the triple point is nonsingular for the values of $p$ studied numerically ($\det\mathsf H\ne0$; for example, $\det\mathsf H=-4.010$ at $p=3$), and consequently Eq.~\eqref{eq:envelope2} applies:
\begin{equation}
\frac{\partial^2\hat\varphi_{\FM}}{\partial\beta^2}
=\varphi_{\beta\beta}-\bm v^{\!\top}\mathsf H^{-1}\bm v,\qquad
\bm v=(\varphi_{m\beta},\varphi_{q\beta})^{\!\top},\qquad
\mathsf H=\begin{pmatrix}\varphi_{mm}&\varphi_{mq}\\ \varphi_{mq}&\varphi_{qq}\end{pmatrix}.
\label{eq:d2FM}
\end{equation}
All partial derivatives below are evaluated at the triple point, where $m=q=\mu$, $H=\Lam=\Lam_c$, $\beta=\bc$, and $j_0=\jM$ is held fixed. The NL identity~\eqref{eq:tilted} gives the following useful averages:
\begin{equation}
\begin{split}
\int Dz\,\mathrm{sech}^2h_c&=1-\mu,\qquad
\int Dz\,\mathrm{sech}^2h_c\,\tanh h_c=\mu-\tau, \\
\int Dz\,\bigl(1-4\tanh^2h_c+3\tanh^4h_c\bigr)&=1-4\mu+3\tau.
\end{split}
\label{eq:momid}
\end{equation}
In this subsection, derivatives are taken through the FM effective field $\eta$ [Eq.~\eqref{eq:FMfield}], which depends on $(m,q,\beta)$ through $H$ and $\Lam$. All Gaussian averages after differentiation are evaluated at the triple point. Since $H=\Lam_c$ there, $\eta$ has the same distribution as $h_c$, and we write the field inside the evaluated averages as $h_c$.

\subsubsection*{(a) Hessian}
We start from Eq.~\eqref{eq:statm}. At a stationary point the factor $(\int Dz\,\tanh-m)$ vanishes, so derivatives of the prefactor drop out:
\begin{align}
\varphi_{mm}
&=p(p-1)\bc\jM\,\mu^{p-2}\Bigl(\frac{\partial\,\int Dz\,\tanh \eta}{\partial m}-1\Bigr)
\nonumber\\
&=\Omega\bigl(\Omega\,\int Dz\,\mathrm{sech}^2h_c-1\bigr)
=\Omega\bigl((1-\mu)\Omega-1\bigr),
\label{eq:hessian}\\
\varphi_{mq}
&=\Omega\,\frac{\partial\,\int Dz\,\tanh \eta}{\partial q}
=\Omega\cdot\frac{\Omega}{2}\int Dz\,(\tanh)''(h_c)
=-\Omega^2(\mu-\tau),
\label{eq:hessianmq}\\
\varphi_{qq}
&=\frac{\Omega}{2}\Bigl(1-\frac{\partial\,\int Dz\,\tanh^2\eta}{\partial q}\Bigr)
=\frac{\Omega}{2}\Bigl(1-\frac{\Omega}{2}\int Dz\,(\tanh^2)''(h_c)\Bigr)
\nonumber\\
&=\frac{\Omega}{2}\bigl(1-\Omega(1-4\mu+3\tau)\bigr).
\label{eq:hessianqq}
\end{align}
Here, because $\jM=\bc/2$,
\begin{equation}
p(p-1)\bc\jM\mu^{p-2}=\frac{p(p-1)}{2}\bc^2\mu^{p-2}=\Omega.
\label{eq:Omegaid}
\end{equation}
We have also used $\partial H/\partial m=\Omega$, $\partial\Lam/\partial q=\Omega$, and Eq.~\eqref{eq:gaussdiff}. In the last equality of each line, $(\tanh)''=-2\,\mathrm{sech}^2\tanh$ and $(\tanh^2)''=2(1-4\tanh^2+3\tanh^4)$ were substituted and the three Gaussian averages were reduced using Eq.~\eqref{eq:momid}. In $\varphi_{qq}$, differentiation of the prefactor drops out because the stationarity factor $(q-\int Dz\,\tanh^2)=0$ in Eq.~\eqref{eq:statq}.

\subsubsection*{(b) Mixed-derivative vector and its symmetry}
We now show that $\varphi_{q\beta}=-\varphi_{m\beta}$. We again use Eqs.~\eqref{eq:gaussdiff} and \eqref{eq:momid}, together with $H=\Lam_c$ at the triple point. At fixed $(m,q,j_0)$, $\partial_\beta H=H/\beta$ and $\partial_\beta\Lam=2\Lam/\beta$, so
\begin{equation}
\begin{split}
\frac{\partial\,\int Dz\,\tanh \eta}{\partial\beta}
&=\frac{H}{\beta}\,\int Dz\,\mathrm{sech}^2h_c
+\frac{2\Lam}{\beta}\cdot\frac12\int Dz\,(\tanh)''(h_c) \\
&=\frac{\Lam_c}{\bc}\bigl((1-\mu)-2(\mu-\tau)\bigr)
=\frac{\Lam_c}{\bc}\,B
\end{split}
\label{eq:dtanhb}
\end{equation}
\begin{equation}
\frac{\partial\,\int Dz\,\tanh^2\eta}{\partial\beta}
=\frac{\Lam_c}{\bc}\cdot2(\mu-\tau)+\frac{2\Lam_c}{\bc}\,(1-4\mu+3\tau)
=\frac{2\Lam_c}{\bc}\,B,
\label{eq:dtanh2b}
\end{equation}
where $B=1-3\mu+2\tau$ was defined in Eq.~\eqref{eq:paramalg}. Equations~\eqref{eq:statm} and \eqref{eq:statq} then give
\begin{equation}
\varphi_{m\beta}=\Omega\,\frac{\partial\,\int Dz\,\tanh \eta}{\partial\beta}
=\frac{\Omega\Lam_c}{\bc}\,B,\qquad
\varphi_{q\beta}=-\frac{\Omega}{2}\,\frac{\partial\,\int Dz\,\tanh^2\eta}{\partial\beta}
=-\frac{\Omega\Lam_c}{\bc}\,B=-\varphi_{m\beta}.
\label{eq:vsym}
\end{equation}
Thus $\bm v=\varphi_{m\beta}(1,-1)^{\!\top}$.

\subsubsection*{(c) Evaluation of $\varphi_{\beta\beta}$}
We first compute the first derivative. Differentiating Eq.~\eqref{eq:phiFMapp} with respect to $\beta$ at fixed $(m,q,j_0)$, and using $\partial_\beta H=H/\beta$, $\partial_\beta\Lam=2\Lam/\beta$, and Eq.~\eqref{eq:gaussdiff}, gives
\begin{equation}
\partial_\beta\varphi_{\FM}
=\frac{\beta}{2}+\frac{(p-1)\beta}{2}q^p-(p-1)j_0m^p-\frac{\Lam}{\beta}
+\frac{H}{\beta}\int Dz\,\tanh \eta+\frac{\Lam}{\beta}\int Dz\,\mathrm{sech}^2\eta.
\label{eq:phibeta}
\end{equation}
Differentiate once more with respect to $\beta$ at fixed $(m,q,j_0)$. Note that $H/\beta=pj_0m^{p-1}$ is independent of $\beta$, while $\partial_\beta(\Lam/\beta)=\Lam/\beta^2$. We obtain
\begin{align}
\varphi_{\beta\beta}
&=\frac12+\frac{(p-1)}{2}\mu^p
+\frac{H}{\beta}\,\frac{\partial\int Dz\,\tanh \eta}{\partial\beta}
+\frac{\Lam}{\beta^2}\,\int Dz\,\mathrm{sech}^2h_c
+\frac{\Lam}{\beta}\,\frac{\partial\int Dz\,\mathrm{sech}^2\eta}{\partial\beta}
-\frac{\Lam}{\beta^2}\\
&=\frac12+\frac{(p-1)}{2}\mu^p
+\frac{\Lam_c^2}{\bc^2}B
+\frac{\Lam_c}{\bc^2}(1-\mu)
-\frac{2\Lam_c^2}{\bc^2}B
-\frac{\Lam_c}{\bc^2}
=\frac12+\frac{(p-1)}{2}\mu^p-\frac{\Lam_c}{\bc^2}\mu-\frac{\Lam_c^2}{\bc^2}B.
\label{eq:pbb}
\end{align}
In the second equality, Eq.~\eqref{eq:dtanhb} was used for the third term, the first identity in Eq.~\eqref{eq:momid} for the fourth, and Eq.~\eqref{eq:dtanh2b} together with $\mathrm{sech}^2=1-\tanh^2$ for the fifth, giving $\partial_\beta\int Dz\,\mathrm{sech}^2=-\partial_\beta\int Dz\,\tanh^2=-2\Lam_cB/\bc$.

\subsubsection*{(d) Assembly and simplification}
Substituting Eq.~\eqref{eq:pbb} into Eq.~\eqref{eq:d2FM} and using $A_{\FM}=1/2-\partial^2\hat\varphi_{\FM}/\partial\beta^2$ [definition~\eqref{eq:Acoef}; $1/2=\partial^2\varphi_{\PM}/\partial\beta^2$] gives
\begin{equation}
A_{\FM}
=\frac12-\varphi_{\beta\beta}+\bm v^{\!\top}\mathsf H^{-1}\bm v
=-\frac{(p-1)}{2}\mu^p+\frac{\Lam_c}{\bc^2}\mu+\frac{\Lam_c^2}{\bc^2}B
+\bm v^{\!\top}\mathsf H^{-1}\bm v .
\end{equation}
Using $\Lam_c=(p/2)\bc^2\mu^{p-1}$, which implies $\Lam_c\mu/\bc^2=(p/2)\mu^p$, the first two terms combine as $-(p-1)\mu^p/2+p\mu^p/2=\mu^p/2$. We therefore obtain Eqs.~\eqref{eq:AFMmain} and \eqref{eq:AFM} of the main text,
\begin{equation}
A_{\FM}=\frac{\mu^p}{2}+\frac{\Lam_c^2}{\bc^2}\,B+\bm v^{\!\top}\mathsf H^{-1}\bm v,
\qquad
\bm v^{\!\top}\mathsf H^{-1}\bm v
=\varphi_{m\beta}^2\,\frac{\varphi_{mm}+2\varphi_{mq}+\varphi_{qq}}{\det\mathsf H}.
\end{equation}
The quadratic form follows from $\bm v=\varphi_{m\beta}(1,-1)^{\!\top}$ [Eq.~\eqref{eq:vsym}] and the inverse of the $2\times2$ Hessian, $\mathsf H^{-1}=(\det\mathsf H)^{-1}\bigl(\begin{smallmatrix}\varphi_{qq}&-\varphi_{mq}\\ -\varphi_{mq}&\varphi_{mm}\end{smallmatrix}\bigr)$, which gives $(1,-1)\mathsf H^{-1}(1,-1)^{\!\top}=(\varphi_{qq}+2\varphi_{mq}+\varphi_{mm})/\det\mathsf H$. Since the FM solution and the PM solution have the same $\varphi$ and the same first derivative at the triple point [Sec.~\ref{sec:mpoint} and Eq.~\eqref{eq:firstder}], it follows that $\hat\varphi_{\FM}(\beta,\jM)=\varphi_{\PM}-(A_{\FM}/2)b^2+O(b^3)$.

\subsection{$p\to2^+$ asymptotic expansion}\label{app:p2}

The purpose of this subsection is to determine how the triple point $\Mpt$ approaches the SK limit as $p\to2^+$ and to obtain the first nonvanishing term in the specific-heat difference that controls the reentrant curvature. We use the abbreviations
\begin{equation}
 s\equiv p-2,\qquad L\equiv\mu^{p-2},\qquad
 \Lam\equiv\Lam_c,\qquad \beta\equiv\bc
\end{equation}
and find
\begin{equation}
 \mu=\frac32(p-2)+O\bigl((p-2)^2\bigr),\qquad
 C_{\SG}-C_{\FM}=\frac{27}{32}(p-2)^2\bigl(1+O(p-2)\bigr),
\end{equation}
\begin{equation}
 \lambda(\Mpt)=\frac{\mu}{3}+O(\mu^2),\qquad
 K\longrightarrow\frac{3}{16}.
\end{equation}
Thus, for finite $p>2$ sufficiently close to 2, $C_{\SG}-C_{\FM}>0$ and hence $K>0$.

\subsubsection*{(a) Small-$\Lam$ expansions}
Applying standard Gaussian moments, $\int Dz\,(\sqrt\Lam z)^{2k}=(2k-1)!!\,\Lam^k$, together with Taylor expansions to $h=\Lam+\sqrt\Lam z$, gives the two functions that characterize the triple point:
\begin{equation}
 \mu(\Lam)=\int Dz\,\tanh h
 =\Lam-\Lam^2+\frac53\Lam^3+O(\Lam^4),
\label{eq:muLam}
\end{equation}
\begin{equation}
 W(\Lam)=\int Dz\,\lc h-\ln2-\frac{\Lam}{2}
 =\frac{\Lam^2}{4}-\frac{\Lam^3}{6}
  +\frac{5}{24}\Lam^4+O(\Lam^5).
\label{eq:WLam}
\end{equation}
Only these coefficients are needed below.

\subsubsection*{(b) Expansions of $V$ and $\tau$}
Expanding $V=\operatorname{Var}(\lc h)$ in the same way, the $+\Lam^3$ contribution from the variance of $h^2/2$ cancels the $-\Lam^3$ contribution from the covariance of $h^2/2$ with $-h^4/12$. Apart from this nontrivial cancellation, the calculation is direct and gives
\begin{equation}
 V(\Lam)=\frac{\Lam^2}{2}-\frac{\Lam^4}{3}+O(\Lam^5),
\label{eq:VLam}
\end{equation}
\begin{equation}
 \tau(\Lam)=\int Dz\,\tanh^3h
 =3\Lam^2-14\Lam^3+O(\Lam^4)
\label{eq:tauLam}
\end{equation}
as required.

\subsubsection*{(c) Softening on the SG side}
On the triple-point manifold, $\varphi_{xx}=V-2W$ [Appendix~\ref{app:A4}(d)]. Using Eqs.~\eqref{eq:WLam} and \eqref{eq:VLam},
\begin{equation}
 \varphi_{xx}
 =\frac{\Lam^3}{3}-\frac34\Lam^4+O(\Lam^5).
\label{eq:phixxLam}
\end{equation}
The $O(\Lam^2)$ terms in Eqs.~\eqref{eq:WLam} and \eqref{eq:VLam} cancel, so $\varphi_{xx}$ starts only at order $\Lam^3$. Since $\varphi_{xx}$ is the curvature of the free energy in the breakpoint direction $x$, the SG free energy becomes unusually flat in this direction. This softening determines the leading term of $A_{\SG}$ in part (h).

\subsubsection*{(d) Change of variable from $\Lam$ to $\mu$}
Inverting Eq.~\eqref{eq:muLam} gives
\begin{equation}
 \Lam=\mu+\mu^2+\frac{\mu^3}{3}+O(\mu^4),\qquad
 \frac{\Lam}{\mu}=1+\mu+\frac{\mu^2}{3}+O(\mu^3),\qquad
 \frac{\mu^3}{\Lam^2}=\mu\bigl(1-2\mu+O(\mu^2)\bigr).
\label{eq:Laminv}
\end{equation}
We henceforth organize all expansions in the small order parameter $\mu$.

\subsubsection*{(e) Relation between $p-2$ and $\mu$}
Dividing the one-variable triple-point condition~\eqref{eq:onevar} by $\Lam\mu$ and substituting Eqs.~\eqref{eq:WLam} and \eqref{eq:Laminv} gives
\begin{equation}
 \frac{p-1}{2p}
 =\frac14+\frac{\mu}{12}-\frac{\mu^2}{24}+O(\mu^3).
\label{eq:pminus2intermediate}
\end{equation}
Expanding the left-hand side in $s=p-2$ and matching coefficients yields
\begin{equation}
 s=p-2=\frac23\mu-\frac{\mu^2}{9}+O(\mu^3)
 \quad\Longleftrightarrow\quad
 \mu=\frac32s+\frac38s^2+O(s^3).
\label{eq:smu}
\end{equation}
Thus the jump of the order parameter vanishes linearly with $p-2$. The inverse temperature also approaches the SK value. Solving the triple-point definition $\Lam=(p/2)\beta^2\mu^{p-1}$ for $\beta$ gives $\beta^2=2\Lam\mu^{1-p}/p$ [Eq.~\eqref{eq:paramalg}]. With $\Lam=\mu+O(\mu^2)$ from Eq.~\eqref{eq:Laminv}, $\mu^{2-p}\to1$, and $p\to2$, one obtains $\beta^2\to1$. Hence $\bc\to1$, continuously connecting to the SK transition temperature.

\subsubsection*{(f) Separation of the logarithmic factor $L$}
Although $L=\mu^s=e^{s\ln\mu}\to1$, it contains an $O(\mu\ln\mu)$ correction. To keep this nonanalytic factor separate from the ordinary power series in $\mu$, we factor $L$ out of both $A_{\SG}$ and $A_{\FM}$ and expand $A/L$. For this purpose we use the relations that follow from the triple-point definition
$\Lam=(p/2)\beta^2\mu^{p-1}$:
\begin{equation}
 \Omega=(p-1)\frac{\Lam}{\mu},\qquad
 \beta^2L=\frac{2}{p}\frac{\Lam}{\mu},\qquad
 \frac{\Lam^2}{\beta^2}=\frac{p}{2}\Lam\mu L,
 \qquad
 \varphi_{x\beta}^2=\frac{\Lam\mu^3L}{2p}.
\label{eq:exactrel}
\end{equation}
These relations will be used below.

\subsubsection*{(g) Basic quantities needed below}
Using Eqs.~\eqref{eq:Laminv}, \eqref{eq:smu}, and \eqref{eq:tauLam}, together with $B=1-3\mu+2\tau$ defined in Eq.~\eqref{eq:paramalg}, we obtain
\begin{align}
 \Omega&=1+\frac53\mu+\frac89\mu^2+O(\mu^3),
\label{eq:Omegaexp}\\
 \tau&=3\mu^2-8\mu^3+O(\mu^4),
\label{eq:taumuexp}\\
 B&=1-3\mu+6\mu^2+O(\mu^3),
 \qquad 1-2\mu+\tau=1-2\mu+3\mu^2+O(\mu^3),
\label{eq:Bexp}\\
 1-4\mu+3\tau&=1-4\mu+9\mu^2+O(\mu^3),
 \qquad \mu-\tau=\mu-3\mu^2+O(\mu^3),
 \qquad \frac p2=1+\frac{\mu}{3}+O(\mu^2)
\label{eq:auxexp}
\end{align}
for the required quantities.

\subsubsection*{(h) Quadratic coefficient on the SG side}
Substituting Eqs.~\eqref{eq:phixxLam}, \eqref{eq:Laminv}, and \eqref{eq:exactrel} into $A_{\SG}=\varphi_{x\beta}^2/\varphi_{xx}$ gives
\begin{equation}
 \frac{A_{\SG}}{L}
 =\frac{\Lam\mu^3/(2p)}{\varphi_{xx}}
 =\frac34\mu\left(1+
 \left(-\frac13-2+\frac94\right)\mu+O(\mu^2)\right),
\label{eq:ASGintermediate}
\end{equation}
and hence
\begin{equation}
 \frac{A_{\SG}}{L}
 =\frac34\mu\left(1-\frac{\mu}{12}\right)+O(\mu^3).
\label{eq:ASGp2asym}
\end{equation}

\subsubsection*{(i) Hessian on the FM side}
Substituting the expansions in part (g) into Eqs.~\eqref{eq:hessian}--\eqref{eq:hessianqq}, we find to the required order
\begin{align}
 \varphi_{mm}&=\frac23\mu+\frac13\mu^2+O(\mu^3),
\label{eq:hmmexp}\\
 \varphi_{qq}&=\frac76\mu+\frac13\mu^2+O(\mu^3),
\label{eq:hqqexp}\\
 \varphi_{mq}&=-\mu-\frac13\mu^2+O(\mu^3).
\label{eq:hmqexp}
\end{align}
The combinations entering the numerator and denominator of the response term in Eq.~\eqref{eq:AFM} are therefore
\begin{equation}
 S\equiv\varphi_{mm}+2\varphi_{mq}+\varphi_{qq}
 =-\frac{\mu}{6}+O(\mu^3),
\label{eq:Sexp}
\end{equation}
\begin{equation}
 \det\mathsf H
 =-\frac29\mu^2\left(1+\frac{\mu}{4}\right)+O(\mu^4).
\label{eq:detHexp}
\end{equation}
Each component in Eqs.~\eqref{eq:hmmexp}--\eqref{eq:hmqexp} is $O(\mu)$, whereas $\det\mathsf H=O(\mu^2)$. Thus the FM side also develops a soft direction. The ratio needed below is
\begin{equation}
 \frac{S}{\det\mathsf H}
 =\frac{3}{4\mu}\left(1-\frac{\mu}{4}\right)+O(\mu),
 \qquad
 \Omega^2B=1+\frac{\mu}{3}+O(\mu^2),
\label{eq:responseparts}
\end{equation}
and consequently
\begin{equation}
 1+\Omega^2B\frac{S}{\det\mathsf H}
 =\frac{3}{4\mu}\left(1+\frac{17}{12}\mu\right)+O(\mu).
\label{eq:bracket}
\end{equation}

\subsubsection*{(j) Quadratic coefficient on the FM side}
Dividing $A_{\FM}$ in Eqs.~\eqref{eq:AFMmain} and \eqref{eq:AFM} by $L$ and using Eq.~\eqref{eq:exactrel}, we obtain
\begin{equation}
 \frac{A_{\FM}}{L}
 =\frac{\mu^2}{2}
 +\frac p2\Lam\mu B
 \left(1+\Omega^2B\frac{S}{\det\mathsf H}\right).
\label{eq:AFMassembly}
\end{equation}
Here
\begin{equation}
 \frac p2\Lam\mu B
 \left(1+\Omega^2B\frac{S}{\det\mathsf H}\right)
 =\frac34\mu\left(1-\frac{\mu}{4}\right)+O(\mu^3),
\label{eq:AFMsecondterm}
\end{equation}
so that
\begin{equation}
 \frac{A_{\FM}}{L}
 =\frac34\mu\left(1+\frac{5\mu}{12}\right)+O(\mu^3).
\label{eq:AFMp2asym}
\end{equation}

\subsubsection*{(k) Main result}
In Eqs.~\eqref{eq:ASGp2asym} and \eqref{eq:AFMp2asym}, the leading terms $3\mu/4$ are identical and cancel. The first nonvanishing difference is
\begin{equation}
 G(\mu)\equiv\frac{A_{\FM}-A_{\SG}}{L}
 =\frac38\mu^2+O(\mu^3).
\label{eq:Gmu}
\end{equation}
Using the thermodynamic relation
$C_{\SG}-C_{\FM}=\beta^2(A_{\FM}-A_{\SG})$
and $\beta^2L=(2/p)(\Lam/\mu)$, we find
\begin{equation}
 C_{\SG}-C_{\FM}\Big|_{\Mpt}
 =\frac{2}{p}\frac{\Lam}{\mu}\,G(\mu)
 =\frac38\mu^2\bigl(1+O(\mu)\bigr)
 =\frac{27}{32}(p-2)^2\bigl(1+O(p-2)\bigr)>0.
\label{eq:CCasym}
\end{equation}
The factor $L$ disappears in the combination $\beta^2L$, and no term of the form $\mu\ln\mu$ remains in the specific-heat difference.

The same expansion gives the common replicon eigenvalue at the triple point as
\begin{equation}
 \lambda(\Mpt)=1-\Omega(1-2\mu+\tau)
 =\frac{\mu}{3}-\frac59\mu^2+O(\mu^3)
 \longrightarrow0^+
\label{eq:lamasym}
\end{equation}
while the curvature is
\begin{equation}
 K=\frac{\beta^3(A_{\FM}-A_{\SG})}{2\mu^p}
 =\frac{3}{16}
 \left(\frac2p\frac{\Lam}{\mu}\right)^{3/2}
 L^{-3/2}\bigl(1+O(\mu)\bigr)
 \longrightarrow\frac{3}{16}.
\label{eq:Kasym}
\end{equation}
Here we used $\mu^p=\mu^2L$. Since $L^{-3/2}=1+O(\mu|\ln\mu|)$, the curvature contains a logarithmic correction but has a finite limit. This occurs because both the specific-heat difference and the first-order jump $\mu^p$ vanish as $O(\mu^2)$.

\subsubsection*{(l) Numerical check}
The numerical evaluation clearly approaches the asymptotic coefficients derived above. Representative values close to the endpoint are listed in Table~\ref{tab:p2}.
\begin{table}[htb]
\centering
\caption{Numerical check of the $p\to2^+$ asymptotic forms. Each ratio is normalized by the corresponding leading term.}
\label{tab:p2}
\begin{tabular}{ccccc}
\toprule
$p$ & $\mu/[(3/2)(p-2)]$
& $(C_{\SG}-C_{\FM})/[(27/32)(p-2)^2]$
& $K$ & $\lambda(\Mpt)/(\mu/3)$\\
\midrule
2.05  & 1.0043 & 0.796 & 0.18413 & 0.907\\
2.02  & 1.0036 & 0.909 & 0.18980 & 0.956\\
2.005 & 1.0012 & 0.976 & 0.18980 & 0.988\\
2.002 & 1.0005 & 0.990 & 0.18891 & 0.995\\
\bottomrule
\end{tabular}
\end{table}
The order parameter $\mu$, the specific-heat difference, and the replicon all approach their predicted leading forms, while $K$ approaches $3/16=0.1875$ with the expected logarithmic correction.

\subsubsection*{(m) Significance of the SK limit and range of validity}
As $p\to2^+$, the thermodynamic driving force of reentrance, $C_{\SG}-C_{\FM}$, and the strength of the first-order transition, $\mu^p$, both vanish as $O(\mu^2)$, leaving their ratio and hence the curvature finite at $3/16$. In contrast, the replicon stability margin vanishes as $O(\mu)$. The limit is therefore singular, and the curvature cannot be extrapolated to $p=2$ itself. At $p=2$, the 1RSB stability window closes and the fRSB solution controls the phase boundary, as discussed in Sec.~\ref{sec:endpoints}. The behavior of the replicons along the two branches away from the triple point is summarized in Appendix~\ref{app:B2}.

What has been established analytically in this subsection is the leading specific-heat difference in Eq.~\eqref{eq:CCasym}, including its positivity, together with the limiting forms of $\mu$, $\lambda(\Mpt)$, and $K$. 

\subsection{Large-$p$ asymptotic expansion}\label{app:largep}

The purpose of this subsection is to determine how rapidly the triple point $\Mpt$ approaches the REM limit as $p\to\infty$ and to use the first nonvanishing correction to determine the sign of $C_{\SG}-C_{\FM}$. We write
\begin{equation}
 h_c=\Lam_c+\sqrt{\Lam_c}\,z,\qquad
 \bar\rho\equiv\frac{e^{-\Lam_c/2}}{\sqrt{2\pi\Lam_c}}
\end{equation}
for later use.
The calculation proceeds in three steps.
\begin{enumerate}
\item We first show from the triple-point condition that $\Lam_c\to\infty$ and
$\Lam_c=2p\ln2\,[1+o(1)]$.
\item We then use an asymptotic expansion in $1/\Lam_c$ to obtain the quadratic coefficients on the SG and FM sides,
\begin{equation}
 A_{\SG}=\frac12-\pi\bar\rho+\cdots,
 \qquad
 A_{\FM}=\frac12-\frac{\pi^3}{16\ln2}\bar\rho+\cdots
\end{equation}
respectively.
\item Finally, we convert their difference into the specific-heat difference.
\end{enumerate}
The result is
\begin{equation}
C_{\SG}-C_{\FM}
=\Bigl(4\pi\ln2-\frac{\pi^3}{4}\Bigr)
\frac{e^{-\Lam_c/2}}{\sqrt{2\pi\Lam_c}}
\Bigl(1+O\bigl(\ln\Lam_c/\Lam_c\bigr)\Bigr).
\label{eq:Dasym}
\end{equation}
The coefficient on the right-hand side is positive. Hence $C_{\SG}-C_{\FM}>0$ and therefore $K>0$ for all sufficiently large but finite $p$.

\subsubsection*{(a) Proof of $\Lam_c\to\infty$}
Dividing the one-variable triple-point condition~\eqref{eq:onevar} by $\Lam_c\mu$ gives
\begin{equation}
\frac{W(\Lam_c)}{\Lam_c\mu(\Lam_c)}
=\frac12-\frac{1}{2p}.
\label{eq:onevardiv}
\end{equation}
Write $\lc h=h+r(h)$ with $r(h)\equiv\ln(1+e^{-2h})$. Then, for any $\Lam>0$,
\begin{equation}
W(\Lam)=\frac{\Lam}{2}-\ln2+R(\Lam),
\qquad
R(\Lam)\equiv\int Dz\,r(\Lam+\sqrt\Lam z)
\label{eq:Wsplit}
\end{equation}
holds. Introducing $\delta_1(\Lam)\equiv1-\mu(\Lam)$, Eq.~\eqref{eq:onevardiv} becomes
\begin{equation}
\frac{\Lam_c\mu}{2p}
=\ln2-R(\Lam_c)-\frac{\Lam_c\delta_1}{2}
\label{eq:Lamcid}
\end{equation}
in exact form.

Define $F(\Lam)\equiv R(\Lam)+\Lam\delta_1(\Lam)/2$. The triple-point condition is then equivalent to
\begin{equation}
\Phi(\Lam_c)\equiv
\frac12-\frac{W(\Lam_c)}{\Lam_c\mu(\Lam_c)}
=\frac{\ln2-F(\Lam_c)}{\Lam_c\mu(\Lam_c)}
=\frac{1}{2p}
\label{eq:Phicond}
\end{equation}
as written above.

We now use the differentiation formula for a Gaussian variable
$h=\Lam+\sqrt\Lam z$ whose mean and variance are both $\Lam$ [the combined form of the two identities in Eq.~\eqref{eq:gaussdiff}]:
\begin{equation}
\frac{d}{d\Lam}\int Dz\,f(h)
=\int Dz\left(f'(h)+\frac12 f''(h)\right).
\end{equation}
Since $r'=\tanh h-1$ and $r''=\mathrm{sech}^2h$, while the NL identity~\eqref{eq:tilted} gives $\int Dz\,\mathrm{sech}^2h=1-\mu=\delta_1$,
we obtain
\begin{equation}
R'(\Lam)
=\int Dz\left(r'+\frac12r''\right)
=-\delta_1+\frac12\delta_1=-\frac12\delta_1.
\end{equation}
Applying the same formula to $f(h)=\tanh h$ gives
\begin{equation}
\mu'(\Lam)
=\int Dz\,\mathrm{sech}^2h\,(1-\tanh h),
\qquad
\delta_1'(\Lam)=-\mu'(\Lam).
\end{equation}
Differentiating $F$, the two terms proportional to $\delta_1/2$ from $R'$ and $(\Lam\delta_1/2)'$ cancel, and we obtain
the monotonicity relation
\begin{equation}
F'(\Lam)
=\frac{\Lam}{2}\delta_1'(\Lam)
=-\frac{\Lam}{2}\int Dz\,\mathrm{sech}^2h\,(1-\tanh h)<0
\qquad(\Lam>0).
\label{eq:Fmono}
\end{equation}
On the other hand, as $\Lam\to0^+$, $R(\Lam)\to\ln2$ and $\Lam\delta_1(\Lam)/2\to0$, so $F(0^+)=\ln2$. Therefore $F(\Lam)<\ln2$ for every $\Lam>0$, and hence $\Phi(\Lam)>0$.
In addition, the small-$\Lam$ expansion in Appendix~\ref{app:p2}, Eqs.~\eqref{eq:muLam} and \eqref{eq:WLam}, gives
\begin{equation}
\mu(\Lam)=\Lam-\Lam^2+O(\Lam^3),
\qquad
W(\Lam)=\frac{\Lam^2}{4}-\frac{\Lam^3}{6}+O(\Lam^4)
\end{equation}
and therefore
\begin{equation}
\frac{W(\Lam)}{\Lam\mu(\Lam)}
=\frac14+O(\Lam),
\qquad
\Phi(\Lam)=\frac14+O(\Lam).
\end{equation}
In particular, $\Phi(0^+)=1/4$. Since $\Phi$ is continuous on $(0,\infty)$ and $\mu=\int Dz\,\tanh^2h>0$, its positivity together with the positive limit at the origin implies that $\Phi$ has a positive lower bound on every finite interval $0<\Lam\le\Lam_0$. The right-hand side of Eq.~\eqref{eq:Phicond}, however, tends to zero as $p\to\infty$. Thus a solution of the triple-point equation cannot remain in any finite interval, and
\begin{equation}
\Lam_c\longrightarrow\infty\qquad(p\to\infty)
\end{equation}
follows.

\subsubsection*{(b) Asymptotic formula for the Gaussian averages and the scale of $\Lam_c$}
Changing variables from $z$ to $h=\Lam+\sqrt\Lam z$, the Gaussian density becomes
\begin{equation}
\frac{dz}{\sqrt{2\pi}}e^{-z^2/2}
=\bar\rho\,e^h e^{-h^2/(2\Lam)}\,dh,
\qquad
\bar\rho=\frac{e^{-\Lam/2}}{\sqrt{2\pi\Lam}}.
\end{equation}
For fixed $\Lam$, $h$ runs over the entire real axis. Let $G$ be such that $e^hh^mG(h)$ is integrable for every $m\ge0$. Then
\begin{equation}
\int Dz\,G(h)
=\bar\rho\sum_{k\ge0}
\frac{(-1)^k}{2^k k!\Lam^k}\,I_{2k}[G],
\qquad
I_m[G]\equiv\int_{-\infty}^{\infty}dh\,e^h h^mG(h).
\label{eq:master}
\end{equation}
This is an asymptotic expansion in $1/\Lam$, and only the required orders will be retained below.

Let $G_1\equiv1-\tanh h$ and $G_3\equiv1-\tanh^3h$. With the substitution $u=e^h$, integration by parts using $r'=-G_1$, and standard beta-function integrals, the required moments are
\begin{equation}
\begin{gathered}
I_0[G_1]=\pi,\quad I_2[G_1]=\frac{\pi^3}{4},\quad
I_4[G_1]=\frac{5\pi^5}{16},\quad
I_1[G_1]=I_3[G_1]=0,\\
I_0[G_3]=\frac{3\pi}{2},\quad
I_2[G_3]=\frac{3\pi^3}{8}+\pi,\quad
I_4[G_3]=\frac{15\pi^5}{32}+\frac{3\pi^3}{2},\\
I_0[r]=\pi,\quad I_2[r]=\frac{\pi^3}{4}+2\pi,\quad
I_0[hr]=-\pi,\quad
I_2[hr]=-\Bigl(\frac{3\pi^3}{4}+6\pi\Bigr),\quad
I_0[r^2]=4\pi\ln2.
\end{gathered}
\label{eq:Itable}
\end{equation}
The odd moments $I_{2n-1}[G_1]$ vanish by antisymmetry under $u\to1/u$. Applying Eq.~\eqref{eq:master} to $G=r$ and using $I_0[r]=\pi$ and $I_2[r]=\pi^3/4+2\pi$ from Eq.~\eqref{eq:Itable},
we find
\begin{equation}
R(\Lam)
=\bar\rho\left(I_0[r]-\frac{I_2[r]}{2\Lam}
 +O(\Lam^{-2})\right)
=\bar\rho\left(\pi-\frac{\pi^3+8\pi}{8\Lam}
 +O(\Lam^{-2})\right)
=\bar\rho[\pi+O(\Lam^{-1})].
\label{eq:Rasym}
\end{equation}
An exponential bound also holds for $\delta_1=1-\mu$. For all real $h$, $1-\tanh h=2/(1+e^{2h})\le2e^{-2h}$, while trivially $1-\tanh h\le2$. Split the Gaussian average at $h=0$ ($z=-\sqrt\Lam$). On the negative side, the second bound gives $2\Phi(-\sqrt\Lam)\le e^{-\Lam/2}$, where $\Phi$ is the standard normal cumulative distribution and $\Phi(-u)\le e^{-u^2/2}/2$. On the positive side, the first bound and completion of the square, $e^{-2h}e^{-z^2/2}=e^{-(z+2\sqrt\Lam)^2/2}$, give the same upper bound, $2\Phi(-\sqrt\Lam)\le e^{-\Lam/2}$. Combining the two regions gives
\begin{equation}
\delta_1=1-\mu=\int Dz\,\bigl(1-\tanh h\bigr)\le 2\,e^{-\Lam/2}
\label{eq:mubound}
\end{equation}
and hence $\delta_1=O(e^{-\Lam/2})$.

Substituting Eqs.~\eqref{eq:Rasym} and \eqref{eq:mubound} into Eq.~\eqref{eq:Lamcid} yields
\begin{equation}
\Lam_c=2p\ln2
\Bigl(1-\frac{\pi}{2\ln2}\Lam_c\bar\rho+O(\bar\rho)\Bigr)
=2p\ln2\bigl(1+O(\sqrt p\,2^{-p})\bigr),
\quad
e^{-\Lam_c/2}=2^{-p}\bigl(1+O(p^{3/2}2^{-p})\bigr).
\label{eq:Lamcasym}
\end{equation}
Only at this stage do we evaluate the limit of $\mu^{p-1}$. Equations~\eqref{eq:Lamcasym} and \eqref{eq:mubound} imply $\delta_1=O(2^{-p})$, so
\begin{equation}
(p-1)|\ln\mu|
\le\frac{(p-1)\delta_1}{\mu}=O(p2^{-p})\to0,
\qquad
\mu^{p-1}\to1,
\qquad
\bc^2=\frac{2\Lam_c}{p}\mu^{1-p}\to4\ln2.
\label{eq:mupm1}
\end{equation}
Thus the known REM value $\bc=2\sqrt{\ln2}$ is recovered from the triple-point condition itself, without assuming the REM result in advance.

\subsubsection*{(c) Required moments}
We now set $\Lam=\Lam_c$ for brevity. With $\delta_1=1-\mu$ and $\delta_3=1-\tau$, Eqs.~\eqref{eq:master} and \eqref{eq:Itable} give
\begin{equation}
\delta_1=\bar\rho\Bigl(\pi-\frac{\pi^3}{8\Lam}+O(\Lam^{-2})\Bigr),
\qquad
\delta_3=\bar\rho\Bigl(\frac{3\pi}{2}
-\frac{3\pi^3+8\pi}{16\Lam}+O(\Lam^{-2})\Bigr).
\label{eq:d1d3}
\end{equation}
It follows that $B=1-3\mu+2\tau$, defined in Eq.~\eqref{eq:paramalg}, is
\begin{equation}
B=3\delta_1-2\delta_3
=\bar\rho\Bigl(\frac{\pi}{\Lam}
-\frac{3\pi^3}{8\Lam^2}+O(\Lam^{-3})\Bigr).
\label{eq:Basym}
\end{equation}
The constant term cancels, $3\pi-3\pi=0$, and therefore $B$ starts at order $\bar\rho/\Lam$. This cancellation is crucial on the FM side.

Using $\lc h=h+r(h)$, we also write
\begin{equation}
V=\operatorname{Var}(h+r)
=\Lam+2\operatorname{Cov}(h,r)+\operatorname{Var}(r)
\end{equation}
and evaluate the terms successively from Eqs.~\eqref{eq:master} and \eqref{eq:Itable}:
\begin{align}
\int Dz\,r(h_c)
&=\bar\rho\left(\pi-\frac{\pi^3+8\pi}{8\Lam}
 +O(\Lam^{-2})\right),\\
\int Dz\,h_c r(h_c)
&=\bar\rho\left(-\pi
 +\frac{3\pi^3+24\pi}{8\Lam}
 +O(\Lam^{-2})\right),\\
\int Dz\,r(h_c)^2
&=4\pi\ln2\,\bar\rho+O(\bar\rho/\Lam).
\end{align}
Since $\int Dz\,h_c=\Lam$, the first two equations give
\begin{align}
\operatorname{Cov}(h_c,r)
&=\int Dz\,h_c r(h_c)
 -\Lam\int Dz\,r(h_c)\\
&=-\pi\Lam\bar\rho
 +\frac{\pi^3}{8}\bar\rho
 +O(\bar\rho/\Lam).
\end{align}
Furthermore, $(\int Dz\,r)^2=O(\bar\rho^2)$ is exponentially smaller, so
\begin{equation}
\operatorname{Var}(r)
=4\pi\ln2\,\bar\rho
 +O(\bar\rho/\Lam)+O(\bar\rho^2).
\end{equation}
Combining these results in $V$ gives
\begin{equation}
\int Dz\,r(h_c)
=\bar\rho\Bigl(\pi-\frac{\pi^3+8\pi}{8\Lam}+O(\Lam^{-2})\Bigr),
\qquad
V=\Lam-2\pi\Lam\bar\rho
+\Bigl(\frac{\pi^3}{4}+4\pi\ln2\Bigr)\bar\rho
+O(\bar\rho/\Lam).
\label{eq:Vasym}
\end{equation}
Intermediate terms of order $\Lam\bar\rho$ appear, whereas the final specific-heat difference is only of order $\bar\rho$. We therefore keep track explicitly of the cancellation of these larger terms on both the SG and FM sides. Terms of order $O(\Lam^2\bar\rho^2)$ are exponentially smaller and will be omitted.

\subsubsection*{(d) Corrections to $\bc^2$ and $\mu^p$ from the triple-point equation}
Combining Eq.~\eqref{eq:Lamcid} with $\bc^2=(2\Lam_c/p)\mu^{1-p}$ gives
\begin{equation}
\bc^2
=4\mu^{-p}\Bigl(\ln2-\int Dz\,r(h_c)-\frac{\Lam_c\delta_1}{2}\Bigr).
\label{eq:betacid}
\end{equation}
Using $\mu^{-p}=1+p\delta_1+O(\Lam^2\bar\rho^2)$ and
$p\delta_1=\Lam_c\delta_1/(2\ln2)+O(\Lam^2\bar\rho^2)$, the $\Lam\bar\rho$ terms cancel and we obtain
\begin{equation}
\bc^2=4\ln2-4\int Dz\,r(h_c)+O(\Lam^2\bar\rho^2),
\qquad
\mu^p=1-\frac{\Lam_c\delta_1}{2\ln2}+O(\Lam^2\bar\rho^2)
\label{eq:betacasym}
\end{equation}
as required.

\subsubsection*{(e) Quadratic coefficient on the SG side}
We use Eq.~\eqref{eq:ASGa}, namely
$A_{\SG}=\varphi_{x\beta}^2/\varphi_{xx}$, with
$\varphi_{x\beta}=\bc\mu^p/2$ and
$\varphi_{xx}=V-2W=V-\Lam_c+2\ln2-2\int Dz\,r$, where the last form follows by substituting Eq.~\eqref{eq:Wsplit} into the identity $\varphi_{xx}=V-2W$ of Appendix~\ref{app:A4}(d). Equations~\eqref{eq:Vasym} and \eqref{eq:betacasym} give
\begin{align}
\varphi_{x\beta}^2
&=\ln2-\Lam_c\delta_1-\int Dz\,r+O(\Lam^2\bar\rho^2),\\
\varphi_{xx}
&=2\ln2-2\pi\Lam\bar\rho
+\Bigl(\frac{\pi^3}{4}+4\pi\ln2-2\pi\Bigr)\bar\rho
+O(\bar\rho/\Lam).
\end{align}
Using $\Lam_c\delta_1=\pi\Lam\bar\rho-(\pi^3/8)\bar\rho+O(\bar\rho/\Lam)$, the terms of order $\Lam\bar\rho$ cancel in the ratio, leaving
\begin{equation}
A_{\SG}=\frac12-\pi\bar\rho
+O\bigl(\bar\rho\ln\Lam_c/\Lam_c\bigr)
\label{eq:ASGasym}
\end{equation}
as claimed.

\subsubsection*{(f) Quadratic coefficient on the FM side}
We separate Eq.~\eqref{eq:AFMmain} into the three contributions
\begin{equation}
A_{\FM}=\frac{\mu^p}{2}
+\frac{\Lam_c^2}{\bc^2}B
+\bm v^{\!\top}\mathsf H^{-1}\bm v
\end{equation}
and evaluate them individually.

First, the response term does not contribute at leading order. Indeed,
$\varphi_{m\beta}=(\Omega\Lam_c/\bc)B$,
and $\Omega=(p-1)\Lam_c/\mu$, so
$\varphi_{m\beta}=O(\Lam^2\bar\rho)$. Matching the scale of the Hessian then gives
$\bm v^{\!\top}\mathsf H^{-1}\bm v=O(\Lam^2\bar\rho^2)$.

For the remaining two terms, Eqs.~\eqref{eq:d1d3}, \eqref{eq:Basym}, and \eqref{eq:betacasym} give
\begin{align}
\frac{\mu^p}{2}
&=\frac12-\frac{\pi\Lam\bar\rho}{4\ln2}
+\frac{\pi^3}{32\ln2}\bar\rho+O(\bar\rho/\Lam),\\
\frac{\Lam_c^2}{\bc^2}B
&=\frac{\pi\Lam\bar\rho}{4\ln2}
-\frac{3\pi^3}{32\ln2}\bar\rho+O(\bar\rho/\Lam).
\end{align}
Again the two $\Lam\bar\rho$ terms cancel exactly, and hence
\begin{equation}
A_{\FM}=\frac12-\frac{\pi^3}{16\ln2}\bar\rho
+O\bigl(\bar\rho\ln\Lam_c/\Lam_c\bigr)
\label{eq:AFMasym}
\end{equation}
follows.

\subsubsection*{(g) Specific-heat difference, curvature, and range of validity}
Taking the difference of Eqs.~\eqref{eq:ASGasym} and \eqref{eq:AFMasym} gives
\begin{equation}
A_{\FM}-A_{\SG}
=\Bigl(\pi-\frac{\pi^3}{16\ln2}\Bigr)\bar\rho
+O\bigl(\bar\rho\ln\Lam_c/\Lam_c\bigr),
\qquad
C_{\SG}-C_{\FM}=\bc^2(A_{\FM}-A_{\SG}).
\label{eq:dAasym}
\end{equation}
Using $\bc^2=4\ln2+O(\bar\rho)$ reproduces Eq.~\eqref{eq:Dasym} stated at the beginning of this subsection. Thus $K>0$ for all sufficiently large but finite $p$, and the curvature behaves as
\begin{equation}
K\sim\sqrt{\ln2}\Bigl(4\pi\ln2-\frac{\pi^3}{4}\Bigr)
\frac{e^{-\Lam_c/2}}{\sqrt{2\pi\Lam_c}}
\end{equation}
and tends to zero in the REM limit. Equation~\eqref{eq:Lamcasym} shows that the exponential factor is $2^{-p}$ and that the full scale, including the algebraic prefactor, is $2^{-p}/\sqrt p$. Therefore the ratio of successive terms when $p$ is increased by one approaches 2 asymptotically. This proves the nearly geometric decay observed in Table~\ref{tab:coeffs}.

Numerically, one also finds $(1/2-A_{\SG})/\bar\rho\to\pi$,
$(1/2-A_{\FM})/\bar\rho\to\pi^3/(16\ln2)$,
$(A_{\FM}-A_{\SG})/\bar\rho\to
\pi-\pi^3/(16\ln2)=0.34581\ldots$, in agreement with the asymptotic result.

What has been established analytically in this subsection is the leading asymptotic term~\eqref{eq:Dasym} and its positivity. We have not derived a uniform bound on the error $O[(\ln\Lam_c)/\Lam_c]$ or an explicit threshold for ``sufficiently large $p$.'' The positive coefficient $4\pi\ln2-\pi^3/4=0.9588\ldots$ is the difference of two comparable terms, $8.7103\ldots$ and $7.7516\ldots$, and is therefore reduced by a factor of less than one order of magnitude by cancellation. The smallness of $C_{\SG}-C_{\FM}$ at finite $p$ is governed primarily by the exponential factor $2^{-p}/\sqrt p$; the cancellation in the prefactor provides an additional, but secondary, suppression.

\section{Stability: Replicon eigenvalues}\label{app:replicon}

This appendix summarizes the basis for the statements about replicon eigenvalues and stability in Sec.~\ref{sec:mpoint}, using the compact form of Eq.~(41) of Ref.~\cite{Nishimori1999}.

\subsection{AT eigenvalue of the RS-FM solution}\label{app:B1}

Equation~(41) of Ref.~\cite{Nishimori1999} gives the innermost-block replicon eigenvalue of a 1RSB solution, which reduces to the AT eigenvalue in the RS limit, in the form
\begin{equation}
\lambda_1=1-\frac{p(p-1)}{2}\,\beta^2\,q_1^{p-2}\;\bigl\langle\mathrm{sech}^4\Xi\bigr\rangle,
\label{eq:NW41}
\end{equation}
where $\Xi$ is the effective field of the corresponding block and $\langle\cdot\rangle$ denotes the Gaussian average with the appropriate block weight. For the 1RSB-SG solution ($m=q_0=0$), $\Xi=\sqrt{\Lam_1}\,v$ and Eq.~\eqref{eq:NW41} reduces to Eq.~\eqref{eq:repdefSG}. Since the RS solution is contained as the degenerate limit $q_0\to q_1$, the AT condition for the RS-FM solution is simply
\begin{equation}
\lambda_{\rm RS}=1-\frac{p(p-1)}{2}\,\beta^2q^{p-2}\;\int Dz\,\mathrm{sech}^4\eta.
\label{eq:ATRS}
\end{equation}
This is the $p$-body version of the de Almeida--Thouless condition \cite{deAlmeida1978}, with the notation of Eq.~\eqref{eq:FMfield}. This expression is Eq.~\eqref{eq:repdefRS} of the main text.

\subsection{Value at the triple point and equality for the two solutions}\label{app:B2}

At the triple point $\Mpt$, the field becomes the Gaussian variable $h_c=\Lam_c+\sqrt{\Lam_c}\,z$. Using $\mathrm{sech}^4=(1-\tanh^2)^2=1-2\tanh^2+\tanh^4$ together with the NL identity~\eqref{eq:tilted},
\begin{equation}
\int Dz\,\mathrm{sech}^4h_c=1-2\mu+\tau
\quad\Longrightarrow\quad
\lambda(\Mpt)=1-\Omega\,(1-2\mu+\tau),\qquad
\Omega=\frac{p(p-1)}{2}\bc^2\mu^{p-2},
\label{eq:lamM}
\end{equation}
which is Eq.~\eqref{eq:lamMmain} of the main text. For the SG-side 1RSB solution, the innermost replicon in Eq.~\eqref{eq:repdefSG} has, as $x\to1$, the same Gaussian distribution by Eq.~\eqref{eq:tiltshift}. It therefore takes the same value~\eqref{eq:lamM} at the triple point. This is the equality stated in Sec.~\ref{sec:mpoint}. As shown in Tables~\ref{tab:mpoint} and \ref{tab:coeffs}, $\lambda(\Mpt)>0$ throughout the range evaluated numerically (integers $3\le p\le12$ and the continuous-$p$ scan), with $\lambda(\Mpt)=0.3305$ for $p=3$. These results strongly suggest positivity for every finite $p>2$. As $p\to2^+$, $\lambda(\Mpt)$ approaches zero from above, as discussed in Sec.~\ref{sec:endpoints}, consistent with marginal stability in the SK limit.

This equality concerns only the value at the single point $\Mpt$ and does not imply that the stability structures of the two branches are the same.

For $T<T_c$, $x^*<1$ and $q_1\ne m$, and the correspondence used above is immediately lost: the $(2\cosh)^x$ average is no longer the same Gaussian average as the FM effective field on the NL. The two replicon eigenvalues therefore split. For $p=3$, at $T/T_c=0.95$ the 1RSB-SG value is $\lambda_1=0.325$, while the RS-FM value on the NL is $\lambda_{\rm RS}=0.486$; at $T/T_c=0.70$ the values are $0.270$ and $0.903$, respectively. The SG-side replicon decreases slowly and reaches zero only at the lower Gardner temperature $T_G$ ($T_G\simeq0.24$ for $p=3$; Sec.~\ref{sec:chaos}), whereas the FM-side replicon on the NL increases and the solution becomes more stable. The latter behavior is consistent with the absence of RSB on the NL discussed in Sec.~\ref{sec:constraints}.

\subsection{Stability along the phase boundary and provisional low-temperature continuation}\label{app:B3}

Evaluating Eq.~\eqref{eq:ATRS} along the phase boundary for $p=3$, the RS replicon changes sign at $T_{\rm AT}\simeq0.435$. Thus the RS solution on the FM side becomes unstable a finite distance below the triple point. The 1RSB solution on the SG side remains replicon stable down to the Gardner temperature $T_G\simeq0.24$. Consequently, both solutions entering the comparison are stable only for $T\gtrsim0.435$, and both the expansion in Sec.~\ref{sec:expansion} and the numerical comparison in Appendix~\ref{app:numeric} are self-contained within this stability window.

As shown by Nishimori and Wong \cite{Nishimori1999}, the RS solution on the FM side becomes unstable at the AT line. One may provisionally continue the phase boundary below $T\simeq0.435$ by using the 1RSB mixed-phase solution of Ref.~\cite{Nishimori1999} and equating its free energy to that of the 1RSB-SG solution. However, the innermost replicon~\eqref{eq:NW41} of this mixed solution is also negative along the resulting boundary. This continuous instability is of the same type as that below the AT line in the SK model \cite{deAlmeida1978,Parisi1979,Parisi1980a}, in the mixed phase \cite{Toulouse1980,GabayToulouse1981}, and at the Gardner transition of the Ising $p$-spin model \cite{Gardner1985}. It suggests that higher-step RSB and ultimately a marginal fRSB solution are required. The 1RSB phase diagram of Ref.~\cite{ZdeborovaKrzakala2010PRB} for $p=3$ draws this boundary down to zero temperature and likewise regards the low-temperature mixed phase as requiring an fRSB description. What is established here is only that the continuous 1RSB mixed-phase branch of Ref.~\cite{Nishimori1999} followed in this work is replicon unstable near the low-temperature part of the boundary and therefore cannot be the stable physical solution. We do not exclude other 1RSB stationary branches. The RSB structure of the stable mixed phase and the corresponding low-temperature phase boundary remain open. The main conclusions of this paper, Sec.~\ref{sec:expansion}--\ref{sec:chaos}, do not depend on this low-temperature continuation.

\section{Numerical determination of the phase boundary from free-energy comparison}\label{app:numeric}

In this appendix, we determine the FM--SG phase boundary directly by numerical comparison of the two free energies and compare it with the analytical result of Sec.~\ref{sec:expansion}. The boundary is followed not only close to the triple point but throughout the range in which both branches are replicon stable.

\subsection{1RSB free energies of the two phases}\label{app:E1}

We use the 1RSB functional as a common variational framework for comparing the two phases. The SG phase is simply the specialization $m=q_0=0$ of the same functional, as explained in Sec.~\ref{sec:mpoint}; the RS solution is included as the degenerate limit $q_0\to q_1$. The phase diagram of Ref.~\cite{Nishimori1999} (their Fig.~7) contains an RSB ferromagnetic, or mixed, phase at lower temperatures. However, the 1RSB mixed-phase solution of Ref.~\cite{Nishimori1999} adjacent to the 1RSB-SG phase is replicon unstable and cannot be regarded as a stable equilibrium branch. We therefore do not use it in the main comparison near the triple point. It is referred to only for the provisional low-temperature continuation below $T\simeq0.435$ discussed in Appendix~\ref{app:E2}; Appendix~\ref{app:B3} explicitly notes that this continuation does not affect the main results. Near the triple point, the stable ferromagnetic phase adjacent to the SG phase is the RS solution, as shown by the stability analysis in Appendix~\ref{app:replicon}. This is consistent with Sec.~\ref{sec:mpoint}, Sec.~\ref{sec:expansion}, and the treatment of the stability window in Appendix~\ref{app:E2}.

\subsection{Numerical phase boundary: Monotonic reentrance}\label{app:E2}

For each $T\in(T_G,T_c)$, we solve $-\beta f_{\FM}(T,j_0)=-\beta f_{\SG}(T)$ for $j_0^*(T)$. The results for $p=3$ are listed in Table~\ref{tab:boundary}.

\begin{table}[htb]
\centering
\caption{FM--SG phase boundary for $p=3$, shown as the deviation from $\jM=1/(2T_c)=0.767595$. On the FM side we use the RS-FM branch for $T\gtrsim0.435$ and the 1RSB mixed-phase branch at lower temperatures. The portion below $T\simeq0.435$ is provisional because the mixed-phase solution is replicon unstable (Appendix~\ref{app:replicon}).}
\label{tab:boundary}
\begin{tabular}{lccccccccccc}
\toprule
$T$ & 0.28 & 0.30 & 0.33 & 0.36 & 0.40 & 0.44 & 0.48 & 0.52 & 0.56 & 0.60 & 0.63 \\
\midrule
$(j_0^*-\jM)\times10^3$ & 1.37 & 1.35 & 1.32 & 1.26 & 1.14 & 0.97 & 0.75 & 0.50 & 0.28 & 0.10 & 0.02 \\
\bottomrule
\end{tabular}
\end{table}

The boundary has three clear properties: (i) $j_0^*\ge\jM$ throughout, (ii) $j_0^*\to\jM$ as $T\to T_c$, and (iii) $j_0^*$ increases monotonically upon cooling, reaching a deviation of $1.37\times10^{-3}$ at $T=0.28$. This small, monotonic bending is consistent with the slight bending toward larger bias reported, without quantitative detail, in the 1RSB phase diagram of Ref.~\cite{ZdeborovaKrzakala2010PRB}. This is precisely the reentrant shape allowed by gauge symmetry, and it is fully consistent with constraint (iii) of Sec.~\ref{sec:constraints}. For a fixed $j_0$ in the interval $\jM<j_0<\jM+0.97\times10^{-3}$, comparison of the two stable branches alone gives the sequence PM $\to$ FM $\to$ SG upon cooling. Within the stability window $T\gtrsim0.435$, the FM branch is the AT-stable RS solution and the SG branch is a replicon-stable 1RSB solution. This comparison by itself does not prove that no other globally stable branch intervenes. The numerical phase boundary in this stable window joins smoothly onto the analytical formula~\eqref{eq:boundary} derived in Sec.~\ref{sec:expansion}; see Fig.~\ref{fig:phase}(b). A quantitative comparison is given in Appendix~\ref{app:E3}.

Below $T\simeq0.435$, the RS-FM solution is AT unstable, and we use the 1RSB mixed-phase solution for a provisional continuation. Since that solution itself is replicon unstable (Appendix~\ref{app:replicon}), the low-temperature part of Table~\ref{tab:boundary} is provisional. None of the main conclusions of the paper depends on this part.

The same stability criterion can be applied to $p>3$. Using accurate values of $T_c(p)$ and fixing the reduced temperature at $T=0.70\,T_c(p)$, we locate the boundary by equating the free energies of the AT-stable RS-FM solution and the replicon-stable 1RSB-SG solution. The results are shown in Table~\ref{tab:ptrend}; the replicon eigenvalues were checked explicitly at every point. The boundary deviation at this fixed reduced temperature decreases monotonically as $0.885\to0.305\to0.116\to0.048$ in units of $10^{-3}$ for $p=3\to6$, showing a clear trend toward the exactly vertical boundary of the REM limit, $p=\infty$ (Sec.~\ref{sec:endpoints}). This is consistent with the decrease of $K(p)$ in Table~\ref{tab:coeffs}.

This is the boundary deviation at a fixed reduced temperature, not the maximum deviation, or the full width of reentrance, for each value of $p$.

\begin{table}[htb]
\centering
\caption{$p$ dependence of the FM--SG phase-boundary deviation at fixed reduced temperature $T=0.70\,T_c(p)$. Here $\lambda_{\rm RS}$ is the AT (replicon) eigenvalue of the RS solution on the FM side and $\lambda_{\SG}$ is the innermost replicon eigenvalue of the 1RSB solution on the SG side. Both are positive at every point shown.}
\label{tab:ptrend}
\begin{tabular}{cccccc}
\toprule
$p$ & $T_c(p)$ & $\jM=1/(2T_c)$ & $(j_0^*-\jM)\times10^3$ & $\lambda_{\rm RS}$ & $\lambda_{\SG}$ \\
\midrule
3 & 0.651385 & 0.767595 & 0.885478 & $+0.05$ & $+0.27$ \\
4 & 0.616883 & 0.810526 & 0.305201 & $+0.41$ & $+0.53$ \\
5 & 0.606952 & 0.823789 & 0.115729 & $+0.64$ & $+0.71$ \\
6 & 0.603296 & 0.828781 & 0.047815 & $+0.78$ & $+0.82$ \\
\bottomrule
\end{tabular}
\end{table}

\subsection{Quantitative comparison between the direct boundary calculation and the analytical formula}\label{app:E3}

Within the stability window, we follow the point $j_0^*(T)$ at which the free energies of the AT-stable RS-FM and replicon-stable 1RSB-SG solutions are equal as $T\to T_c$. We compare $\delta/t^2$, where $\delta=j_0^*-\jM$ and $t=T_c-T$, with the analytical curvature $K$ in Eq.~\eqref{eq:boundary}. The $p=3$ results are listed in Table~\ref{tab:direct}.

\begin{table}[htb]
\centering
\caption{Comparison for $p=3$ between direct numerical determination of the phase boundary and the analytical formula. The last column is normalized by $K=0.043330$ from Table~\ref{tab:coeffs}. A linear extrapolation to $t\to0$ gives $K=0.04333$, in agreement with the analytical value to five digits. The $t$ dependence of the residual, with slope $\simeq-0.12$, is the contribution from the cubic term.}
\label{tab:direct}
\begin{tabular}{ccccc}
\toprule
$T$ & $t=T_c-T$ & $\delta=j_0^*-\jM$ & $\delta/t^2$ & $(\delta/t^2)/K$ \\
\midrule
0.650 & 0.001385 & $8.281\times10^{-8}$ & 0.04316 & 0.9961 \\
0.648 & 0.003385 & $4.917\times10^{-7}$ & 0.04291 & 0.9903 \\
0.644 & 0.007385 & $2.313\times10^{-6}$ & 0.04241 & 0.9788 \\
0.635 & 0.016385 & $1.109\times10^{-5}$ & 0.04132 & 0.9536 \\
0.630 & 0.021385 & $1.862\times10^{-5}$ & 0.04073 & 0.9400 \\
0.615 & 0.036385 & $5.162\times10^{-5}$ & 0.03899 & 0.8998 \\
\bottomrule
\end{tabular}
\end{table}

For $p=4$, the same procedure gives $\delta/t^2=0.013139,\,0.013086,\,0.012982$ at $t=0.002,\,0.004,\,0.008$, respectively. Extrapolation to $t\to0$ gives $0.013192$, in agreement with the analytical value $K=0.013191$ in Table~\ref{tab:coeffs}.

\section{Two-temperature replica calculation}
\label{app:twotemp}

This appendix independently tests the logical consequence obtained in the main text, i.e., the existence of a pair of distinct temperatures in the finite-temperature SG phase for which $P_2(x\mid\beta_1,\beta_2)=\delta(x)$, by placing two species of replicas at different temperatures in the same disorder realization, as in Refs.~\cite{Rizzo2001,Rizzo2003}. Whereas Refs.~\cite{Rizzo2001,Rizzo2003} study a constrained free energy at fixed cross overlap, we impose the structure of matched pairs as an ansatz. This distinction is discussed in Appendix~\ref{app:F5}(iv). A different approach to the same question is the state-following method of Refs.~\cite{KrzakalaZdeborova2010EPL,ZdeborovaKrzakala2010PRB}, which computes by a generalized cavity method the free energy of the state in equilibrium at one temperature when the temperature is changed, and compares it with the equilibrium free energy at the new temperature. Here, instead, the two temperatures are treated symmetrically within the replica method, and the cost of a one-to-one correspondence of states is evaluated. We translate the physical picture required by the absence of temperature chaos, a one-to-one correspondence between pure states at the two temperatures, into a replica saddle-point structure and evaluate its cost.

\subsection{Two-temperature functional and matched-pair sector}\label{app:F2}

\subsubsection*{(a) General two-temperature replicated action}

For the same realization of the disorder $\{J\}$, introduce replicas $\sigma^a$ ($a=1,\dots,n$) at inverse temperature $\beta_1$ and replicas $\tau^\alpha$ ($\alpha=1,\dots,m$) at inverse temperature $\beta_2$. In this appendix $m$ denotes the number of replicas at temperature 2 and is unrelated to the magnetization $m$ used in the main text. The disorder average of the replicated partition function for the Hamiltonian~\eqref{eq:H} is
\begin{align}
\bigl[Z_1^nZ_2^m\bigr]
&=\int\Bigl[\prod_{i_1<\cdots<i_p}dJ_{i_1\cdots i_p}\,P(J_{i_1\cdots i_p})\Bigr]
 \operatorname{Tr}_{\sigma,\tau}\exp\Bigl[\sum_{i_1<\cdots<i_p}J_{i_1\cdots i_p}A_{i_1\cdots i_p}\Bigr],
\label{eq:Z2rep}\\
A_{i_1\cdots i_p}&=\beta_1\sum_{a}\sigma^a_{i_1}\cdots\sigma^a_{i_p}
 +\beta_2\sum_{\alpha}\tau^\alpha_{i_1}\cdots\tau^\alpha_{i_p}.
\label{eq:Adef}
\end{align}
The Gaussian integral over each $p$-tuple of couplings can be performed independently, and the distribution~\eqref{eq:dist} gives
\begin{equation}
\bigl[Z_1^nZ_2^m\bigr]
=\operatorname{Tr}_{\sigma,\tau}\exp\sum_{i_1<\cdots<i_p}
 \Bigl(\frac{j_0\,p!}{N^{p-1}}A_{i_1\cdots i_p}
 +\frac{p!}{4N^{p-1}}A_{i_1\cdots i_p}^2\Bigr).
\label{eq:afterJ}
\end{equation}

We next expand $A$ and $A^2$ and carry out the sum over $p$-tuples. Define the overlaps
\begin{equation}
\begin{split}
&m_a=\frac1N\sum_i\sigma^a_i,\quad
\tilde m_\alpha=\frac1N\sum_i\tau^\alpha_i,\quad
q_{ab}=\frac1N\sum_i\sigma^a_i\sigma^b_i, \\
&r_{\alpha\beta}=\frac1N\sum_i\tau^\alpha_i\tau^\beta_i,\quad
s_{a\alpha}=\frac1N\sum_i\sigma^a_i\tau^\alpha_i.
\end{split}
\label{eq:overlapdef}
\end{equation}
Then, for large $N$, the exponent per spin in Eq.~\eqref{eq:afterJ} is
\begin{equation}
\frac14\bigl(n\beta_1^2+m\beta_2^2\bigr)
+\frac14\Bigl(\beta_1^2\sum_{a\ne b}q_{ab}^p
 +\beta_2^2\sum_{\alpha\ne\beta}r_{\alpha\beta}^p
 +2\beta_1\beta_2\sum_{a,\alpha}s_{a\alpha}^p\Bigr)
+j_0\Bigl(\beta_1\sum_am_a^p+\beta_2\sum_\alpha\tilde m_\alpha^p\Bigr).
\label{eq:expbeforeOP}
\end{equation}
In $A^2$, terms with $a=b$ give $(\sigma^a_{i_1}\cdots\sigma^a_{i_p})^2=1$ and hence the constant $n$; terms with $\alpha=\beta$ similarly give $m$. By contrast, $\sigma$ and $\tau$ belong to different replica sets, so there is no diagonal component in the cross term and all $nm$ pairs $(a,\alpha)$ contribute nontrivially. In the SG sector the magnetizations vanish, $m_a=\tilde m_\alpha=0$, and the last term in Eq.~\eqref{eq:expbeforeOP}, containing $j_0$, drops out.

The quantities $q_{ab}$, $r_{\alpha\beta}$, and $s_{a\alpha}$ in Eq.~\eqref{eq:expbeforeOP} are still functions of the spin configurations. We promote them to independent variational variables by inserting delta functions for every pair,
\begin{equation}
1=N\!\int dq_{ab}\;\delta\Bigl(Nq_{ab}-\sum_i\sigma^a_i\sigma^b_i\Bigr)
\end{equation}
and similarly for $r_{\alpha\beta}$ and $s_{a\alpha}$. Their Fourier representations introduce conjugate variables $\lambda^q_{ab}$, $\lambda^r_{\alpha\beta}$, and $\lambda^s_{a\alpha}$. Taking the conjugate term in the form $\sum_{a<b}\lambda^q_{ab}(\sum_i\sigma_i^a\sigma_i^b-Nq_{ab})$ and analogously for the other sectors (with $\sum_{a,\alpha}$ for the cross term), the full exponent is proportional to $N$ and can be evaluated by the saddle-point method. Stationarity with respect to the overlap variables $q_{ab}$, $r_{\alpha\beta}$, and $s_{a\alpha}$, i.e., differentiation of Eq.~\eqref{eq:expbeforeOP} with respect to them, gives
\begin{equation}
\lambda^q_{ab}=\frac{p\beta_1^2}{2}q_{ab}^{p-1},\qquad
\lambda^r_{\alpha\beta}=\frac{p\beta_2^2}{2}r_{\alpha\beta}^{p-1},\qquad
\lambda^s_{a\alpha}=\frac{p\beta_1\beta_2}{2}s_{a\alpha}^{p-1}
\label{eq:conjsaddle}
\end{equation}
for the conjugate fields. The trace over spins factorizes over sites, so the terms containing $\lambda$ produce $\ln\operatorname{Tr}e^{\Xi'}$ per site, where $\operatorname{Tr}$ is over the $n+m$ one-site replica variables. At the same time, the subtraction terms $-\sum\lambda q$ remove $p$ times the polynomial contributions through Eq.~\eqref{eq:conjsaddle}; consequently the coefficients of $q^p$, $r^p$, and $s^p$ change from $1$ to $1-p$.

Collecting the terms, the replicated action per spin, after eliminating the conjugate variables at the saddle point and treating the intratemperature overlaps $q_{ab}$ and $r_{\alpha\beta}$ and the intertemperature overlap $s_{a\alpha}$ (hereafter the cross overlap) as variational variables, is
\begin{align}
G={}&\frac14\bigl(n\beta_1^2+m\beta_2^2\bigr)
-\frac{p-1}{4}\Bigl(
 \beta_1^2\sum_{a\ne b}q_{ab}^p
 +\beta_2^2\sum_{\alpha\ne\beta}r_{\alpha\beta}^p
 +2\beta_1\beta_2\sum_{a,\alpha}s_{a\alpha}^p
 \Bigr)
 +\ln\operatorname{Tr}e^{\Xi'},
\label{eq:G2T}\\
\Xi'={}&\frac{p}{4}\Bigl(
 \beta_1^2\sum_{a\ne b}q_{ab}^{p-1}\sigma^a\sigma^b
 +\beta_2^2\sum_{\alpha\ne\beta}r_{\alpha\beta}^{p-1}\tau^\alpha\tau^\beta
 \Bigr)
 +\frac{p}{2}\beta_1\beta_2
 \sum_{a,\alpha}s_{a\alpha}^{p-1}\sigma^a\tau^\alpha .
\label{eq:Xi2T}
\end{align}
The two-temperature overlap distribution $P_2$ is the disorder-averaged distribution of the overlap between a configuration $\sigma$ drawn from the Gibbs measure at $\beta_1$ and a configuration $\tau$ drawn from that at $\beta_2$:
\begin{equation}
P_2(\zeta)=\Bigl[\sum_{\sigma,\tau}
 \frac{e^{-\beta_1H(\sigma)}}{Z_1}\frac{e^{-\beta_2H(\tau)}}{Z_2}\,
 \delta\Bigl(\zeta-\frac1N\sum_i\sigma_i\tau_i\Bigr)\Bigr].
\label{eq:P2def}
\end{equation}
The factors $1/Z_1$ and $1/Z_2$ in the denominator can be removed with $1/Z_1=\lim_{n\to0}Z_1^{n-1}$ and $1/Z_2=\lim_{m\to0}Z_2^{m-1}$, so the right-hand side can be represented by the same $n+m$ replicas as Eq.~\eqref{eq:Z2rep}. The overlap in the argument of the delta function is one particular $s_{a\alpha}$, say $(a,\alpha)=(1,1)$. Since replica labels are arbitrary, we may average over all $nm$ pairs. Therefore
\begin{equation}
P_2(\zeta)=\lim_{n,m\to0}\frac1{nm}
 \sum_{a,\alpha}\delta(\zeta-s_{a\alpha})
\label{eq:P2rep}
\end{equation}
follows.

\subsubsection*{(b) Independent equilibrium clusters}

Before introducing the matched-pair sector, we restrict Eqs.~\eqref{eq:G2T} and \eqref{eq:Xi2T} to one temperature and substitute the 1RSB structure in order to isolate the contribution of a single independent equilibrium cluster. This is the standard 1RSB replica calculation, but we give the steps explicitly to make the comparison with a matched pair in part (c) transparent.

Setting $s_{a\alpha}=0$ completely decouples the two temperatures. Keeping only the $n$ replicas at temperature 1 gives the action
\begin{equation}
G_1=\frac14n\beta_1^2
-\frac{(p-1)\beta_1^2}{4}\sum_{a\ne b}q_{ab}^p
+\ln\operatorname{Tr}\exp\Bigl[\frac{p\beta_1^2}{4}\sum_{a\ne b}q_{ab}^{p-1}\sigma^a\sigma^b\Bigr].
\label{eq:G1T}
\end{equation}
We now impose the 1RSB ansatz in the SG sector: divide the $n$ replicas into $n/x$ blocks of size $x$ and set
\begin{equation}
q_{ab}=
\begin{cases}
 q_1 & (a\ne b\ \text{in the same block}),\\[2pt]
 q_0=0 & (\text{in different blocks})
\end{cases}
\label{eq:1RSBansatz}
\end{equation}
as usual.

In the polynomial part, there are $x(x-1)$ ordered pairs within each block and $n/x$ blocks, so
\begin{equation}
\sum_{a\ne b}q_{ab}^p=\frac nx\cdot x(x-1)\cdot q_1^p=n(x-1)q_1^p .
\label{eq:polyblock}
\end{equation}
For the exponential term, correcting for the diagonal contribution $(\sigma^a)^2=1$, we write
\begin{equation}
\frac{p\beta_1^2}{4}\sum_{a\ne b}q_{ab}^{p-1}\sigma^a\sigma^b
=\frac{\Lambda_1}{2}\sum_{B}\Bigl(\Bigl(\sum_{a\in B}\sigma^a\Bigr)^2-x\Bigr),
\qquad
\Lambda_1=\frac p2\beta_1^2q_1^{p-1}
\label{eq:Lam1}
\end{equation}
where $\sum_B$ runs over the blocks. The variance $\Lambda_1$ appears naturally at this stage as the coefficient of the conjugate field.

Because $q_0=0$, different blocks are uncoupled. Linearizing the square in each block by a Gaussian integral, the one-site trace factorizes completely over blocks:
\begin{equation}
\operatorname{Tr}\exp\Bigl[\frac{\Lambda_1}{2}\sum_B\Bigl\{\Bigl(\sum_{a\in B}\sigma^a\Bigr)^2-x\Bigr\}\Bigr]
=\prod_B\Bigl[e^{-\Lambda_1x/2}\int Dz\,\bigl(2\cosh\sqrt{\Lambda_1}z\bigr)^{x}\Bigr]
\equiv\bigl(I_1(x)\bigr)^{n/x}
\label{eq:blockfact}
\end{equation}
as shown.

Substituting Eqs.~\eqref{eq:polyblock} and \eqref{eq:blockfact} into Eq.~\eqref{eq:G1T}, $G_1$ separates into a term proportional to $n$ and a term proportional to the number of blocks $n/x$:
\begin{equation}
G_1=\frac14n\beta_1^2+\frac nx\,g_1(x),
\label{eq:G1block}
\end{equation}
\begin{equation}
g_1(x)=-\frac{(p-1)\beta_1^2}{4}x(x-1)q_1^p
+\ln I_1(x),\qquad
I_1(x)=e^{-\Lambda_1x/2}
 \int Dz\,\bigl(2\cosh\sqrt{\Lambda_1}z\bigr)^x.
\label{eq:g1}
\end{equation}
Thus $g_1(x)$ is the contribution of one 1RSB cluster at temperature 1 to the replicated action. In the limit $n\to0$, the free energy per spin is
$\varphi=\lim_{n\to0}G_1/n=\beta_1^2/4+g_1(x)/x$,
which agrees with $\varphi_{\SG}$ in Appendix~\ref{app:A1}(a), Eq.~\eqref{eq:gdef}. The corresponding quantity $g_2(x)$ at temperature 2 is obtained by the replacements $\beta_1\to\beta_2$ and $q_1\to q_2$.

We call the solution $(q_i,x_i)$ of the isolated one-temperature 1RSB equations $\partial\varphi/\partial q_1=\partial\varphi/\partial x=0$ the equilibrium values, and a block carrying these values an equilibrium cluster (equilibrium 1RSB cluster). The equilibrium values are determined independently at each temperature before the matched-pair sector is varied. The notation $g_1(x_1)$ and $g_2(x_2)$ below always refers to these on-shell equilibrium values.

\subsubsection*{(c) Matched pairs}

A matched pair consists of a block $B_1$ of $\tilde x_1$ replicas at temperature 1 paired with a block $B_2$ of $\tilde x_2$ replicas at temperature 2. We form $k$ such pairs; all remaining blocks stay as unpaired equilibrium clusters. By the ``interior of a pair'' we mean the set of $\tilde x_1+\tilde x_2$ replicas belonging to one such pair. Extending the one-temperature 1RSB ansatz~\eqref{eq:1RSBansatz}, we partition the full replica set into unpaired blocks and paired block pairs, and inside one pair set
\begin{equation}
q_{ab}=\tilde q_1\ \ (a\ne b\in B_1),\qquad
r_{\alpha\beta}=\tilde r_1\ \ (\alpha\ne\beta\in B_2),\qquad
s_{a\alpha}=s_1\ \ (a\in B_1,\ \alpha\in B_2)
\label{eq:pairansatz}
\end{equation}
with all overlaps to replicas outside the pair equal to zero ($q_0=r_0=0$ and $s=0$). Tildes emphasize that these are variational values defined only inside a pair and may differ from the equilibrium values $(q_1,x_1)$ and $(q_2,x_2)$ of unpaired blocks. We also allow $\tilde x_1\ne x_1$ and $\tilde x_2\ne x_2$. Thus all five quantities $(\tilde q_1,\tilde x_1;\,\tilde r_1,\tilde x_2;\,s_1)$ inside a pair are varied.

The structure is most transparent in the $(n+m)\times(n+m)$ overlap matrix
\begin{equation}
\begin{pmatrix} Q & S\\ S^{\mathsf T} & R\end{pmatrix},
\qquad Q:n\times n,\quad R:m\times m,\quad S:n\times m,
\label{eq:blockmat}
\end{equation}
where $Q$ contains $q_{ab}$, $R$ contains $r_{\alpha\beta}$, and $S$ contains the cross overlaps $s_{a\alpha}$. Compared with the one-temperature 1RSB matrix $Q$, the essential new information is which subblocks of $S$ are nonzero; these determine which blocks are paired.

As an example, take $n=m=6$, all block sizes equal to 3, and $k=1$. Divide temperature 1 into $B_1=\{\sigma^1,\sigma^2,\sigma^3\}$ and $B_1'=\{\sigma^4,\sigma^5,\sigma^6\}$, and temperature 2 into $B_2=\{\tau^1,\tau^2,\tau^3\}$ and $B_2'=\{\tau^4,\tau^5,\tau^6\}$. Pairing $B_1$ with $B_2$ gives
{\setlength{\arraycolsep}{2.2pt}
\begin{equation}
\left(\begin{array}{ccc|ccc||ccc|ccc}
0&\tilde q_1&\tilde q_1&&&&s_1&s_1&s_1&&&\\
\tilde q_1&0&\tilde q_1&&0&&s_1&s_1&s_1&&0&\\
\tilde q_1&\tilde q_1&0&&&&s_1&s_1&s_1&&&\\ \hline
&&&0&q_1&q_1&&&&&&\\
&0&&q_1&0&q_1&&0&&&0&\\
&&&q_1&q_1&0&&&&&&\\ \hline\hline
s_1&s_1&s_1&&&&0&\tilde r_1&\tilde r_1&&&\\
s_1&s_1&s_1&&0&&\tilde r_1&0&\tilde r_1&&0&\\
s_1&s_1&s_1&&&&\tilde r_1&\tilde r_1&0&&&\\ \hline
&&&&&&&&&0&r_1&r_1\\
&0&&&0&&&0&&r_1&0&r_1\\
&&&&&&&&&r_1&r_1&0
\end{array}\right).
\label{eq:examplemat}
\end{equation}}
All blank entries are zero, and the double lines separate $Q$, $R$, and $S$. The matrix is read as follows.
\begin{itemize}
\item The upper-left block of $Q$ (inside $B_1$) is paired and therefore has overlap $\tilde q_1$; the lower-right block (inside $B_1'$) is unpaired and retains the equilibrium value $q_1$; the overlap between the two blocks is $q_0=0$. The $R$ sector is analogous, with $\tilde r_1$, $r_1$, and $r_0=0$.
\item In $S$, only the $\tilde x_1\tilde x_2$ entries in $B_1\times B_2$ equal $s_1$; all other entries vanish. There is no diagonal zero inside this block because $\sigma$ and $\tau$ are different replica sets: $\sigma^1$ and $\tau^1$ are independent spin variables. This is the matrix version of the statement in part (a) that the cross term has no diagonal component.
\end{itemize}
The block sizes were chosen equal only for illustration. In general we allow $\tilde x_1\ne x_1$, $\tilde x_2\ne x_2$, and $\tilde x_1\ne\tilde x_2$.

An unpaired block is a block such as $B_1'$ or $B_2'$ in this example, for which the corresponding rows or columns of $S$ vanish identically, i.e., a block with no partner. Since it is not constrained to match a state at the other temperature, both its internal overlap and its block size remain at the one-temperature equilibrium values $(q_1,x_1)$, and $g_1(x_1)$ from part (b) is the contribution of one such block. By contrast, $B_1$ is constrained to have a partner at the other temperature, which biases the state selection and shifts the internal overlap to $\tilde q_1\ne q_1$ and the block size to $\tilde x_1\ne x_1$.

Which block is paired with which is arbitrary, and pairings related by a permutation of replica labels have the same action. Their number contributes only as a multiplicity. The dependence on this combinatorial factor will appear as a limitation of the criterion~\eqref{eq:criterion} below. Increasing $k$ simply places $k$ copies of the $\tilde q_1$ and $s_1$ block structure in Eq.~\eqref{eq:examplemat}; the action is therefore linear in $k$. The difference between the replicated action with $k$ matched pairs and that with no pairs can be written as $k\Delta$. We call $\Delta$ the action difference per matched pair; its explicit form is derived below in Eq.~\eqref{eq:Delta}.

The sign of $\Delta$ is the central quantity in this Appendix. Since the action is defined per spin, the statistical weight of a configuration containing one matched pair relative to the unpaired configuration is proportional to $e^{N\Delta}$. If the stationary value $\Delta^*$ is zero, a finite-overlap configuration with $s_1\ne0$ that pairs pure states at the two temperatures can occur without an exponential penalty, and $P_2(\zeta)$ may retain a peak at $\zeta=s_1^*>0$. If $\Delta^*<0$, its weight vanishes in the thermodynamic limit and the cross overlap collapses to $\zeta=0$. Thus, within this ansatz, the sign of $\Delta^*$ distinguishes correlated from uncorrelated configurations at the two temperatures and serves as a diagnostic of temperature chaos.

The five internal quantities of a pair, $(\tilde q_1,\tilde x_1)$, $(\tilde r_1,\tilde x_2)$, and $s_1$, are not prescribed. They are determined self-consistently by the stationarity conditions derived in part (d); in particular, $s_1$ is not chosen by hand. In the replica limit $n,m\to0$, the number of blocks $n/x_1$ itself tends to zero, and Eq.~\eqref{eq:examplemat} is only a formal illustration. The quantity that survives analytic continuation is the coefficient linear in $k$.

Setting $u_1=\sum_{a\in B_1}\sigma^a$ and $u_2=\sum_{\alpha\in B_2}\tau^\alpha$ and restricting Eq.~\eqref{eq:Xi2T} to a single pair, the same algebra as in Eq.~\eqref{eq:Lam1} gives
\begin{equation}
\Xi'\big|_{\text{pair}}
=\frac{\tilde\Lambda_1}{2}\bigl(u_1^2-\tilde x_1\bigr)
+\frac{\tilde\Lambda_2}{2}\bigl(u_2^2-\tilde x_2\bigr)
+\tilde\Lambda_{12}\,u_1u_2,
\label{eq:Xipair}
\end{equation}
where the coefficients are precisely the conjugate-field coefficients in Eq.~\eqref{eq:Xi2T}:
\begin{equation}
 \tilde\Lambda_1=\frac p2\beta_1^2\tilde q_1^{p-1},\qquad
 \tilde\Lambda_2=\frac p2\beta_2^2\tilde r_1^{p-1},\qquad
 \tilde\Lambda_{12}=\frac p2\beta_1\beta_2s_1^{p-1}.
\label{eq:Lamtilde}
\end{equation}
The cross term has no factor $1/2$ because $\sum_{a,\alpha}$ counts each pair only once, whereas $\sum_{a\ne b}$ counts an intratemperature pair twice.

To linearize this quadratic form, we extend the one-dimensional Gaussian transformation in part (b) to two variables:
\begin{equation}
\exp\Bigl[\frac12(u_1,u_2)\,\tilde\Lambda\binom{u_1}{u_2}\Bigr]
=\int D[h_1,h_2]\;e^{h_1u_1+h_2u_2},
\qquad
\tilde\Lambda\equiv
\begin{pmatrix}\tilde\Lambda_1&\tilde\Lambda_{12}\\[2pt] \tilde\Lambda_{12}&\tilde\Lambda_2\end{pmatrix}.
\label{eq:2dHS}
\end{equation}
The measure is the zero-mean two-dimensional Gaussian measure with covariance matrix $\tilde\Lambda$:
\begin{equation}
D[h_1,h_2]
=\frac{dh_1\,dh_2}{2\pi\sqrt{\det\tilde\Lambda}}\,
\exp\Bigl[-\frac12(h_1,h_2)\,\tilde\Lambda^{-1}\binom{h_1}{h_2}\Bigr],
\qquad
\det\tilde\Lambda=\tilde\Lambda_1\tilde\Lambda_2-\tilde\Lambda_{12}^2.
\label{eq:Dmeasure}
\end{equation}
Equivalently,
$\langle h_1\rangle=\langle h_2\rangle=0$,
$\langle h_1^2\rangle=\tilde\Lambda_1$,
$\langle h_2^2\rangle=\tilde\Lambda_2$,
$\langle h_1h_2\rangle=\tilde\Lambda_{12}$.
Using independent standard Gaussian variables $z_1,z_2$, one may write
\begin{equation}
h_1=\sqrt{\tilde\Lambda_1}\,z_1,\qquad
h_2=\sqrt{\tilde\Lambda_2}\bigl(\rho\,z_1+\sqrt{1-\rho^2}\,z_2\bigr),
\qquad
\rho=\frac{\tilde\Lambda_{12}}{\sqrt{\tilde\Lambda_1\tilde\Lambda_2}}
=\Bigl(\frac{s_1}{\sqrt{\tilde q_1\tilde r_1}}\Bigr)^{p-1}
\label{eq:rho}
\end{equation}
and $\int D[h_1,h_2]=\int Dz_1\!\int Dz_2$. The parameter $\rho$ is the correlation coefficient between the two Gaussian fields. Tracing over the spins in $B_1$ and $B_2$ produces $(2\cosh h_1)^{\tilde x_1}(2\cosh h_2)^{\tilde x_2}$, while the terms $-\tilde\Lambda_i\tilde x_i/2$ in Eq.~\eqref{eq:Xipair} give the prefactor $e^{-(\tilde\Lambda_1\tilde x_1+\tilde\Lambda_2\tilde x_2)/2}$.

For Eq.~\eqref{eq:2dHS} to hold, the covariance matrix must be positive semidefinite: $\det\tilde\Lambda\ge0$, or equivalently $|\rho|\le1$. Since Eq.~\eqref{eq:Lamtilde} gives $\tilde\Lambda\propto(\text{overlap})^{p-1}$, this is equivalent to $s_1^{2(p-1)}\le(\tilde q_1\tilde r_1)^{p-1}$, i.e., $|s_1|\le\sqrt{\tilde q_1\tilde r_1}$. The meaning of this upper bound is clearer from the definition of the overlaps in Eq.~\eqref{eq:overlapdef}. If $\langle\sigma_i\rangle_1$ and $\langle\tau_i\rangle_2$ are the site magnetizations within the two pure states forming a pair, then for replicas inside the pair Eq.~\eqref{eq:overlapdef} gives
\begin{equation}
\tilde q_1=\frac1N\sum_i\langle\sigma_i\rangle_1^2,\qquad
\tilde r_1=\frac1N\sum_i\langle\tau_i\rangle_2^2,\qquad
s_1=\frac1N\sum_i\langle\sigma_i\rangle_1\langle\tau_i\rangle_2.
\label{eq:overlapPS}
\end{equation}
The sums run over all $N$ sites, as in Eq.~\eqref{eq:overlapdef}; the factor $1/N$ is the normalization per spin and is unrelated to the number of matched pairs. Applying the Cauchy--Schwarz inequality to this form of $s_1$ gives the same upper bound. We refer to it below as the Cauchy--Schwarz bound.

The sign of $s_1$ may be restricted to the positive branch for the following reason. For odd $p$, $p-1$ is even, and hence $\rho$ in Eq.~\eqref{eq:rho} is nonnegative irrespective of the sign of $s_1$; hence $h_1$ and $h_2$ are always nonnegatively correlated. The right-hand side $\langle\tanh h_1\tanh h_2\rangle$ of the overlap equation~\eqref{eq:stat2T} is then nonnegative, so any stationary solution satisfies $s_1\ge0$ and restricting the search to $s_1\ge0$ is self-consistent. Moreover, for odd $p$ the Hamiltonian~\eqref{eq:H} is not invariant under the global reversal $\sigma\to-\sigma$, so there is no $\pm s_1$ degeneracy to begin with. For even $p$, solutions with $\rho<0$ also exist, but the polynomial part of Eq.~\eqref{eq:g12} depends on $s_1$ only through $s_1^p$ and is therefore even in $s_1$, while $\ln\tilde I$ is invariant under $\rho\to-\rho$ because $(2\cosh h_2)^{\tilde x_2}$ is even. Thus the action is invariant under $s_1\to-s_1$ by global spin-reversal symmetry, and the two signs represent the same solution. In either case it is sufficient to consider
\begin{equation}
0\le s_1\le\sqrt{\tilde q_1\tilde r_1}
\label{eq:CS}
\end{equation}
as the allowed range.

The contribution of one matched pair is therefore
\begin{align}
g_{12}={}&-\frac{p-1}{4}\Bigl(
 \beta_1^2\tilde x_1(\tilde x_1-1)\tilde q_1^p
 +\beta_2^2\tilde x_2(\tilde x_2-1)\tilde r_1^p
 +2\beta_1\beta_2\tilde x_1\tilde x_2s_1^p
 \Bigr)+\ln\tilde I,
\label{eq:g12}\\
\tilde I={}&e^{-(\tilde\Lambda_1\tilde x_1+\tilde\Lambda_2\tilde x_2)/2}
 \int D[h_1,h_2]\,
 (2\cosh h_1)^{\tilde x_1}(2\cosh h_2)^{\tilde x_2}.
\label{eq:Itilde}
\end{align}
We call all terms in $g_{12}$ other than $\ln\tilde I$, i.e., the algebraic terms containing powers of the overlaps, the polynomial part.

Because the action is additive over clusters, the action difference can be obtained by keeping only the pieces that change when a pair is formed. In the reference state with $k=0$, temperature 1 contains $n/x_1$ equilibrium clusters and temperature 2 contains $m/x_2$ equilibrium clusters; the two sets are independent. The cluster-dependent part of the action is therefore
\begin{equation}
G(0)=\frac{n}{x_1}g_1(x_1)+\frac{m}{x_2}g_2(x_2).
\label{eq:G0}
\end{equation}
Now form $k$ matched pairs. The number $n$ of replicas at temperature 1 is fixed. Of these, $k\tilde x_1$ belong to paired blocks, while the remaining $n-k\tilde x_1$ belong to equilibrium clusters of size $x_1$. Thus the number of equilibrium clusters on side 1 decreases from $n/x_1$ to $(n-k\tilde x_1)/x_1$, a loss of $\tilde x_1/x_1$ clusters per matched pair. The same reasoning gives $\tilde x_2/x_2$ on side 2. Each matched pair contributes $g_{12}$, so
\begin{equation}
G(k)=k\,g_{12}
+\Bigl(\frac{n}{x_1}-k\frac{\tilde x_1}{x_1}\Bigr)g_1(x_1)
+\Bigl(\frac{m}{x_2}-k\frac{\tilde x_2}{x_2}\Bigr)g_2(x_2).
\label{eq:Gk}
\end{equation}
Taking the difference cancels all terms proportional to $n$ and $m$, giving $G(k)-G(0)=k\Delta$ with
\begin{equation}
\Delta=g_{12}-\frac{\tilde x_1}{x_1}g_1(x_1)
              -\frac{\tilde x_2}{x_2}g_2(x_2)
\label{eq:Delta}
\end{equation}
as claimed.

The subtraction has a simple bookkeeping meaning. If one merely added $g_{12}$, the replicas used in the matched pair would be counted twice. Before pairing, those replicas contributed as parts of equilibrium clusters through $g_1$ and $g_2$, and these contributions must be removed. The factors $\tilde x_i/x_i$ arise because a pair uses $\tilde x_i$ replicas while one equilibrium cluster contains $x_i$ replicas. Thus $\Delta$ measures the change in the action produced by forming one matched pair relative to the reference state in which the two temperatures are treated independently. The quantity $g_{12}$ by itself is only the absolute contribution of one pair; without a reference state, its sign has no meaning.

Since $G(k)-G(0)$ is linear in $k$, only the coefficient $\Delta$ matters and there is no need to specify a particular value of $k$. This is the direct consequence of the linearity of the action in the number of matched pairs.

In terms of states, $s_1$ is the overlap between a configuration drawn from a pure state at $\beta_1$ and one drawn from a pure state at $\beta_2$. Its relation to the overlap distribution can be followed directly. Equation~\eqref{eq:P2rep} shows that $P_2$ is the histogram of the $nm$ cross overlaps $s_{a\alpha}$. In the matched-pair ansatz, among the entries of the $S$ block in Eq.~\eqref{eq:examplemat}, exactly $k\tilde x_1\tilde x_2$ are equal to $s_1$ and all others are zero. Hence
\begin{equation}
P_2(\zeta)=\bigl(1-w_{\rm pair}\bigr)\,\delta(\zeta)
 +w_{\rm pair}\,\delta(\zeta-s_1),
\qquad
w_{\rm pair}=\lim_{n,m\to0}\frac{\tilde x_1\tilde x_2}{nm}\,\langle k\rangle.
\label{eq:P2ansatz}
\end{equation}
Here $k$ is the number of matched pairs in a particular replica configuration and $\langle k\rangle$ denotes the average with respect to the sum over $k$ described below. Configurations with all possible values of $k$ contribute to $[Z_1^nZ_2^m]$, so $k$ need not be fixed. Since the action is linear in $k$ as in Eq.~\eqref{eq:Gk}, let $M(k)$ be the number of ways of choosing $k$ matched pairs, a combinatorial factor independent of $N$. Then
\begin{equation}
\bigl[Z_1^nZ_2^m\bigr]\simeq\sum_kM(k)\,e^{N(G(0)+k\Delta)}
\label{eq:ksum}
\end{equation}
and the distribution of $k$ is weighted by $e^{Nk\Delta}$. For $\Delta<0$, all $k\ge1$ sectors are exponentially suppressed and $\langle k\rangle\simeq[M(1)/M(0)]e^{N\Delta}$. Equation~\eqref{eq:P2ansatz} then suggests the exponential identification
$w_{\rm pair}\sim e^{N\Delta^*}$
for the weight of the nonzero-overlap peak.

This identification concerns only the leading exponential dependence. We do not prove that the prefactor $\lim_{n,m\to0}[M(1)/M(0)]\,\tilde x_1\tilde x_2/(nm)$ remains finite and nonzero. For example, a formal block-level counting would choose one block out of the $n/x_1$ blocks at temperature 1 and one out of the $m/x_2$ blocks at temperature 2, giving $M(1)/M(0)=nm/(x_1x_2)$. The factors $nm$ then cancel and the prefactor becomes the finite quantity $\tilde x_1\tilde x_2/(x_1x_2)$. In the replica limit, however, the block numbers $n/x_1$ and $m/x_2$ themselves tend to zero, and thus this counting is formal and requires an analytic continuation of the combinatorial coefficients $M(k)$, which we do not justify here. We therefore use $\Delta^*$ only as the exponential suppression rate of the nonzero-overlap peak within this restricted sector. See also Appendix~\ref{app:F5}(iv).

The criterion within the present ansatz is therefore
\begin{equation}
\begin{split}
&s_1^*>0,\quad \Delta^*=0
   \quad\Longrightarrow\quad
   w_{\rm pair}\ \text{is not exponentially suppressed},\\
&\Delta^*<0
   \quad\Longrightarrow\quad
   w_{\rm pair}\sim e^{N\Delta^*}\longrightarrow0
\end{split}
\label{eq:criterion}
\end{equation}
as written. We call the case $\Delta^*=0$ zero cost.

If $\Delta^*=0$, a configuration that pairs pure states at $\beta_1$ and $\beta_2$ with a finite overlap $s_1^*$ carries no exponential penalty, and the overlap distribution $P_2$ can retain a peak at $\zeta=s_1^*>0$. This is the picture expected in the absence of temperature chaos. If $\Delta^*<0$, the weight of this peak vanishes as $N\to\infty$ and, within the present ansatz, the weight of $P_2$ is driven to $\zeta=0$, as expected in the presence of temperature chaos. The conclusion $\Delta^*<0$, however, excludes only the peak associated with the particular type of matching explored by this ansatz. The limitations of the ansatz and of identifying $\Delta^*$ with a large-deviation rate for $P_2$ are discussed in Appendix~\ref{app:F5}.

Zero cost is a necessary but not a sufficient condition for the absence of chaos. If $\Delta=0$ in Eq.~\eqref{eq:ksum}, all $k$ sectors have the same exponential weight, and $w_{\rm pair}$ is controlled entirely by the prefactor above, which need not be $O(1)$. The REM is an explicit example: only when the exact result that the nonzero-overlap peak has $O(1)$ weight \cite{Derrida2021} is combined with $\Delta^*=0$ in the $p\to\infty$ limit can one conclude that temperature chaos is absent.

\subsubsection*{(d) Stationarity equations}

The stationarity conditions follow by differentiating $\Delta$ with respect to the five variational parameters $(\tilde q_1,\tilde r_1,s_1,\tilde x_1,\tilde x_2)$. For later use, define the unnormalized integral with respect to the Gaussian measure $D[h_1,h_2]$ and the normalized average weighted by $(2\cosh h_1)^{\tilde x_1}(2\cosh h_2)^{\tilde x_2}$ as
\begin{equation}
\mathcal J[F]\equiv\int D[h_1,h_2]\,F(h_1,h_2),
\qquad
\langle F\rangle\equiv
\frac{\mathcal J\bigl[(2\cosh h_1)^{\tilde x_1}(2\cosh h_2)^{\tilde x_2}F\bigr]}
     {\mathcal J\bigl[(2\cosh h_1)^{\tilde x_1}(2\cosh h_2)^{\tilde x_2}\bigr]}.
\label{eq:avgdef}
\end{equation}
The denominator in $\langle\cdot\rangle$ is the integral in $\tilde I$ [Eq.~\eqref{eq:Itilde}] without the prefactor $e^{-(\tilde\Lambda_1\tilde x_1+\tilde\Lambda_2\tilde x_2)/2}$. In terms of the independent standard Gaussian variables in Eq.~\eqref{eq:rho},
\begin{equation}
\langle F\rangle=
\frac{\displaystyle\int\! Dz_1\!\int\! Dz_2\,
 (2\cosh h_1)^{\tilde x_1}(2\cosh h_2)^{\tilde x_2}F}
     {\displaystyle\int\! Dz_1\!\int\! Dz_2\,
 (2\cosh h_1)^{\tilde x_1}(2\cosh h_2)^{\tilde x_2}}
\label{eq:avgexplicit}
\end{equation}
with $h_1=\sqrt{\tilde\Lambda_1}z_1$ and $h_2=\sqrt{\tilde\Lambda_2}(\rho z_1+\sqrt{1-\rho^2}z_2)$.

Consider first the overlap variables $(\tilde q_1,\tilde r_1,s_1)$. The logarithm $\ln\tilde I$ depends on them only through the covariance parameters $(\tilde\Lambda_1,\tilde\Lambda_2,\tilde\Lambda_{12})$, because the measure~\eqref{eq:Dmeasure} is specified entirely by the covariance and the integrand has no additional explicit dependence on the overlaps. Thus
\begin{equation}
\frac{\partial\Delta}{\partial\tilde q_1}
=\frac{\partial(\text{polynomial part})}{\partial\tilde q_1}
+\frac{\partial\ln\tilde I}{\partial\tilde\Lambda_1}
 \frac{\partial\tilde\Lambda_1}{\partial\tilde q_1},
\label{eq:chain}
\end{equation}
where the factors are
\begin{equation}
\frac{\partial(\text{polynomial part})}{\partial\tilde q_1}
=-\frac{p(p-1)\beta_1^2}{4}\,\tilde x_1(\tilde x_1-1)\,\tilde q_1^{p-1},
\qquad
\frac{\partial\tilde\Lambda_1}{\partial\tilde q_1}
=\frac{p(p-1)}{2}\beta_1^2\tilde q_1^{p-2},
\label{eq:chainfactors}
\end{equation}
and similarly for $\tilde r_1$ and $s_1$; the latter enters through $\tilde\Lambda_{12}$. Derivatives with respect to the covariance can be evaluated with Gaussian integration by parts for $\mathcal J$:
\begin{equation}
\frac{\partial\mathcal J[F]}{\partial\tilde\Lambda_1}
=\frac12\,\mathcal J\Bigl[\frac{\partial^2F}{\partial h_1^2}\Bigr],
\qquad
\frac{\partial\mathcal J[F]}{\partial\tilde\Lambda_{12}}
=\mathcal J\Bigl[\frac{\partial^2F}{\partial h_1\partial h_2}\Bigr].
\label{eq:gaussIBP}
\end{equation}
To see this directly, write the measure~\eqref{eq:Dmeasure} in Fourier form. With $\boldsymbol h=(h_1,h_2)^{\mathsf T}$ and $\boldsymbol k=(k_1,k_2)^{\mathsf T}$,
\begin{equation}
D[h_1,h_2]=dh_1dh_2\int\frac{d^2k}{(2\pi)^2}
 \exp\Bigl[-\frac12\boldsymbol k^{\mathsf T}\tilde\Lambda\boldsymbol k
  +i\boldsymbol k\!\cdot\!\boldsymbol h\Bigr],
\qquad
\boldsymbol k^{\mathsf T}\tilde\Lambda\boldsymbol k
 =\tilde\Lambda_1k_1^2+2\tilde\Lambda_{12}k_1k_2+\tilde\Lambda_2k_2^2.
\label{eq:Dfourier}
\end{equation}
This is the inverse transform of the characteristic function $e^{-\boldsymbol k^{\mathsf T}\tilde\Lambda\boldsymbol k/2}$ of the Gaussian measure. In the exponent, $\tilde\Lambda_1$ appears only as $-\tilde\Lambda_1k_1^2/2$, so $\partial/\partial\tilde\Lambda_1$ brings down $-k_1^2/2$, whereas $\partial^2/\partial h_1^2$ brings down $(ik_1)^2=-k_1^2$. Thus the Gaussian measure itself satisfies the diffusion equations
\begin{equation}
\frac{\partial}{\partial\tilde\Lambda_1}D[h_1,h_2]
=\frac12\frac{\partial^2}{\partial h_1^2}D[h_1,h_2],
\qquad
\frac{\partial}{\partial\tilde\Lambda_{12}}D[h_1,h_2]
=\frac{\partial^2}{\partial h_1\partial h_2}D[h_1,h_2].
\label{eq:heateq}
\end{equation}
There is no factor $1/2$ for the cross covariance because $\tilde\Lambda_{12}$ appears in the exponent as $-\tilde\Lambda_{12}k_1k_2$. Substituting these identities into $\mathcal J[F]$ and integrating by parts twice transfers the derivatives to $F$ and gives Eq.~\eqref{eq:gaussIBP}. The boundary terms vanish because the Gaussian measure decays quadratically in the exponent while $F$ grows at most exponentially. For $F=(2\cosh h_1)^{\tilde x_1}(2\cosh h_2)^{\tilde x_2}$,
\begin{equation}
\frac{\partial^2F}{\partial h_1^2}
=\bigl(\tilde x_1\,\mathrm{sech}^2h_1+\tilde x_1^2\tanh^2h_1\bigr)F,
\qquad
\frac{\partial^2F}{\partial h_1\partial h_2}
=\tilde x_1\tilde x_2\tanh h_1\tanh h_2\,F.
\label{eq:d2F}
\end{equation}
Dividing by $\mathcal J[F]$ then produces the normalized averages in Eq.~\eqref{eq:avgdef}. Adding the derivative $-\tilde x_1/2$ of the prefactor in $\ln\tilde I=-(\tilde\Lambda_1\tilde x_1+\tilde\Lambda_2\tilde x_2)/2+\ln\mathcal J[F]$, and using $\mathrm{sech}^2=1-\tanh^2$, gives
\begin{equation}
\frac{\partial\ln\tilde I}{\partial\tilde\Lambda_1}
=-\frac{\tilde x_1}{2}
+\frac{\tilde x_1}{2}\bigl(1-\langle\tanh^2h_1\rangle\bigr)
+\frac{\tilde x_1^2}{2}\langle\tanh^2h_1\rangle
=\frac{\tilde x_1(\tilde x_1-1)}{2}\langle\tanh^2h_1\rangle
\label{eq:dlnI}
\end{equation}
where the $\tilde x_1/2$ terms cancel exactly. For the cross covariance there is no prefactor term, so
\begin{equation}
\frac{\partial\ln\tilde I}{\partial\tilde\Lambda_{12}}
=\tilde x_1\tilde x_2\,\langle\tanh h_1\tanh h_2\rangle
\label{eq:dlnI12}
\end{equation}
follows. Substituting Eq.~\eqref{eq:dlnI} into Eq.~\eqref{eq:chain} factors out the same prefactor from the polynomial and Gaussian terms and gives
\begin{equation}
\frac{\partial\Delta}{\partial\tilde q_1}
=\frac{p(p-1)\beta_1^2}{4}\,\tilde x_1(\tilde x_1-1)\,\tilde q_1^{p-2}
\bigl(\,\langle\tanh^2h_1\rangle-\tilde q_1\,\bigr).
\label{eq:dDelta_dq}
\end{equation}
The derivatives with respect to $\tilde r_1$ and $s_1$ are obtained in the same way. Since $s_1$ enters through $\tilde\Lambda_{12}$, Eq.~\eqref{eq:dlnI12} is used; the common prefactor is then $p(p-1)\beta_1\beta_2\tilde x_1\tilde x_2s_1^{p-2}/2$. The three overlap stationarity equations therefore close as
\begin{equation}
\tilde q_1=\langle\tanh^2h_1\rangle,
\qquad
\tilde r_1=\langle\tanh^2h_2\rangle,
\qquad
s_1=\langle\tanh h_1\tanh h_2\rangle.
\label{eq:stat2T}
\end{equation}

By contrast, $\tilde x_1$ and $\tilde x_2$ appear directly in the exponent and prefactor of $\tilde I$, not through the covariance, so their derivatives have a different form. The resulting equations balance the breakpoint-type derivative of $g_{12}$ against the subtraction terms in Eq.~\eqref{eq:Delta}. On the temperature-1 side,
\begin{equation}
-\frac{p-1}{4}\Bigl(
 \beta_1^2(2\tilde x_1-1)\tilde q_1^p
 +2\beta_1\beta_2\tilde x_2s_1^p
 \Bigr)
-\frac{\tilde\Lambda_1}{2}
+\langle\ln(2\cosh h_1)\rangle
=\frac{g_1(x_1)}{x_1},
\label{eq:breakpoint2T}
\end{equation}
and the temperature-2 equation is obtained by exchanging $1\leftrightarrow2$. In what follows, only points satisfying all five stationarity conditions, i.e., the three equations in Eq.~\eqref{eq:stat2T}, Eq.~\eqref{eq:breakpoint2T}, and its $1\leftrightarrow2$ counterpart, are considered as candidates for $\Delta^*$.

The solution $s_1=0$ is simply the unpaired configuration itself. Then $(\tilde q_1,\tilde x_1,\tilde r_1,\tilde x_2)=(q_1,x_1,q_2,x_2)$, so each block returns to an independent equilibrium cluster and $\Delta=0$ by definition. For $p\ge3$, $\partial\Delta/\partial s_1\propto s_1^{p-2}$ also vanishes at $s_1=0$, so this is a stationary point. But this configuration corresponds precisely to uncorrelated states at the two temperatures, i.e., to the chaotic situation. Therefore the zero cost at $s_1=0$ is not evidence for the absence of chaos. What is required for a nonchaotic interpretation is a zero-cost stationary point with nonzero overlap, which is why the criterion~\eqref{eq:criterion} explicitly requires $s_1^*>0$.

\subsection{Equal-temperature limit}\label{app:F3}

Set $\beta_1=\beta_2$ and split a single equilibrium 1RSB cluster into two parts according to
\begin{equation}
 \tilde q_1=\tilde r_1=s_1=q_1,
 \qquad
 \tilde x_1+\tilde x_2=x_1.
\end{equation}
The covariance saturates and $h_1=h_2$, while the spin-trace weight becomes
$(2\cosh h_1)^{\tilde x_1+\tilde x_2}=(2\cosh h_1)^{x_1}$
which is the original one-temperature weight. In addition,
\begin{equation}
\tilde x_1(\tilde x_1-1)+\tilde x_2(\tilde x_2-1)
+2\tilde x_1\tilde x_2=x_1(x_1-1)
\end{equation}
and consequently the polynomial part also reduces to that of the equilibrium cluster. Hence
\begin{equation}
 g_{12}=g_1(x_1),\qquad
 \Delta=g_1(x_1)\left(1-\frac{\tilde x_1+\tilde x_2}{x_1}\right)=0
\end{equation}
holds algebraically, and all five stationarity equations reduce to the ordinary equilibrium 1RSB equations. As long as $\tilde x_1+\tilde x_2=x_1$ is kept fixed, $\Delta=0$ independently of the splitting ratio $\tilde x_1/x_1$; this ratio is a flat direction of $\Delta$. Physically, it expresses that assigning the replicas of the same state to the two replica species in different proportions has no effect when the temperatures are equal. Since this zero-cost structure requires $\tilde x_1+\tilde x_2=x_1$, fixing $\tilde x_1$ and $\tilde x_2$ separately to the equilibrium values $x_1$ and $x_2$ would fail even in the equal-temperature limit. It is therefore essential to treat $\tilde x_1$ and $\tilde x_2$ as variational variables.

For $p=3$, the numerical implementation reproduces this zero-cost structure with $|\Delta|\simeq4\times10^{-10}$. This provides a numerical consistency check in the equal-temperature limit. The values $|\Delta^*|\sim10^{-5}$--$10^{-4}$ obtained for unequal-temperature pairs at the same $p=3$ (Table~\ref{tab:delta}) are more than four orders of magnitude larger. The negative cost at unequal temperatures is therefore not a numerical artifact but a finite effect that separates continuously from the zero-cost equal-temperature limit.

\subsection{Finite-$p$ results}\label{app:F4}

For $p=3$, we choose both temperatures inside the 1RSB-SG phase and solve the five stationarity equations to residuals below $10^{-13}$. The results are shown in Table~\ref{tab:delta}. A high-overlap stationary point with $s_1^*>0$ is found for every temperature pair, and $s_1^*$ lies just below the Cauchy--Schwarz bound. The stationary cost, however, is always negative and its magnitude increases with the temperature difference.

\begin{table}[htb]
\centering
\caption{Stationary points in the matched-pair sector for $p=3$. For unequal temperatures, $\Delta^*<0$ is obtained systematically. The cross overlap $s_1^*$ lies just below the Cauchy--Schwarz bound $\sqrt{\tilde q_1\tilde r_1}$.}
\label{tab:delta}
\begin{tabular}{ccccc}
\toprule
$T_1$ & $T_2$ & $s_1^*$ & $\sqrt{\tilde q_1\tilde r_1}$ & $\Delta^*$ \\
\midrule
0.45 & 0.40 & 0.92029 & 0.92083 & $-6.79\times10^{-6}$ \\
0.45 & 0.35 & 0.92569 & 0.92796 & $-2.20\times10^{-5}$ \\
0.50 & 0.35 & 0.91356 & 0.91828 & $-5.71\times10^{-5}$ \\
0.55 & 0.35 & 0.89932 & 0.90721 & $-1.15\times10^{-4}$ \\
0.60 & 0.40 & 0.87972 & 0.88716 & $-1.59\times10^{-4}$ \\
0.60 & 0.35 & 0.88273 & 0.89454 & $-2.03\times10^{-4}$ \\
\bottomrule
\end{tabular}
\end{table}

According to the criterion~\eqref{eq:criterion}, this matched-pair stationary point has a negative action difference relative to the independent sector, and the nonzero-overlap peak is exponentially suppressed within the present ansatz. If one uses only the exponential factor $e^{N\Delta^*}$ as a guide, the system size at which the suppression becomes appreciable is, for $p=3$,
\begin{equation}
 N^*\sim\frac1{|\Delta^*|}\sim10^4\text{--}10^5,
\end{equation}
which is extremely large on the exponential scale. At the sizes accessible to ordinary numerical studies, the matched-pair peak may therefore remain almost intact and the system may appear nonchaotic. The matched-pair sector is nevertheless on the chaotic side of the criterion, with an exceptionally small cost $|\Delta^*|$. This is qualitatively similar to the very weak temperature chaos of the SK model, which first appears at ninth order in perturbation theory \cite{Rizzo2003}.

Table~\ref{tab:deltap} shows the result when $p$ is increased at fixed reduced temperatures $(T_1/T_c,T_2/T_c)=(0.70,0.50)$. The value of $\Delta^*$ approaches zero from below, decreasing by nearly three orders of magnitude over the displayed range, while $s_1^*\to1$ ($s_1^*=0.99997$ at $p=12$).

\begin{table}[htb]
\centering
\caption{$p$ dependence of the stationary cost $\Delta^*$ at fixed reduced temperatures $(T_1/T_c,T_2/T_c)=(0.70,0.50)$.}
\label{tab:deltap}
\begin{tabular}{cc}
\toprule
$p$ & $\Delta^*$ \\
\midrule
3 & $-3.35\times10^{-5}$ \\
4 & $-2.13\times10^{-5}$ \\
5 & $-9.83\times10^{-6}$ \\
6 & $-4.42\times10^{-6}$ \\
8 & $-9.29\times10^{-7}$ \\
10 & $-2.05\times10^{-7}$ \\
12 & $-4.65\times10^{-8}$ \\
\bottomrule
\end{tabular}
\end{table}

The present model reduces to the REM as $p\to\infty$ \cite{Derrida1981,Gross1984}, and the REM is known to have no temperature chaos \cite{Derrida2021}. The trends $\Delta^*\to0^-$ and $s_1^*\to1$ therefore describe a continuous disappearance of the weak exponential suppression found at finite $p$ as the nonchaotic REM limit is approached. This behavior suggests a noncommutativity of the limits $N\to\infty$ and $p\to\infty$, with a crossover size $N^*(p)=1/|\Delta_p^*|$ that grows rapidly with $p$.

The negative cost found here is consistent with temperature chaos at finite $p$. The conclusion, however, applies only to the high-overlap one-to-one matching sector. No zero-cost point is found even when the low-overlap stationary branch mentioned below is included, but we have not exhausted all hierarchical cross-overlap structures.

For cross-overlap structures not explored here, the main result of the paper gives one further constraint. If one accepts the phase-boundary calculation of Sec.~\ref{sec:expansion} together with the existence of a finite-temperature SG phase, the contrapositive of the proposition requires at least one pair of distinct temperatures in the SG phase for which $P_2(x\mid\beta_1,\beta_2)=\delta(x)$. For that pair, no structure, including hierarchical cross blocks, can support a nonzero-overlap peak with $O(1)$ weight. Thus the main result itself guarantees, for at least one temperature pair, the absence of any missed zero-cost structure capable of carrying a finite-weight nonzero-overlap peak. This is nevertheless the nonconstructive existence statement emphasized in Sec.~\ref{sec:chaos}: it does not identify the temperature pair and does not apply to every pair of distinct temperatures. See Sec.~\ref{sec:chaoslogic} for the physically natural, more uniform interpretation.

\subsection{Limitations and qualifications of the two-temperature calculation}\label{app:F5}

Four limitations remain concerning the completeness of the saddle-point search and its interpretation. (i) Stationary points other than the high-overlap branch exist; for $(T_1,T_2)=(0.50,0.35)$, for example, a low-overlap branch with $s_1=0.359$ and $\Delta^*=-1.64\times10^{-2}$ is found, but it is even more strongly suppressed and does not provide a zero-cost peak. (ii) The present ansatz assumes a one-to-one correspondence of states; state splitting or merging would require hierarchical cross blocks. (iii) Fluctuations of the block size that appear in the exact REM analysis \cite{Derrida2021} are not included in the finite-$p$ calculation. (iv) The relation between $\Delta^*$ and $P_2$ passes through a formal analytic continuation of the combinatorial factor for matched pairs, as explained in Appendix~\ref{app:F2}(c), and therefore does not constitute a complete replica derivation identifying $\Delta^*$ with the large-deviation rate of the nonzero-overlap peak. A direct derivation would require constructing a constrained two-temperature free energy at fixed cross overlap and establishing a large-deviation relation of the form $P_N(\zeta)\sim e^{N\Delta(\zeta)}$. This remains an open problem.

Accordingly, this appendix establishes the following limited result: for finite $p$, the most natural matched-pair stationary point representing the absence of chaos has nonzero overlap but a negative action difference and is therefore exponentially suppressed in the thermodynamic limit. This is numerical evidence supporting the conclusion of the main text, but it is not a classification of all possible two-temperature replica saddle points. It is also consistent with the free-energy level crossings obtained for $p=3$ by the state-following method of Ref.~\cite{ZdeborovaKrzakala2010PRB}. These qualifications concern the completeness of the saddle-point search; the existence statement about chaos itself follows independently from the contrapositive of the reentrance result, as explained in Sec.~\ref{sec:chaos}.

\makeatletter\expandafter\def\csname b@apsrev42Control\endcsname{}\makeatother
\nocite{apsrev42Control}
\bibliographystyle{apsrev4-2}
%

\end{document}